\documentclass[twocolumn,tighten]{aastex61}
\usepackage[]{txfonts}
\usepackage{natbib, twoopt}
\usepackage{float}
\usepackage{graphicx}
\usepackage{epstopdf}
\usepackage{enumitem}
\usepackage[caption=false]{subfig}
\usepackage{color}\definecolor{AliceBlue}{rgb}{0.94,0.97,1.00}
\definecolor{AntiqueWhite1}{rgb}{1.00,0.94,0.86}
\definecolor{AntiqueWhite2}{rgb}{0.93,0.87,0.80}
\definecolor{AntiqueWhite3}{rgb}{0.80,0.75,0.69}
\definecolor{AntiqueWhite4}{rgb}{0.55,0.51,0.47}
\definecolor{AntiqueWhite}{rgb}{0.98,0.92,0.84}
\definecolor{BlanchedAlmond}{rgb}{1.00,0.92,0.80}
\definecolor{BlueViolet}{rgb}{0.54,0.17,0.89}
\definecolor{CadetBlue1}{rgb}{0.60,0.96,1.00}
\definecolor{CadetBlue2}{rgb}{0.56,0.90,0.93}
\definecolor{CadetBlue3}{rgb}{0.48,0.77,0.80}
\definecolor{CadetBlue4}{rgb}{0.33,0.53,0.55}
\definecolor{CadetBlue}{rgb}{0.37,0.62,0.63}
\definecolor{CornflowerBlue}{rgb}{0.39,0.58,0.93}
\definecolor{DarkBlue}{rgb}{0.00,0.00,0.55}
\definecolor{DarkCyan}{rgb}{0.00,0.55,0.55}
\definecolor{DarkGoldenrod1}{rgb}{1.00,0.73,0.06}
\definecolor{DarkGoldenrod2}{rgb}{0.93,0.68,0.05}
\definecolor{DarkGoldenrod3}{rgb}{0.80,0.58,0.05}
\definecolor{DarkGoldenrod4}{rgb}{0.55,0.40,0.03}
\definecolor{DarkGoldenrod}{rgb}{0.72,0.53,0.04}
\definecolor{DarkGray}{rgb}{0.66,0.66,0.66}
\definecolor{DarkGreen}{rgb}{0.00,0.39,0.00}
\definecolor{DarkGrey}{rgb}{0.66,0.66,0.66}
\definecolor{DarkKhaki}{rgb}{0.74,0.72,0.42}
\definecolor{DarkMagenta}{rgb}{0.55,0.00,0.55}
\definecolor{DarkOliveGreen1}{rgb}{0.79,1.00,0.44}
\definecolor{DarkOliveGreen2}{rgb}{0.74,0.93,0.41}
\definecolor{DarkOliveGreen3}{rgb}{0.64,0.80,0.35}
\definecolor{DarkOliveGreen4}{rgb}{0.43,0.55,0.24}
\definecolor{DarkOliveGreen}{rgb}{0.33,0.42,0.18}
\definecolor{DarkOrange1}{rgb}{1.00,0.50,0.00}
\definecolor{DarkOrange2}{rgb}{0.93,0.46,0.00}
\definecolor{DarkOrange3}{rgb}{0.80,0.40,0.00}
\definecolor{DarkOrange4}{rgb}{0.55,0.27,0.00}
\definecolor{DarkOrange}{rgb}{1.00,0.55,0.00}
\definecolor{DarkOrchid1}{rgb}{0.75,0.24,1.00}
\definecolor{DarkOrchid2}{rgb}{0.70,0.23,0.93}
\definecolor{DarkOrchid3}{rgb}{0.60,0.20,0.80}
\definecolor{DarkOrchid4}{rgb}{0.41,0.13,0.55}
\definecolor{DarkOrchid}{rgb}{0.60,0.20,0.80}
\definecolor{DarkRed}{rgb}{0.55,0.00,0.00}
\definecolor{DarkSalmon}{rgb}{0.91,0.59,0.48}
\definecolor{DarkSeaGreen1}{rgb}{0.76,1.00,0.76}
\definecolor{DarkSeaGreen2}{rgb}{0.71,0.93,0.71}
\definecolor{DarkSeaGreen3}{rgb}{0.61,0.80,0.61}
\definecolor{DarkSeaGreen4}{rgb}{0.41,0.55,0.41}
\definecolor{DarkSeaGreen}{rgb}{0.56,0.74,0.56}
\definecolor{DarkSlateBlue}{rgb}{0.28,0.24,0.55}
\definecolor{DarkSlateGray1}{rgb}{0.59,1.00,1.00}
\definecolor{DarkSlateGray2}{rgb}{0.55,0.93,0.93}
\definecolor{DarkSlateGray3}{rgb}{0.47,0.80,0.80}
\definecolor{DarkSlateGray4}{rgb}{0.32,0.55,0.55}
\definecolor{DarkSlateGray}{rgb}{0.18,0.31,0.31}
\definecolor{DarkSlateGrey}{rgb}{0.18,0.31,0.31}
\definecolor{DarkTurquoise}{rgb}{0.00,0.81,0.82}
\definecolor{DarkViolet}{rgb}{0.58,0.00,0.83}
\definecolor{DeepPink1}{rgb}{1.00,0.08,0.58}
\definecolor{DeepPink2}{rgb}{0.93,0.07,0.54}
\definecolor{DeepPink3}{rgb}{0.80,0.06,0.46}
\definecolor{DeepPink4}{rgb}{0.55,0.04,0.31}
\definecolor{DeepPink}{rgb}{1.00,0.08,0.58}
\definecolor{DeepSkyBlue1}{rgb}{0.00,0.75,1.00}
\definecolor{DeepSkyBlue2}{rgb}{0.00,0.70,0.93}
\definecolor{DeepSkyBlue3}{rgb}{0.00,0.60,0.80}
\definecolor{DeepSkyBlue4}{rgb}{0.00,0.41,0.55}
\definecolor{DeepSkyBlue}{rgb}{0.00,0.75,1.00}
\definecolor{DimGray}{rgb}{0.41,0.41,0.41}
\definecolor{DimGrey}{rgb}{0.41,0.41,0.41}
\definecolor{DodgerBlue1}{rgb}{0.12,0.56,1.00}
\definecolor{DodgerBlue2}{rgb}{0.11,0.53,0.93}
\definecolor{DodgerBlue3}{rgb}{0.09,0.45,0.80}
\definecolor{DodgerBlue4}{rgb}{0.06,0.31,0.55}
\definecolor{DodgerBlue}{rgb}{0.12,0.56,1.00}
\definecolor{FloralWhite}{rgb}{1.00,0.98,0.94}
\definecolor{ForestGreen}{rgb}{0.13,0.55,0.13}
\definecolor{GhostWhite}{rgb}{0.97,0.97,1.00}
\definecolor{GreenYellow}{rgb}{0.68,1.00,0.18}
\definecolor{HotPink1}{rgb}{1.00,0.43,0.71}
\definecolor{HotPink2}{rgb}{0.93,0.42,0.65}
\definecolor{HotPink3}{rgb}{0.80,0.38,0.56}
\definecolor{HotPink4}{rgb}{0.55,0.23,0.38}
\definecolor{HotPink}{rgb}{1.00,0.41,0.71}
\definecolor{IndianRed1}{rgb}{1.00,0.42,0.42}
\definecolor{IndianRed2}{rgb}{0.93,0.39,0.39}
\definecolor{IndianRed3}{rgb}{0.80,0.33,0.33}
\definecolor{IndianRed4}{rgb}{0.55,0.23,0.23}
\definecolor{IndianRed}{rgb}{0.80,0.36,0.36}
\definecolor{LavenderBlush1}{rgb}{1.00,0.94,0.96}
\definecolor{LavenderBlush2}{rgb}{0.93,0.88,0.90}
\definecolor{LavenderBlush3}{rgb}{0.80,0.76,0.77}
\definecolor{LavenderBlush4}{rgb}{0.55,0.51,0.53}
\definecolor{LavenderBlush}{rgb}{1.00,0.94,0.96}
\definecolor{LawnGreen}{rgb}{0.49,0.99,0.00}
\definecolor{LemonChiffon1}{rgb}{1.00,0.98,0.80}
\definecolor{LemonChiffon2}{rgb}{0.93,0.91,0.75}
\definecolor{LemonChiffon3}{rgb}{0.80,0.79,0.65}
\definecolor{LemonChiffon4}{rgb}{0.55,0.54,0.44}
\definecolor{LemonChiffon}{rgb}{1.00,0.98,0.80}
\definecolor{LightBlue1}{rgb}{0.75,0.94,1.00}
\definecolor{LightBlue2}{rgb}{0.70,0.87,0.93}
\definecolor{LightBlue3}{rgb}{0.60,0.75,0.80}
\definecolor{LightBlue4}{rgb}{0.41,0.51,0.55}
\definecolor{LightBlue}{rgb}{0.68,0.85,0.90}
\definecolor{LightCoral}{rgb}{0.94,0.50,0.50}
\definecolor{LightCyan1}{rgb}{0.88,1.00,1.00}
\definecolor{LightCyan2}{rgb}{0.82,0.93,0.93}
\definecolor{LightCyan3}{rgb}{0.71,0.80,0.80}
\definecolor{LightCyan4}{rgb}{0.48,0.55,0.55}
\definecolor{LightCyan}{rgb}{0.88,1.00,1.00}
\definecolor{LightGoldenrod1}{rgb}{1.00,0.93,0.55}
\definecolor{LightGoldenrod2}{rgb}{0.93,0.86,0.51}
\definecolor{LightGoldenrod3}{rgb}{0.80,0.75,0.44}
\definecolor{LightGoldenrod4}{rgb}{0.55,0.51,0.30}
\definecolor{LightGoldenrodYellow}{rgb}{0.98,0.98,0.82}
\definecolor{LightGoldenrod}{rgb}{0.93,0.87,0.51}
\definecolor{LightGray}{rgb}{0.83,0.83,0.83}
\definecolor{LightGreen}{rgb}{0.56,0.93,0.56}
\definecolor{LightGrey}{rgb}{0.83,0.83,0.83}
\definecolor{LightPink1}{rgb}{1.00,0.68,0.73}
\definecolor{LightPink2}{rgb}{0.93,0.64,0.68}
\definecolor{LightPink3}{rgb}{0.80,0.55,0.58}
\definecolor{LightPink4}{rgb}{0.55,0.37,0.40}
\definecolor{LightPink}{rgb}{1.00,0.71,0.76}
\definecolor{LightSalmon1}{rgb}{1.00,0.63,0.48}
\definecolor{LightSalmon2}{rgb}{0.93,0.58,0.45}
\definecolor{LightSalmon3}{rgb}{0.80,0.51,0.38}
\definecolor{LightSalmon4}{rgb}{0.55,0.34,0.26}
\definecolor{LightSalmon}{rgb}{1.00,0.63,0.48}
\definecolor{LightSeaGreen}{rgb}{0.13,0.70,0.67}
\definecolor{LightSkyBlue1}{rgb}{0.69,0.89,1.00}
\definecolor{LightSkyBlue2}{rgb}{0.64,0.83,0.93}
\definecolor{LightSkyBlue3}{rgb}{0.55,0.71,0.80}
\definecolor{LightSkyBlue4}{rgb}{0.38,0.48,0.55}
\definecolor{LightSkyBlue}{rgb}{0.53,0.81,0.98}
\definecolor{LightSlateBlue}{rgb}{0.52,0.44,1.00}
\definecolor{LightSlateGray}{rgb}{0.47,0.53,0.60}
\definecolor{LightSlateGrey}{rgb}{0.47,0.53,0.60}
\definecolor{LightSteelBlue1}{rgb}{0.79,0.88,1.00}
\definecolor{LightSteelBlue2}{rgb}{0.74,0.82,0.93}
\definecolor{LightSteelBlue3}{rgb}{0.64,0.71,0.80}
\definecolor{LightSteelBlue4}{rgb}{0.43,0.48,0.55}
\definecolor{LightSteelBlue}{rgb}{0.69,0.77,0.87}
\definecolor{LightYellow1}{rgb}{1.00,1.00,0.88}
\definecolor{LightYellow2}{rgb}{0.93,0.93,0.82}
\definecolor{LightYellow3}{rgb}{0.80,0.80,0.71}
\definecolor{LightYellow4}{rgb}{0.55,0.55,0.48}
\definecolor{LightYellow}{rgb}{1.00,1.00,0.88}
\definecolor{LimeGreen}{rgb}{0.20,0.80,0.20}
\definecolor{MediumAquamarine}{rgb}{0.40,0.80,0.67}
\definecolor{MediumBlue}{rgb}{0.00,0.00,0.80}
\definecolor{MediumOrchid1}{rgb}{0.88,0.40,1.00}
\definecolor{MediumOrchid2}{rgb}{0.82,0.37,0.93}
\definecolor{MediumOrchid3}{rgb}{0.71,0.32,0.80}
\definecolor{MediumOrchid4}{rgb}{0.48,0.22,0.55}
\definecolor{MediumOrchid}{rgb}{0.73,0.33,0.83}
\definecolor{MediumPurple1}{rgb}{0.67,0.51,1.00}
\definecolor{MediumPurple2}{rgb}{0.62,0.47,0.93}
\definecolor{MediumPurple3}{rgb}{0.54,0.41,0.80}
\definecolor{MediumPurple4}{rgb}{0.36,0.28,0.55}
\definecolor{MediumPurple}{rgb}{0.58,0.44,0.86}
\definecolor{MediumSeaGreen}{rgb}{0.24,0.70,0.44}
\definecolor{MediumSlateBlue}{rgb}{0.48,0.41,0.93}
\definecolor{MediumSpringGreen}{rgb}{0.00,0.98,0.60}
\definecolor{MediumTurquoise}{rgb}{0.28,0.82,0.80}
\definecolor{MediumVioletRed}{rgb}{0.78,0.08,0.52}
\definecolor{MidnightBlue}{rgb}{0.10,0.10,0.44}
\definecolor{MintCream}{rgb}{0.96,1.00,0.98}
\definecolor{MistyRose1}{rgb}{1.00,0.89,0.88}
\definecolor{MistyRose2}{rgb}{0.93,0.84,0.82}
\definecolor{MistyRose3}{rgb}{0.80,0.72,0.71}
\definecolor{MistyRose4}{rgb}{0.55,0.49,0.48}
\definecolor{MistyRose}{rgb}{1.00,0.89,0.88}
\definecolor{NavajoWhite1}{rgb}{1.00,0.87,0.68}
\definecolor{NavajoWhite2}{rgb}{0.93,0.81,0.63}
\definecolor{NavajoWhite3}{rgb}{0.80,0.70,0.55}
\definecolor{NavajoWhite4}{rgb}{0.55,0.47,0.37}
\definecolor{NavajoWhite}{rgb}{1.00,0.87,0.68}
\definecolor{NavyBlue}{rgb}{0.00,0.00,0.50}
\definecolor{OldLace}{rgb}{0.99,0.96,0.90}
\definecolor{OliveDrab1}{rgb}{0.75,1.00,0.24}
\definecolor{OliveDrab2}{rgb}{0.70,0.93,0.23}
\definecolor{OliveDrab3}{rgb}{0.60,0.80,0.20}
\definecolor{OliveDrab4}{rgb}{0.41,0.55,0.13}
\definecolor{OliveDrab}{rgb}{0.42,0.56,0.14}
\definecolor{OrangeRed1}{rgb}{1.00,0.27,0.00}
\definecolor{OrangeRed2}{rgb}{0.93,0.25,0.00}
\definecolor{OrangeRed3}{rgb}{0.80,0.22,0.00}
\definecolor{OrangeRed4}{rgb}{0.55,0.15,0.00}
\definecolor{OrangeRed}{rgb}{1.00,0.27,0.00}
\definecolor{PaleGoldenrod}{rgb}{0.93,0.91,0.67}
\definecolor{PaleGreen1}{rgb}{0.60,1.00,0.60}
\definecolor{PaleGreen2}{rgb}{0.56,0.93,0.56}
\definecolor{PaleGreen3}{rgb}{0.49,0.80,0.49}
\definecolor{PaleGreen4}{rgb}{0.33,0.55,0.33}
\definecolor{PaleGreen}{rgb}{0.60,0.98,0.60}
\definecolor{PaleTurquoise1}{rgb}{0.73,1.00,1.00}
\definecolor{PaleTurquoise2}{rgb}{0.68,0.93,0.93}
\definecolor{PaleTurquoise3}{rgb}{0.59,0.80,0.80}
\definecolor{PaleTurquoise4}{rgb}{0.40,0.55,0.55}
\definecolor{PaleTurquoise}{rgb}{0.69,0.93,0.93}
\definecolor{PaleVioletRed1}{rgb}{1.00,0.51,0.67}
\definecolor{PaleVioletRed2}{rgb}{0.93,0.47,0.62}
\definecolor{PaleVioletRed3}{rgb}{0.80,0.41,0.54}
\definecolor{PaleVioletRed4}{rgb}{0.55,0.28,0.36}
\definecolor{PaleVioletRed}{rgb}{0.86,0.44,0.58}
\definecolor{PapayaWhip}{rgb}{1.00,0.94,0.84}
\definecolor{PeachPuff1}{rgb}{1.00,0.85,0.73}
\definecolor{PeachPuff2}{rgb}{0.93,0.80,0.68}
\definecolor{PeachPuff3}{rgb}{0.80,0.69,0.58}
\definecolor{PeachPuff4}{rgb}{0.55,0.47,0.40}
\definecolor{PeachPuff}{rgb}{1.00,0.85,0.73}
\definecolor{PowderBlue}{rgb}{0.69,0.88,0.90}
\definecolor{RosyBrown1}{rgb}{1.00,0.76,0.76}
\definecolor{RosyBrown2}{rgb}{0.93,0.71,0.71}
\definecolor{RosyBrown3}{rgb}{0.80,0.61,0.61}
\definecolor{RosyBrown4}{rgb}{0.55,0.41,0.41}
\definecolor{RosyBrown}{rgb}{0.74,0.56,0.56}
\definecolor{RoyalBlue1}{rgb}{0.28,0.46,1.00}
\definecolor{RoyalBlue2}{rgb}{0.26,0.43,0.93}
\definecolor{RoyalBlue3}{rgb}{0.23,0.37,0.80}
\definecolor{RoyalBlue4}{rgb}{0.15,0.25,0.55}
\definecolor{RoyalBlue}{rgb}{0.25,0.41,0.88}
\definecolor{SaddleBrown}{rgb}{0.55,0.27,0.07}
\definecolor{SandyBrown}{rgb}{0.96,0.64,0.38}
\definecolor{SeaGreen1}{rgb}{0.33,1.00,0.62}
\definecolor{SeaGreen2}{rgb}{0.31,0.93,0.58}
\definecolor{SeaGreen3}{rgb}{0.26,0.80,0.50}
\definecolor{SeaGreen4}{rgb}{0.18,0.55,0.34}
\definecolor{SeaGreen}{rgb}{0.18,0.55,0.34}
\definecolor{SkyBlue1}{rgb}{0.53,0.81,1.00}
\definecolor{SkyBlue2}{rgb}{0.49,0.75,0.93}
\definecolor{SkyBlue3}{rgb}{0.42,0.65,0.80}
\definecolor{SkyBlue4}{rgb}{0.29,0.44,0.55}
\definecolor{SkyBlue}{rgb}{0.53,0.81,0.92}
\definecolor{SlateBlue1}{rgb}{0.51,0.44,1.00}
\definecolor{SlateBlue2}{rgb}{0.48,0.40,0.93}
\definecolor{SlateBlue3}{rgb}{0.41,0.35,0.80}
\definecolor{SlateBlue4}{rgb}{0.28,0.24,0.55}
\definecolor{SlateBlue}{rgb}{0.42,0.35,0.80}
\definecolor{SlateGray1}{rgb}{0.78,0.89,1.00}
\definecolor{SlateGray2}{rgb}{0.73,0.83,0.93}
\definecolor{SlateGray3}{rgb}{0.62,0.71,0.80}
\definecolor{SlateGray4}{rgb}{0.42,0.48,0.55}
\definecolor{SlateGray}{rgb}{0.44,0.50,0.56}
\definecolor{SlateGrey}{rgb}{0.44,0.50,0.56}
\definecolor{SpringGreen1}{rgb}{0.00,1.00,0.50}
\definecolor{SpringGreen2}{rgb}{0.00,0.93,0.46}
\definecolor{SpringGreen3}{rgb}{0.00,0.80,0.40}
\definecolor{SpringGreen4}{rgb}{0.00,0.55,0.27}
\definecolor{SpringGreen}{rgb}{0.00,1.00,0.50}
\definecolor{SteelBlue1}{rgb}{0.39,0.72,1.00}
\definecolor{SteelBlue2}{rgb}{0.36,0.67,0.93}
\definecolor{SteelBlue3}{rgb}{0.31,0.58,0.80}
\definecolor{SteelBlue4}{rgb}{0.21,0.39,0.55}
\definecolor{SteelBlue}{rgb}{0.27,0.51,0.71}
\definecolor{VioletRed1}{rgb}{1.00,0.24,0.59}
\definecolor{VioletRed2}{rgb}{0.93,0.23,0.55}
\definecolor{VioletRed3}{rgb}{0.80,0.20,0.47}
\definecolor{VioletRed4}{rgb}{0.55,0.13,0.32}
\definecolor{VioletRed}{rgb}{0.82,0.13,0.56}
\definecolor{WhiteSmoke}{rgb}{0.96,0.96,0.96}
\definecolor{YellowGreen}{rgb}{0.60,0.80,0.20}
\definecolor{aliceblue}{rgb}{0.94,0.97,1.00}
\definecolor{antiquewhite}{rgb}{0.98,0.92,0.84}
\definecolor{aquamarine1}{rgb}{0.50,1.00,0.83}
\definecolor{aquamarine2}{rgb}{0.46,0.93,0.78}
\definecolor{aquamarine3}{rgb}{0.40,0.80,0.67}
\definecolor{aquamarine4}{rgb}{0.27,0.55,0.45}
\definecolor{aquamarine}{rgb}{0.50,1.00,0.83}
\definecolor{azure1}{rgb}{0.94,1.00,1.00}
\definecolor{azure2}{rgb}{0.88,0.93,0.93}
\definecolor{azure3}{rgb}{0.76,0.80,0.80}
\definecolor{azure4}{rgb}{0.51,0.55,0.55}
\definecolor{azure}{rgb}{0.94,1.00,1.00}
\definecolor{beige}{rgb}{0.96,0.96,0.86}
\definecolor{bisque1}{rgb}{1.00,0.89,0.77}
\definecolor{bisque2}{rgb}{0.93,0.84,0.72}
\definecolor{bisque3}{rgb}{0.80,0.72,0.62}
\definecolor{bisque4}{rgb}{0.55,0.49,0.42}
\definecolor{bisque}{rgb}{1.00,0.89,0.77}
\definecolor{black}{rgb}{0.00,0.00,0.00}
\definecolor{blanchedalmond}{rgb}{1.00,0.92,0.80}
\definecolor{blue1}{rgb}{0.00,0.00,1.00}
\definecolor{blue2}{rgb}{0.00,0.00,0.93}
\definecolor{blue3}{rgb}{0.00,0.00,0.80}
\definecolor{blue4}{rgb}{0.00,0.00,0.55}
\definecolor{blueviolet}{rgb}{0.54,0.17,0.89}
\definecolor{blue}{rgb}{0.00,0.00,1.00}
\definecolor{brown1}{rgb}{1.00,0.25,0.25}
\definecolor{brown2}{rgb}{0.93,0.23,0.23}
\definecolor{brown3}{rgb}{0.80,0.20,0.20}
\definecolor{brown4}{rgb}{0.55,0.14,0.14}
\definecolor{brown}{rgb}{0.65,0.16,0.16}
\definecolor{burlywood1}{rgb}{1.00,0.83,0.61}
\definecolor{burlywood2}{rgb}{0.93,0.77,0.57}
\definecolor{burlywood3}{rgb}{0.80,0.67,0.49}
\definecolor{burlywood4}{rgb}{0.55,0.45,0.33}
\definecolor{burlywood}{rgb}{0.87,0.72,0.53}
\definecolor{cadetblue}{rgb}{0.37,0.62,0.63}
\definecolor{chartreuse1}{rgb}{0.50,1.00,0.00}
\definecolor{chartreuse2}{rgb}{0.46,0.93,0.00}
\definecolor{chartreuse3}{rgb}{0.40,0.80,0.00}
\definecolor{chartreuse4}{rgb}{0.27,0.55,0.00}
\definecolor{chartreuse}{rgb}{0.50,1.00,0.00}
\definecolor{chocolate1}{rgb}{1.00,0.50,0.14}
\definecolor{chocolate2}{rgb}{0.93,0.46,0.13}
\definecolor{chocolate3}{rgb}{0.80,0.40,0.11}
\definecolor{chocolate4}{rgb}{0.55,0.27,0.07}
\definecolor{chocolate}{rgb}{0.82,0.41,0.12}
\definecolor{coral1}{rgb}{1.00,0.45,0.34}
\definecolor{coral2}{rgb}{0.93,0.42,0.31}
\definecolor{coral3}{rgb}{0.80,0.36,0.27}
\definecolor{coral4}{rgb}{0.55,0.24,0.18}
\definecolor{coral}{rgb}{1.00,0.50,0.31}
\definecolor{cornflowerblue}{rgb}{0.39,0.58,0.93}
\definecolor{cornsilk1}{rgb}{1.00,0.97,0.86}
\definecolor{cornsilk2}{rgb}{0.93,0.91,0.80}
\definecolor{cornsilk3}{rgb}{0.80,0.78,0.69}
\definecolor{cornsilk4}{rgb}{0.55,0.53,0.47}
\definecolor{cornsilk}{rgb}{1.00,0.97,0.86}
\definecolor{cyan1}{rgb}{0.00,1.00,1.00}
\definecolor{cyan2}{rgb}{0.00,0.93,0.93}
\definecolor{cyan3}{rgb}{0.00,0.80,0.80}
\definecolor{cyan4}{rgb}{0.00,0.55,0.55}
\definecolor{cyan}{rgb}{0.00,1.00,1.00}
\definecolor{darkblue}{rgb}{0.00,0.00,0.55}
\definecolor{darkcyan}{rgb}{0.00,0.55,0.55}
\definecolor{darkgoldenrod}{rgb}{0.72,0.53,0.04}
\definecolor{darkgray}{rgb}{0.66,0.66,0.66}
\definecolor{darkgreen}{rgb}{0.00,0.39,0.00}
\definecolor{darkgrey}{rgb}{0.66,0.66,0.66}
\definecolor{darkkhaki}{rgb}{0.74,0.72,0.42}
\definecolor{darkmagenta}{rgb}{0.55,0.00,0.55}
\definecolor{darkolive}{rgb}{0.33,0.42,0.18}
\definecolor{darkorange}{rgb}{1.00,0.55,0.00}
\definecolor{darkorchid}{rgb}{0.60,0.20,0.80}
\definecolor{darkred}{rgb}{0.55,0.00,0.00}
\definecolor{darksalmon}{rgb}{0.91,0.59,0.48}
\definecolor{darksea}{rgb}{0.56,0.74,0.56}
\definecolor{darkslate}{rgb}{0.18,0.31,0.31}
\definecolor{darkslate}{rgb}{0.18,0.31,0.31}
\definecolor{darkslate}{rgb}{0.28,0.24,0.55}
\definecolor{darkturquoise}{rgb}{0.00,0.81,0.82}
\definecolor{darkviolet}{rgb}{0.58,0.00,0.83}
\definecolor{deeppink}{rgb}{1.00,0.08,0.58}
\definecolor{deepsky}{rgb}{0.00,0.75,1.00}
\definecolor{dimgray}{rgb}{0.41,0.41,0.41}
\definecolor{dimgrey}{rgb}{0.41,0.41,0.41}
\definecolor{dodgerblue}{rgb}{0.12,0.56,1.00}
\definecolor{firebrick1}{rgb}{1.00,0.19,0.19}
\definecolor{firebrick2}{rgb}{0.93,0.17,0.17}
\definecolor{firebrick3}{rgb}{0.80,0.15,0.15}
\definecolor{firebrick4}{rgb}{0.55,0.10,0.10}
\definecolor{firebrick}{rgb}{0.70,0.13,0.13}
\definecolor{floralwhite}{rgb}{1.00,0.98,0.94}
\definecolor{forestgreen}{rgb}{0.13,0.55,0.13}
\definecolor{gainsboro}{rgb}{0.86,0.86,0.86}
\definecolor{ghostwhite}{rgb}{0.97,0.97,1.00}
\definecolor{gold1}{rgb}{1.00,0.84,0.00}
\definecolor{gold2}{rgb}{0.93,0.79,0.00}
\definecolor{gold3}{rgb}{0.80,0.68,0.00}
\definecolor{gold4}{rgb}{0.55,0.46,0.00}
\definecolor{goldenrod1}{rgb}{1.00,0.76,0.15}
\definecolor{goldenrod2}{rgb}{0.93,0.71,0.13}
\definecolor{goldenrod3}{rgb}{0.80,0.61,0.11}
\definecolor{goldenrod4}{rgb}{0.55,0.41,0.08}
\definecolor{goldenrod}{rgb}{0.85,0.65,0.13}
\definecolor{gold}{rgb}{1.00,0.84,0.00}
\definecolor{gray0}{rgb}{0.00,0.00,0.00}
\definecolor{gray100}{rgb}{1.00,1.00,1.00}
\definecolor{gray10}{rgb}{0.10,0.10,0.10}
\definecolor{gray11}{rgb}{0.11,0.11,0.11}
\definecolor{gray12}{rgb}{0.12,0.12,0.12}
\definecolor{gray13}{rgb}{0.13,0.13,0.13}
\definecolor{gray14}{rgb}{0.14,0.14,0.14}
\definecolor{gray15}{rgb}{0.15,0.15,0.15}
\definecolor{gray16}{rgb}{0.16,0.16,0.16}
\definecolor{gray17}{rgb}{0.17,0.17,0.17}
\definecolor{gray18}{rgb}{0.18,0.18,0.18}
\definecolor{gray19}{rgb}{0.19,0.19,0.19}
\definecolor{gray1}{rgb}{0.01,0.01,0.01}
\definecolor{gray20}{rgb}{0.20,0.20,0.20}
\definecolor{gray21}{rgb}{0.21,0.21,0.21}
\definecolor{gray22}{rgb}{0.22,0.22,0.22}
\definecolor{gray23}{rgb}{0.23,0.23,0.23}
\definecolor{gray24}{rgb}{0.24,0.24,0.24}
\definecolor{gray25}{rgb}{0.25,0.25,0.25}
\definecolor{gray26}{rgb}{0.26,0.26,0.26}
\definecolor{gray27}{rgb}{0.27,0.27,0.27}
\definecolor{gray28}{rgb}{0.28,0.28,0.28}
\definecolor{gray29}{rgb}{0.29,0.29,0.29}
\definecolor{gray2}{rgb}{0.02,0.02,0.02}
\definecolor{gray30}{rgb}{0.30,0.30,0.30}
\definecolor{gray31}{rgb}{0.31,0.31,0.31}
\definecolor{gray32}{rgb}{0.32,0.32,0.32}
\definecolor{gray33}{rgb}{0.33,0.33,0.33}
\definecolor{gray34}{rgb}{0.34,0.34,0.34}
\definecolor{gray35}{rgb}{0.35,0.35,0.35}
\definecolor{gray36}{rgb}{0.36,0.36,0.36}
\definecolor{gray37}{rgb}{0.37,0.37,0.37}
\definecolor{gray38}{rgb}{0.38,0.38,0.38}
\definecolor{gray39}{rgb}{0.39,0.39,0.39}
\definecolor{gray3}{rgb}{0.03,0.03,0.03}
\definecolor{gray40}{rgb}{0.40,0.40,0.40}
\definecolor{gray41}{rgb}{0.41,0.41,0.41}
\definecolor{gray42}{rgb}{0.42,0.42,0.42}
\definecolor{gray43}{rgb}{0.43,0.43,0.43}
\definecolor{gray44}{rgb}{0.44,0.44,0.44}
\definecolor{gray45}{rgb}{0.45,0.45,0.45}
\definecolor{gray46}{rgb}{0.46,0.46,0.46}
\definecolor{gray47}{rgb}{0.47,0.47,0.47}
\definecolor{gray48}{rgb}{0.48,0.48,0.48}
\definecolor{gray49}{rgb}{0.49,0.49,0.49}
\definecolor{gray4}{rgb}{0.04,0.04,0.04}
\definecolor{gray50}{rgb}{0.50,0.50,0.50}
\definecolor{gray51}{rgb}{0.51,0.51,0.51}
\definecolor{gray52}{rgb}{0.52,0.52,0.52}
\definecolor{gray53}{rgb}{0.53,0.53,0.53}
\definecolor{gray54}{rgb}{0.54,0.54,0.54}
\definecolor{gray55}{rgb}{0.55,0.55,0.55}
\definecolor{gray56}{rgb}{0.56,0.56,0.56}
\definecolor{gray57}{rgb}{0.57,0.57,0.57}
\definecolor{gray58}{rgb}{0.58,0.58,0.58}
\definecolor{gray59}{rgb}{0.59,0.59,0.59}
\definecolor{gray5}{rgb}{0.05,0.05,0.05}
\definecolor{gray60}{rgb}{0.60,0.60,0.60}
\definecolor{gray61}{rgb}{0.61,0.61,0.61}
\definecolor{gray62}{rgb}{0.62,0.62,0.62}
\definecolor{gray63}{rgb}{0.63,0.63,0.63}
\definecolor{gray64}{rgb}{0.64,0.64,0.64}
\definecolor{gray65}{rgb}{0.65,0.65,0.65}
\definecolor{gray66}{rgb}{0.66,0.66,0.66}
\definecolor{gray67}{rgb}{0.67,0.67,0.67}
\definecolor{gray68}{rgb}{0.68,0.68,0.68}
\definecolor{gray69}{rgb}{0.69,0.69,0.69}
\definecolor{gray6}{rgb}{0.06,0.06,0.06}
\definecolor{gray70}{rgb}{0.70,0.70,0.70}
\definecolor{gray71}{rgb}{0.71,0.71,0.71}
\definecolor{gray72}{rgb}{0.72,0.72,0.72}
\definecolor{gray73}{rgb}{0.73,0.73,0.73}
\definecolor{gray74}{rgb}{0.74,0.74,0.74}
\definecolor{gray75}{rgb}{0.75,0.75,0.75}
\definecolor{gray76}{rgb}{0.76,0.76,0.76}
\definecolor{gray77}{rgb}{0.77,0.77,0.77}
\definecolor{gray78}{rgb}{0.78,0.78,0.78}
\definecolor{gray79}{rgb}{0.79,0.79,0.79}
\definecolor{gray7}{rgb}{0.07,0.07,0.07}
\definecolor{gray80}{rgb}{0.80,0.80,0.80}
\definecolor{gray81}{rgb}{0.81,0.81,0.81}
\definecolor{gray82}{rgb}{0.82,0.82,0.82}
\definecolor{gray83}{rgb}{0.83,0.83,0.83}
\definecolor{gray84}{rgb}{0.84,0.84,0.84}
\definecolor{gray85}{rgb}{0.85,0.85,0.85}
\definecolor{gray86}{rgb}{0.86,0.86,0.86}
\definecolor{gray87}{rgb}{0.87,0.87,0.87}
\definecolor{gray88}{rgb}{0.88,0.88,0.88}
\definecolor{gray89}{rgb}{0.89,0.89,0.89}
\definecolor{gray8}{rgb}{0.08,0.08,0.08}
\definecolor{gray90}{rgb}{0.90,0.90,0.90}
\definecolor{gray91}{rgb}{0.91,0.91,0.91}
\definecolor{gray92}{rgb}{0.92,0.92,0.92}
\definecolor{gray93}{rgb}{0.93,0.93,0.93}
\definecolor{gray94}{rgb}{0.94,0.94,0.94}
\definecolor{gray95}{rgb}{0.95,0.95,0.95}
\definecolor{gray96}{rgb}{0.96,0.96,0.96}
\definecolor{gray97}{rgb}{0.97,0.97,0.97}
\definecolor{gray98}{rgb}{0.98,0.98,0.98}
\definecolor{gray99}{rgb}{0.99,0.99,0.99}
\definecolor{gray9}{rgb}{0.09,0.09,0.09}
\definecolor{gray}{rgb}{0.75,0.75,0.75}
\definecolor{green1}{rgb}{0.00,1.00,0.00}
\definecolor{green2}{rgb}{0.00,0.93,0.00}
\definecolor{green3}{rgb}{0.00,0.80,0.00}
\definecolor{green4}{rgb}{0.00,0.55,0.00}
\definecolor{greenyellow}{rgb}{0.68,1.00,0.18}
\definecolor{green}{rgb}{0.00,1.00,0.00}
\definecolor{grey0}{rgb}{0.00,0.00,0.00}
\definecolor{grey100}{rgb}{1.00,1.00,1.00}
\definecolor{grey10}{rgb}{0.10,0.10,0.10}
\definecolor{grey11}{rgb}{0.11,0.11,0.11}
\definecolor{grey12}{rgb}{0.12,0.12,0.12}
\definecolor{grey13}{rgb}{0.13,0.13,0.13}
\definecolor{grey14}{rgb}{0.14,0.14,0.14}
\definecolor{grey15}{rgb}{0.15,0.15,0.15}
\definecolor{grey16}{rgb}{0.16,0.16,0.16}
\definecolor{grey17}{rgb}{0.17,0.17,0.17}
\definecolor{grey18}{rgb}{0.18,0.18,0.18}
\definecolor{grey19}{rgb}{0.19,0.19,0.19}
\definecolor{grey1}{rgb}{0.01,0.01,0.01}
\definecolor{grey20}{rgb}{0.20,0.20,0.20}
\definecolor{grey21}{rgb}{0.21,0.21,0.21}
\definecolor{grey22}{rgb}{0.22,0.22,0.22}
\definecolor{grey23}{rgb}{0.23,0.23,0.23}
\definecolor{grey24}{rgb}{0.24,0.24,0.24}
\definecolor{grey25}{rgb}{0.25,0.25,0.25}
\definecolor{grey26}{rgb}{0.26,0.26,0.26}
\definecolor{grey27}{rgb}{0.27,0.27,0.27}
\definecolor{grey28}{rgb}{0.28,0.28,0.28}
\definecolor{grey29}{rgb}{0.29,0.29,0.29}
\definecolor{grey2}{rgb}{0.02,0.02,0.02}
\definecolor{grey30}{rgb}{0.30,0.30,0.30}
\definecolor{grey31}{rgb}{0.31,0.31,0.31}
\definecolor{grey32}{rgb}{0.32,0.32,0.32}
\definecolor{grey33}{rgb}{0.33,0.33,0.33}
\definecolor{grey34}{rgb}{0.34,0.34,0.34}
\definecolor{grey35}{rgb}{0.35,0.35,0.35}
\definecolor{grey36}{rgb}{0.36,0.36,0.36}
\definecolor{grey37}{rgb}{0.37,0.37,0.37}
\definecolor{grey38}{rgb}{0.38,0.38,0.38}
\definecolor{grey39}{rgb}{0.39,0.39,0.39}
\definecolor{grey3}{rgb}{0.03,0.03,0.03}
\definecolor{grey40}{rgb}{0.40,0.40,0.40}
\definecolor{grey41}{rgb}{0.41,0.41,0.41}
\definecolor{grey42}{rgb}{0.42,0.42,0.42}
\definecolor{grey43}{rgb}{0.43,0.43,0.43}
\definecolor{grey44}{rgb}{0.44,0.44,0.44}
\definecolor{grey45}{rgb}{0.45,0.45,0.45}
\definecolor{grey46}{rgb}{0.46,0.46,0.46}
\definecolor{grey47}{rgb}{0.47,0.47,0.47}
\definecolor{grey48}{rgb}{0.48,0.48,0.48}
\definecolor{grey49}{rgb}{0.49,0.49,0.49}
\definecolor{grey4}{rgb}{0.04,0.04,0.04}
\definecolor{grey50}{rgb}{0.50,0.50,0.50}
\definecolor{grey51}{rgb}{0.51,0.51,0.51}
\definecolor{grey52}{rgb}{0.52,0.52,0.52}
\definecolor{grey53}{rgb}{0.53,0.53,0.53}
\definecolor{grey54}{rgb}{0.54,0.54,0.54}
\definecolor{grey55}{rgb}{0.55,0.55,0.55}
\definecolor{grey56}{rgb}{0.56,0.56,0.56}
\definecolor{grey57}{rgb}{0.57,0.57,0.57}
\definecolor{grey58}{rgb}{0.58,0.58,0.58}
\definecolor{grey59}{rgb}{0.59,0.59,0.59}
\definecolor{grey5}{rgb}{0.05,0.05,0.05}
\definecolor{grey60}{rgb}{0.60,0.60,0.60}
\definecolor{grey61}{rgb}{0.61,0.61,0.61}
\definecolor{grey62}{rgb}{0.62,0.62,0.62}
\definecolor{grey63}{rgb}{0.63,0.63,0.63}
\definecolor{grey64}{rgb}{0.64,0.64,0.64}
\definecolor{grey65}{rgb}{0.65,0.65,0.65}
\definecolor{grey66}{rgb}{0.66,0.66,0.66}
\definecolor{grey67}{rgb}{0.67,0.67,0.67}
\definecolor{grey68}{rgb}{0.68,0.68,0.68}
\definecolor{grey69}{rgb}{0.69,0.69,0.69}
\definecolor{grey6}{rgb}{0.06,0.06,0.06}
\definecolor{grey70}{rgb}{0.70,0.70,0.70}
\definecolor{grey71}{rgb}{0.71,0.71,0.71}
\definecolor{grey72}{rgb}{0.72,0.72,0.72}
\definecolor{grey73}{rgb}{0.73,0.73,0.73}
\definecolor{grey74}{rgb}{0.74,0.74,0.74}
\definecolor{grey75}{rgb}{0.75,0.75,0.75}
\definecolor{grey76}{rgb}{0.76,0.76,0.76}
\definecolor{grey77}{rgb}{0.77,0.77,0.77}
\definecolor{grey78}{rgb}{0.78,0.78,0.78}
\definecolor{grey79}{rgb}{0.79,0.79,0.79}
\definecolor{grey7}{rgb}{0.07,0.07,0.07}
\definecolor{grey80}{rgb}{0.80,0.80,0.80}
\definecolor{grey81}{rgb}{0.81,0.81,0.81}
\definecolor{grey82}{rgb}{0.82,0.82,0.82}
\definecolor{grey83}{rgb}{0.83,0.83,0.83}
\definecolor{grey84}{rgb}{0.84,0.84,0.84}
\definecolor{grey85}{rgb}{0.85,0.85,0.85}
\definecolor{grey86}{rgb}{0.86,0.86,0.86}
\definecolor{grey87}{rgb}{0.87,0.87,0.87}
\definecolor{grey88}{rgb}{0.88,0.88,0.88}
\definecolor{grey89}{rgb}{0.89,0.89,0.89}
\definecolor{grey8}{rgb}{0.08,0.08,0.08}
\definecolor{grey90}{rgb}{0.90,0.90,0.90}
\definecolor{grey91}{rgb}{0.91,0.91,0.91}
\definecolor{grey92}{rgb}{0.92,0.92,0.92}
\definecolor{grey93}{rgb}{0.93,0.93,0.93}
\definecolor{grey94}{rgb}{0.94,0.94,0.94}
\definecolor{grey95}{rgb}{0.95,0.95,0.95}
\definecolor{grey96}{rgb}{0.96,0.96,0.96}
\definecolor{grey97}{rgb}{0.97,0.97,0.97}
\definecolor{grey98}{rgb}{0.98,0.98,0.98}
\definecolor{grey99}{rgb}{0.99,0.99,0.99}
\definecolor{grey9}{rgb}{0.09,0.09,0.09}
\definecolor{grey}{rgb}{0.75,0.75,0.75}
\definecolor{honeydew1}{rgb}{0.94,1.00,0.94}
\definecolor{honeydew2}{rgb}{0.88,0.93,0.88}
\definecolor{honeydew3}{rgb}{0.76,0.80,0.76}
\definecolor{honeydew4}{rgb}{0.51,0.55,0.51}
\definecolor{honeydew}{rgb}{0.94,1.00,0.94}
\definecolor{hotpink}{rgb}{1.00,0.41,0.71}
\definecolor{indianred}{rgb}{0.80,0.36,0.36}
\definecolor{ivory1}{rgb}{1.00,1.00,0.94}
\definecolor{ivory2}{rgb}{0.93,0.93,0.88}
\definecolor{ivory3}{rgb}{0.80,0.80,0.76}
\definecolor{ivory4}{rgb}{0.55,0.55,0.51}
\definecolor{ivory}{rgb}{1.00,1.00,0.94}
\definecolor{khaki1}{rgb}{1.00,0.96,0.56}
\definecolor{khaki2}{rgb}{0.93,0.90,0.52}
\definecolor{khaki3}{rgb}{0.80,0.78,0.45}
\definecolor{khaki4}{rgb}{0.55,0.53,0.31}
\definecolor{khaki}{rgb}{0.94,0.90,0.55}
\definecolor{lavenderblush}{rgb}{1.00,0.94,0.96}
\definecolor{lavender}{rgb}{0.90,0.90,0.98}
\definecolor{lawngreen}{rgb}{0.49,0.99,0.00}
\definecolor{lemonchiffon}{rgb}{1.00,0.98,0.80}
\definecolor{lightblue}{rgb}{0.68,0.85,0.90}
\definecolor{lightcoral}{rgb}{0.94,0.50,0.50}
\definecolor{lightcyan}{rgb}{0.88,1.00,1.00}
\definecolor{lightgoldenrod}{rgb}{0.93,0.87,0.51}
\definecolor{lightgoldenrod}{rgb}{0.98,0.98,0.82}
\definecolor{lightgray}{rgb}{0.83,0.83,0.83}
\definecolor{lightgreen}{rgb}{0.56,0.93,0.56}
\definecolor{lightgrey}{rgb}{0.83,0.83,0.83}
\definecolor{lightpink}{rgb}{1.00,0.71,0.76}
\definecolor{lightsalmon}{rgb}{1.00,0.63,0.48}
\definecolor{lightsea}{rgb}{0.13,0.70,0.67}
\definecolor{lightsky}{rgb}{0.53,0.81,0.98}
\definecolor{lightslate}{rgb}{0.47,0.53,0.60}
\definecolor{lightslate}{rgb}{0.47,0.53,0.60}
\definecolor{lightslate}{rgb}{0.52,0.44,1.00}
\definecolor{lightsteel}{rgb}{0.69,0.77,0.87}
\definecolor{lightyellow}{rgb}{1.00,1.00,0.88}
\definecolor{limegreen}{rgb}{0.20,0.80,0.20}
\definecolor{linen}{rgb}{0.98,0.94,0.90}
\definecolor{magenta1}{rgb}{1.00,0.00,1.00}
\definecolor{magenta2}{rgb}{0.93,0.00,0.93}
\definecolor{magenta3}{rgb}{0.80,0.00,0.80}
\definecolor{magenta4}{rgb}{0.55,0.00,0.55}
\definecolor{magenta}{rgb}{1.00,0.00,1.00}
\definecolor{maroon1}{rgb}{1.00,0.20,0.70}
\definecolor{maroon2}{rgb}{0.93,0.19,0.65}
\definecolor{maroon3}{rgb}{0.80,0.16,0.56}
\definecolor{maroon4}{rgb}{0.55,0.11,0.38}
\definecolor{maroon}{rgb}{0.69,0.19,0.38}
\definecolor{mediumaquamarine}{rgb}{0.40,0.80,0.67}
\definecolor{mediumblue}{rgb}{0.00,0.00,0.80}
\definecolor{mediumorchid}{rgb}{0.73,0.33,0.83}
\definecolor{mediumpurple}{rgb}{0.58,0.44,0.86}
\definecolor{mediumsea}{rgb}{0.24,0.70,0.44}
\definecolor{mediumslate}{rgb}{0.48,0.41,0.93}
\definecolor{mediumspring}{rgb}{0.00,0.98,0.60}
\definecolor{mediumturquoise}{rgb}{0.28,0.82,0.80}
\definecolor{mediumviolet}{rgb}{0.78,0.08,0.52}
\definecolor{midnightblue}{rgb}{0.10,0.10,0.44}
\definecolor{mintcream}{rgb}{0.96,1.00,0.98}
\definecolor{mistyrose}{rgb}{1.00,0.89,0.88}
\definecolor{moccasin}{rgb}{1.00,0.89,0.71}
\definecolor{navajowhite}{rgb}{1.00,0.87,0.68}
\definecolor{navyblue}{rgb}{0.00,0.00,0.50}
\definecolor{navy}{rgb}{0.00,0.00,0.50}
\definecolor{oldlace}{rgb}{0.99,0.96,0.90}
\definecolor{olivedrab}{rgb}{0.42,0.56,0.14}
\definecolor{orange1}{rgb}{1.00,0.65,0.00}
\definecolor{orange2}{rgb}{0.93,0.60,0.00}
\definecolor{orange3}{rgb}{0.80,0.52,0.00}
\definecolor{orange4}{rgb}{0.55,0.35,0.00}
\definecolor{orangered}{rgb}{1.00,0.27,0.00}
\definecolor{orange}{rgb}{1.00,0.65,0.00}
\definecolor{orchid1}{rgb}{1.00,0.51,0.98}
\definecolor{orchid2}{rgb}{0.93,0.48,0.91}
\definecolor{orchid3}{rgb}{0.80,0.41,0.79}
\definecolor{orchid4}{rgb}{0.55,0.28,0.54}
\definecolor{orchid}{rgb}{0.85,0.44,0.84}
\definecolor{palegoldenrod}{rgb}{0.93,0.91,0.67}
\definecolor{palegreen}{rgb}{0.60,0.98,0.60}
\definecolor{paleturquoise}{rgb}{0.69,0.93,0.93}
\definecolor{paleviolet}{rgb}{0.86,0.44,0.58}
\definecolor{papayawhip}{rgb}{1.00,0.94,0.84}
\definecolor{peachpuff}{rgb}{1.00,0.85,0.73}
\definecolor{peru}{rgb}{0.80,0.52,0.25}
\definecolor{pink1}{rgb}{1.00,0.71,0.77}
\definecolor{pink2}{rgb}{0.93,0.66,0.72}
\definecolor{pink3}{rgb}{0.80,0.57,0.62}
\definecolor{pink4}{rgb}{0.55,0.39,0.42}
\definecolor{pink}{rgb}{1.00,0.75,0.80}
\definecolor{plum1}{rgb}{1.00,0.73,1.00}
\definecolor{plum2}{rgb}{0.93,0.68,0.93}
\definecolor{plum3}{rgb}{0.80,0.59,0.80}
\definecolor{plum4}{rgb}{0.55,0.40,0.55}
\definecolor{plum}{rgb}{0.87,0.63,0.87}
\definecolor{powderblue}{rgb}{0.69,0.88,0.90}
\definecolor{purple1}{rgb}{0.61,0.19,1.00}
\definecolor{purple2}{rgb}{0.57,0.17,0.93}
\definecolor{purple3}{rgb}{0.49,0.15,0.80}
\definecolor{purple4}{rgb}{0.33,0.10,0.55}
\definecolor{purple}{rgb}{0.63,0.13,0.94}
\definecolor{red1}{rgb}{1.00,0.00,0.00}
\definecolor{red2}{rgb}{0.93,0.00,0.00}
\definecolor{red3}{rgb}{0.80,0.00,0.00}
\definecolor{red4}{rgb}{0.55,0.00,0.00}
\definecolor{red}{rgb}{1.00,0.00,0.00}
\definecolor{rosybrown}{rgb}{0.74,0.56,0.56}
\definecolor{royalblue}{rgb}{0.25,0.41,0.88}
\definecolor{saddlebrown}{rgb}{0.55,0.27,0.07}
\definecolor{salmon1}{rgb}{1.00,0.55,0.41}
\definecolor{salmon2}{rgb}{0.93,0.51,0.38}
\definecolor{salmon3}{rgb}{0.80,0.44,0.33}
\definecolor{salmon4}{rgb}{0.55,0.30,0.22}
\definecolor{salmon}{rgb}{0.98,0.50,0.45}
\definecolor{sandybrown}{rgb}{0.96,0.64,0.38}
\definecolor{seagreen}{rgb}{0.18,0.55,0.34}
\definecolor{seashell1}{rgb}{1.00,0.96,0.93}
\definecolor{seashell2}{rgb}{0.93,0.90,0.87}
\definecolor{seashell3}{rgb}{0.80,0.77,0.75}
\definecolor{seashell4}{rgb}{0.55,0.53,0.51}
\definecolor{seashell}{rgb}{1.00,0.96,0.93}
\definecolor{sienna1}{rgb}{1.00,0.51,0.28}
\definecolor{sienna2}{rgb}{0.93,0.47,0.26}
\definecolor{sienna3}{rgb}{0.80,0.41,0.22}
\definecolor{sienna4}{rgb}{0.55,0.28,0.15}
\definecolor{sienna}{rgb}{0.63,0.32,0.18}
\definecolor{skyblue}{rgb}{0.53,0.81,0.92}
\definecolor{slateblue}{rgb}{0.42,0.35,0.80}
\definecolor{slategray}{rgb}{0.44,0.50,0.56}
\definecolor{slategrey}{rgb}{0.44,0.50,0.56}
\definecolor{snow1}{rgb}{1.00,0.98,0.98}
\definecolor{snow2}{rgb}{0.93,0.91,0.91}
\definecolor{snow3}{rgb}{0.80,0.79,0.79}
\definecolor{snow4}{rgb}{0.55,0.54,0.54}
\definecolor{snow}{rgb}{1.00,0.98,0.98}
\definecolor{springgreen}{rgb}{0.00,1.00,0.50}
\definecolor{steelblue}{rgb}{0.27,0.51,0.71}
\definecolor{tan1}{rgb}{1.00,0.65,0.31}
\definecolor{tan2}{rgb}{0.93,0.60,0.29}
\definecolor{tan3}{rgb}{0.80,0.52,0.25}
\definecolor{tan4}{rgb}{0.55,0.35,0.17}
\definecolor{tan}{rgb}{0.82,0.71,0.55}
\definecolor{thistle1}{rgb}{1.00,0.88,1.00}
\definecolor{thistle2}{rgb}{0.93,0.82,0.93}
\definecolor{thistle3}{rgb}{0.80,0.71,0.80}
\definecolor{thistle4}{rgb}{0.55,0.48,0.55}
\definecolor{thistle}{rgb}{0.85,0.75,0.85}
\definecolor{tomato1}{rgb}{1.00,0.39,0.28}
\definecolor{tomato2}{rgb}{0.93,0.36,0.26}
\definecolor{tomato3}{rgb}{0.80,0.31,0.22}
\definecolor{tomato4}{rgb}{0.55,0.21,0.15}
\definecolor{tomato}{rgb}{1.00,0.39,0.28}
\definecolor{turquoise1}{rgb}{0.00,0.96,1.00}
\definecolor{turquoise2}{rgb}{0.00,0.90,0.93}
\definecolor{turquoise3}{rgb}{0.00,0.77,0.80}
\definecolor{turquoise4}{rgb}{0.00,0.53,0.55}
\definecolor{turquoise}{rgb}{0.25,0.88,0.82}
\definecolor{violetred}{rgb}{0.82,0.13,0.56}
\definecolor{violet}{rgb}{0.93,0.51,0.93}
\definecolor{wheat1}{rgb}{1.00,0.91,0.73}
\definecolor{wheat2}{rgb}{0.93,0.85,0.68}
\definecolor{wheat3}{rgb}{0.80,0.73,0.59}
\definecolor{wheat4}{rgb}{0.55,0.49,0.40}
\definecolor{wheat}{rgb}{0.96,0.87,0.70}
\definecolor{whitesmoke}{rgb}{0.96,0.96,0.96}
\definecolor{white}{rgb}{1.00,1.00,1.00}
\definecolor{yellow1}{rgb}{1.00,1.00,0.00}
\definecolor{yellow2}{rgb}{0.93,0.93,0.00}
\definecolor{yellow3}{rgb}{0.80,0.80,0.00}
\definecolor{yellow4}{rgb}{0.55,0.55,0.00}
\definecolor{yellowgreen}{rgb}{0.60,0.80,0.20}
\definecolor{yellow}{rgb}{1.00,1.00,0.00}
\usepackage{amsfonts}
\usepackage{bm}
\usepackage{longtable}
\usepackage{pifont}
\usepackage{gensymb}

\usepackage{booktabs} 
 
\newcommand{\vpfont}{
  \fontfamily{pcr}
  \bfseries 
  \color{blue}
}

\newcommand{\gskfont}{
  \color{red}
}
\newcommand{\jcafont}{
  \color{green}
}
\newcommand{\revfont}{
  \color{black}
}

\DeclareTextFontCommand{\vp}{\vpfont}
\DeclareTextFontCommand{\gsk}{\gskfont}
\DeclareTextFontCommand{\jca}{\jcafont}
\DeclareTextFontCommand{\rev}{\revfont}

\shorttitle{{Solar Flare Arcade Modelling}}
\shortauthors{Kerr, Allred, Polito}

\begin{document}


	\title{Solar Flare Arcade Modelling: Bridging the gap from 1D to 3D Simulations of Optically Thin Radiation}
	\author{Graham~S.~Kerr}
	\email{graham.s.kerr@nasa.gov}
	\altaffiliation{Work began while a NASA Postdoctoral Program (NPP) Fellow, administered by USRA}
	\affil{NASA Goddard Space Flight Center, Heliophysics Sciences Division, Code 671, 8800 Greenbelt Rd., Greenbelt, MD 20771, USA}
 	\affil{Department of Physics, Catholic University of America, 620 Michigan Avenue, Northeast, Washington, DC 20064, USA}
 
	 \author{Joel~C.~Allred}
	 \affil{NASA Goddard Space Flight Center, Heliophysics Sciences Division, Code 671, 8800 Greenbelt Rd., Greenbelt, MD 20771, USA}
	
	\author{Vanessa~Polito}
	\affiliation{Bay Area Environmental Research Institute, NASA Research Park,  Moffett Field, CA 94035-0001, USA}
        \affiliation{Lockheed Martin Solar and Astrophysics Laboratory, Building 252, 3251 Hanover Street, Palo Alto, CA 94304, USA}

	\date{Received / Accepted}
	
	
	\begin{abstract}	
       Solar flares are 3D phenomenon but modelling a flare in 3D, including many of the important processes in the chromosphere, is a computational challenge. Accurately modelling the chromosphere is important, even if the transition region and corona are the areas of interest, due to the flow of energy, mass, and radiation through the interconnected layers. We present a solar flare arcade model, that aims to bridge the gap between 1D and 3D modelling. Our approach is limited to the synthesis of optically thin emission. Using observed active region loop structures in a 3D domain we graft simulated 1D flare atmospheres onto each loop, synthesise the emission and then project that emission onto to the 2D observational plane. Emission from SDO/AIA, GOES/XRS, and IRIS/SG Fe \textsc{xxi} 1354.1\AA\ was forward modelled. We analyse the temperatures, durations, mass flows, and line widths associated with the flare, finding qualitative agreement but certain quantitative differences. Compared to observations, the Doppler shifts are of similar magnitude but decay too quickly. They are not as ordered, containing a larger amount of scatter compared to observations. The duration of gradual phase emission from GOES and AIA emission is also too short. Fe \textsc{xxi} lines are broadened, but not sufficiently. These findings suggest that additional physics is required in our model. The arcade model that we show here as a proof-of-concept can be extended to investigate other lines and global aspects of solar flares, providing a means to better test the coronal response to models of flare energy injection.
        	\end{abstract}


\section{Introduction}\label{sec:intro}
Magnetic reconnection in the solar corona can liberate a tremendous amount of magnetic energy. The released energy can intensely heat and ionise the solar atmosphere, leading to a broadband enhancement to the solar radiative output, known as a solar flare. This same process can result in the production of coronal mass ejections (CMEs) and solar energetic particles (SEPs). Flares, CMEs and SEPs drive geoeffective space weather, making understanding the various physical processes involved in energy release and transport of crucial importance. Here we focus on the solar flare component, presenting a new approach to model the coronal flare arcade with radiation hydrodynamic modelling.

Following reconnection, energy is transported along the legs of magnetic loops. In the standard flare model energy is carried by a beamed distribution of non-thermal electrons accelerated out of the ambient corona, which undergo Coulomb collisions, thermalising the electrons in the chromosphere or transition region \citep{1971SoPh...18..489B,2011SSRv..159..107H}. The sudden temperature increase leads to expansion of chromospheric material up into the corona \citep[`chromospheric ablation', also referred to as `evaporation';][]{1985ApJ...289..425F,1985ApJ...289..434F} and down into the deeper atmosphere \citep[`chromospheric condensation';][]{1989ApJ...346.1019F}. Energy transport via non-thermal particles and thermal conduction are field-aligned processes. The resulting dynamics of the flaring plasma are also field-aligned, constrained by the magnetic field. 
 
Plasma heating, ionisation, mass flows, and other physical properties reveal themselves through the emission of both continuum and spectral line radiation from the various layers of the solar atmosphere \citep{2011SSRv..159...19F,2015SoPh..290.3399M}. Optical and ultraviolet (UV) emission typically appear in ribbon like sources, which show substructure, and are due to thermal plasma processes. Hard X-ray emission appears as compact footpoint sources at the base of magnetic loops, and present unambiguous evidence of particle acceleration in flares. Soft X-ray emission is somewhat delayed relative to the hard X-ray peak, and appears in flare loops and loop tops once the density has increased sufficiently following ablation. Extreme ultraviolet emission appears both in footpoint and loop sources, and usually indicates the presence of high temperature plasma. See \cite{2011SSRv..159...19F}, \cite{2011SSRv..159..107H}, \cite{2011SSRv..159..301K}, and \cite{2016JGRA..12111667H} for reviews of flare observations and physical processes.

As reconnection progresses energy is deposited into new loops. Ribbons spread both along and away from the polarity inversion line \citep[e.g.][]{2005ApJ...625L.143G,2012ApJ...744...48C,2010ApJ...725..319Q,2017ApJ...838...17Q}, and loops brighten, so that a flare arcade forms. A quintessential example of a flare arcade is the Bastille Day flare \citep[see images in ][]{2001SoPh..204...69F,2016JGRA..12111667H}. The arcade structure means that along the line of sight we are potentially detecting radiation from multiple ribbons, footpoints, loops, and loop tops. 

Through comparisons of synthetic observables derived from numerical models of flares we can determine if our models of energy transport and the atmospheric response are sufficient, or if additional ingredients are required. For example, do we need to consider the presence of turbulence, non-thermal ions, Alfv\'enic waves, or more sophisticated treatments of the electron beam?

The small spatial scales involved in shocks and steep gradients demand high resolution in numerical models that include an accurate chromosphere and transition region \citep{1999ApJ...521..906A,2005ApJ...630..573A,2013ApJ...770...12B}. Simulations that model the chromosphere's response to a beam of energetic particles, with non-LTE, non-local radiation transport and time-dependent atomic level populations, that feedback to the hydrodynamics, in 3D are presently a computationally difficult (if not intractable) problem. An advanced 3D radiation magnetohydrodynamic model of a flare was recently performed by \cite{2019NatAs...3..160C}. While this impressive achievement managed to model reconnection, the structure of flaring magnetic loops, and energy release in an active region, leading to heating equivalent to a C-class flare (determined from the flux of forward modelled soft X-rays) it did not include non-thermal particle acceleration and their resulting energy deposition profile. Energy transport was solely via thermal conduction in their model. 

Since both energy transport and dynamics within the flare loop are field-aligned processes, we note (and stress) that the 1D modelling approach is actually a reasonable assumption, allowing us to include more detailed physics than would be possible in 3D modelling. 

Magnetic field aligned (1D) loop models of solar flares have thus emerged as vital tools to understand the various aspects of the flare problem, including the response of the atmosphere to non-thermal electron beam heating \citep[examples include][]{1999ApJ...521..906A,2005ApJ...630..573A,2009A&A...499..923K,2013ApJ...778...76R,2015ApJ...809..104A}, investigating alternative energy transport techniques \citep[examples include][]{2016ApJ...818L..20R,2016ApJ...818...44R,2016ApJ...827..101K,2018ApJ...853..101R,2018ApJ...862...76P,2018ApJ...856..178P}, to understanding the detailed formation processes of various observables to aid in the interpretation of observations \citep[][]{2015SoPh..290.3487K,2017ApJ...836...12K,2018ApJ...862...59B,2019ApJ...871...23K,2019ApJ...879...19Z,2019ApJ...883...57K,2019ApJ...885..119K,2020ApJ...895....6G}. 

The impact of omitting optically thick 3D radiation transfer effects (such as radiative heating and cooling) in flares is not well known, and is beyond the scope of this current study. We do know, though, that the 3D nature of flaring structures is important to consider when either interpreting or modelling optically thin emission, since emission is summed along the line of sight. 

When studying emission from the chromosphere or transition region and performing model-data comparisons, it suffices to treat the vertical extent of the 1D model's upper chromosphere and transition as part of the flare ribbon or footpoint, as these layers are fairly narrow. However, if one is interested in coronal emission from the flare loops then this simplification is no longer appropriate. A spectral line could form over an extended portion of a hot flare loop, so summing emission through the loop in that manner would not be realistic. Instead, only portions of the loop should be selected, and the line of sight to the observer accounted for.

The technique of \cite{2011ApJS..194...26B} has been used to model coronal emission in this manner \citep{2015ApJ...811..129B,2016ApJ...816...89P,2016ApJ...827..145R,2018ApJ...856..149R,2020ApJ...891..122M}. In that approach the flare loop simulated is assumed to be semi-circular, at disk-center, and orientated perpendicular to the solar surface, aligned east-west. The line of sight is parallel to the plane of the loop. The spatial emission along the loop can then be binned into a single row of detector pixels. See figure 1 in \cite{2011ApJS..194...26B} for a visual depiction. Another approach to model coronal emission was that of \cite{2019ApJ...879L..17P}, who used a similar method, but included inclination angles of the loops relative to the detector, and superposition of several loops.

Here we present our approach to study optically thin flare emission that aims to account for the spatial extent of the loop relative to a detector pixel, loop geometry, inclination, the superposition of loops, and the location on the solar disk. A flare arcade model was produced as a proof-of-concept, illustrating how we bridge the gap from the state-of-the-art 1D field-aligned detailed loop model of a flare, to a data-constrained 3D arcade structure from which emission is forward modelled and then projected onto a 2D observational plane. We use observed loop structures from an active region for this purpose.  

Forward modelling of several observables is presented to illustrate how the arcade modelling approach can facilitate model-data predictions, and to assess how well the model can reproduce aspects of the flare. Both a qualitative and quantitative comparison is made to observations from the \textsl{Geostationary Operational Environmental Satellite's} X-ray Sensor (GOES/XRS), the \textsl{Solar Dynamics Observatory's} Atmospheric Imaging Assembly \citep[SDO/AIA;][]{2012SoPh..275....3P,2012SoPh..275...17L}, and the \textsl{Interface Region Imaging Spectrograph's} Spectrograph \citep[IRIS/SG;][]{2014SoPh..289.2733D}.


\section{\texttt{RADYN} Field-Aligned Loop modelling}
The 1D, field-aligned, radiation hydrodynamics code \texttt{RADYN} \citep{1992ApJ...397L..59C,1997ApJ...481..500C,2002ApJ...572..626C,2005ApJ...630..573A,2015ApJ...809..104A,FP_inprep} models the solar atmosphere's response to flare energy injection, including the feedback between the non-local, NLTE radiation transfer and hydrodynamics. Radiation from hydrogen, helium, and calcium (species important for energy balance) are treated in detail (including non-equilibrium effects), and other species are included via a radiation loss function. It uses adaptive grid \citep{1987JCoPh..69..175D} to capture shocks and strong gradients that typically form in flares. An important feature of \texttt{RADYN} is its ability to model an accurate chromosphere, which has impacts  the response of other atmospheric layers and the spatio-temporal evolution of flare energy deposition. Thus, the development of flows and coronal plasma properties are impacted by how the lower atmosphere responds. 

Flare energy is injected via a non-thermal electron distribution, with a Fokker-Planck treatment that includes transport effects through the flare loop \citep{2015ApJ...809..104A,FP_inprep}. Thermal conduction is Spitzer with a flux limiter to avoid exceeding the electron free streaming rate \citep{1980ApJ...238.1126S}. It is also possible to inject flare energy via an approximated form of downward propagating Alfv\'enic waves \citep{2016ApJ...827..101K} though we do not model those in this work. 

\texttt{RADYN} has become a commonly used resource to study both energy transport in flares, and the formation of radiation during flares (until recently, typically focussing on chromospheric and transition region radiation). For a more detailed description of the code, and for examples of recent studies using \texttt{RADYN} consult \cite{2015ApJ...809..104A}, \cite{2015SoPh..290.3487K}, \cite{2019ApJ...883...57K}, \cite{2019ApJ...871...23K}, \cite{2019ApJ...879L..17P}, and references therein.

\begin{figure}
	\centering 
	{\includegraphics[width = .5\textwidth, clip = true, trim = 0.cm 0.cm 0.cm 0.cm]{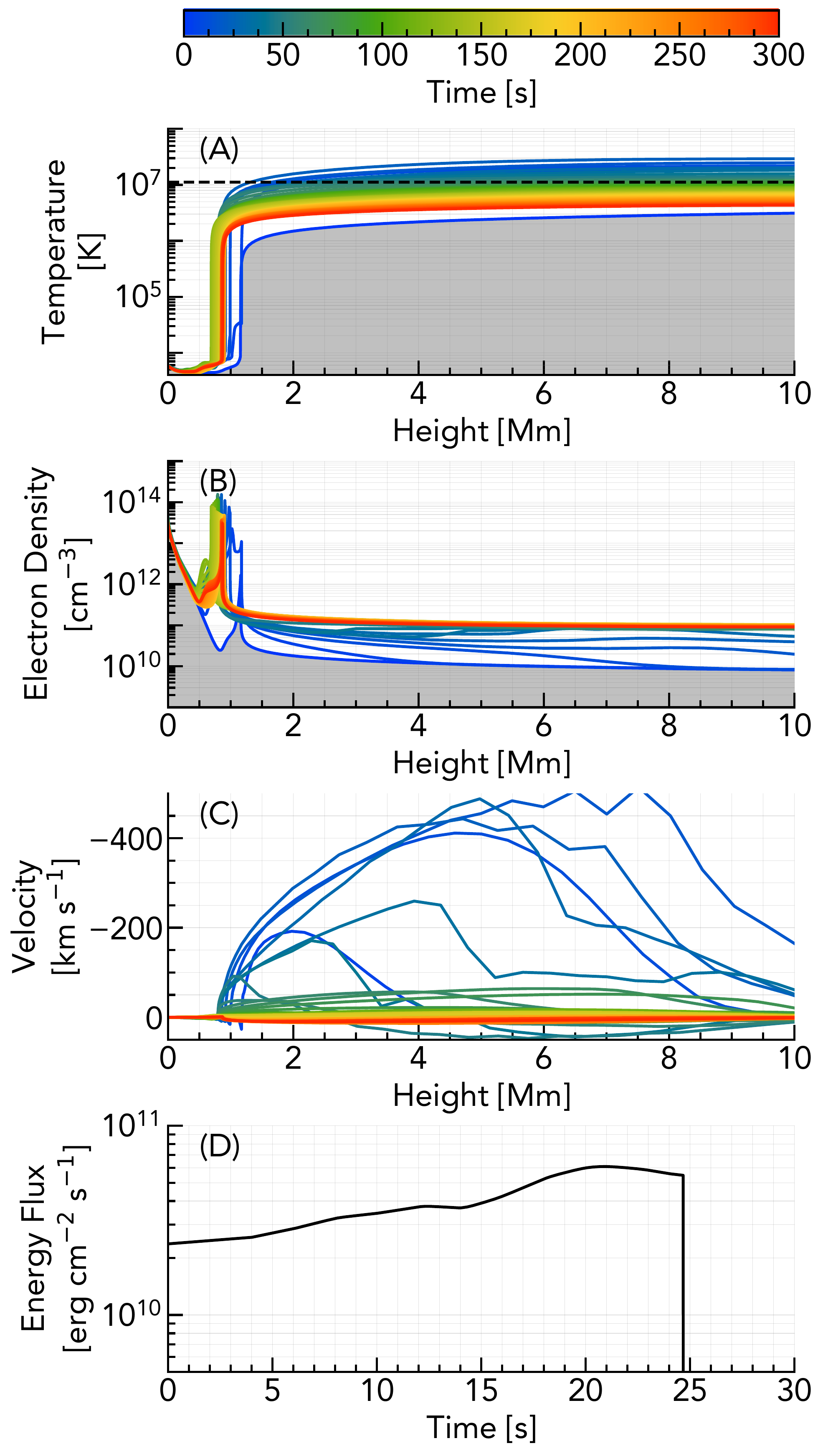}}	
	\caption{\textsl{The flare atmospheres (the temperature, electron density, and atmospheric velocity, where upflows are negative, shown in panels A-C respectively) from \texttt{RADYN} at various times in the simulation. Heating was applied for 25~s, panel (D). The dashed horizontal line indicates the peak formation temperature of Fe~\textsc{xxi}, $T\sim11.2$~MK.}}
	\label{fig:flaresim}
\end{figure}

The pre-flare atmosphere was one half of a symmetric loop, spanning the sub-photosphere, photosphere, chromosphere, transition region (TR), and corona. Energy was injected at the loop apex. The injected non-thermal electron distribution had a flux on the order $1-6\times10^{10}$~erg~cm$^{-2}$~s$^{-1}$ (the actual injected flux varied in time and is shown in Figure~\ref{fig:flaresim}) with a spectral index $\delta = 7.2$, above a low-energy cutoff $E_{c} = 25.3$~keV. Energy was injected for approximately $25$~s, after which the flare loop cooled for several hundred seconds. Figure~\ref{fig:flaresim} shows the response of the atmosphere at several snapshots. For the purposes of this experiment combining \texttt{RADYN} and arcade modelling, we selected a pre-existing \texttt{RADYN} simulation that had a large amount of plasma at $T>11$~MK and large mass flows against gravity (upflows), so that we could explore characteristics of the Fe~\textsc{xxi} 1354.1~\AA\ flare line. \rev{These parameters lie within the typical range for moderate-to-strong flares based on RHESSI hard X-ray observations \citep[e.g.][]{2011SSRv..159..107H}. We do not believe that changing these parameters would affect our overall conclusions.}   

The chromosphere rapidly heats and ionises, increasing the electron density throughout the lower atmosphere. Explosive chromospheric ablation (upflowing material, commonly referred to as `evaporation') results, filling in the coronal portion of the loop, producing a hotter and denser corona. The location of the TR during the flare decreases in altitude. 

These conditions allow for the presence of highly ionised species and the generation of spectral lines that typically only appear during flares or other transient heating events. \texttt{RADYN} tracks non-equilibrium processes for certain chromospheric species, but does not do this as standard for iron. This will feature as a future avenue of investigation but for this current work we rely on the assumption of equilibrium ionisation to obtain the fraction of Fe~\textsc{xxi} present in our flare loop. 

Following cessation of energy injection the mass flows decrease, and radiative losses and thermal conduction efficiently cool the corona which undergoes a rapid catastrophic cooling period. 

\subsection{Fe~\textsc{xxi} 1354.1~\AA\ Emission from the 1D Model}
From the field-aligned model we synthesised Fe~\textsc{xxi} lightcurves. This is appropriate since we are interested in the integrated line intensity. Full line profiles and associated characteristics require imposing a line of sight information for the Doppler shifts, which we do in the arcade modelling. 

Data from the CHIANTI atomic database \citep[version 8.0.7;][]{1997A&AS..125..149D,2015A&A...582A..56D} was used with the physical properties of the plasma to forward model the emissivity of Fe~\textsc{xxi} 1354.1\AA\ spectral line in each grid cell. The contribution functions $G(\lambda, n_{e}, T)$ were built using the standard atomic data in CHIANTI, assuming ionisation equilibrium. These were tabulated with a resolution of $\delta \log T = 0.05$ and $\delta \log n_{e} = 0.5$. In each grid cell $G(\lambda, n_{e}, T)$ was interpolated to the $n_{e}$ and $T$ of the plasma, and the emissivity calculated as:

\begin{equation}\label{eq:emiss}
	j_{\lambda, z} = A_{Fe} G(\lambda,n_{e},T)n_{e}(z)n_{H}(z),
\end{equation}

    \noindent where $n_{H}$ is the hydrogen density, and $A_{Fe}$ is the elemental abundance of iron in the solar atmosphere. We used the abundances value from \cite{2012ApJ...755...33S} $A_{Fe} = 7.85$, defined on the usual logarithmic scale, where $A_{H} = 12$ (the abundance relative to hydrogen, $A_{\mathrm{rel}}$, is obtained using $A_{\mathrm{rel}} = 10^{A_{\mathrm{log}}-A_{H}}$ for an abundance value expressed on the logarithmic scale, $A_{\mathrm{log}}$). 
    There is much debate over which abundance value to use for iron and other low first ionisation potential (FIP) elements during flares. Coronal abundances of low FIP elements are enhanced relative to photospheric values. For iron the photospheric abundance is $A_{Fe} = 7.50$ \citep{2009ARA&A..47..481A}, but the coronal abundance can be as high as $A_{Fe} = 8.10$ \citep{1992PhyS...46..202F}. While some studies have shown that in flares the low FIP elements actually have abundances closer to photospheric \citep[e.g][since ablation carries chromospheric material into the flare loop]{2014ApJ...786L...2W}, studies of the iron abundance in flares have produced a range of values $A_{Fe} = [7.56,~7.72,~7.91,~7.99]$ \citep{2014ApJ...786L...2W,2015ApJ...803...67D,2012ApJ...748...52P,2014SoPh..289.1585N}. In the latter case the abundance varied in times, and in fact sometimes exceeded the canonical coronal value. We chose the value from \cite{2012ApJ...755...33S} as a middle ground and note that the intensity values of the synthetic spectra could be some factor smaller if alternative values of $A_{Fe}$ were used. Our quoted intensties would be a factor $2.2\times$ smaller if the photospheric abundance value from \cite{2009ARA&A..47..481A} was used. 
    
The intensity in each grid cell is then

\begin{equation}\label{eq:emiss2}
	I_{\lambda,z} = j_{\lambda,z} \delta z,
\end{equation}

\noindent where $\delta z$ is the size of the grid cell. The emergent Fe~\textsc{xxi} intensity from either the full 1D field-aligned atmosphere, or some range of heights, was obtained by summing the intensity in each grid cell along the extent of interest, $I_{\lambda} = \Sigma_{z_{1}}^{z_{2}} I_{\lambda,z}$. 

\begin{figure}
	\centering 
	\hbox{
	\subfloat{\includegraphics[width = .5\textwidth, clip = true, trim = 0.cm 0.cm 0.cm 0.cm]{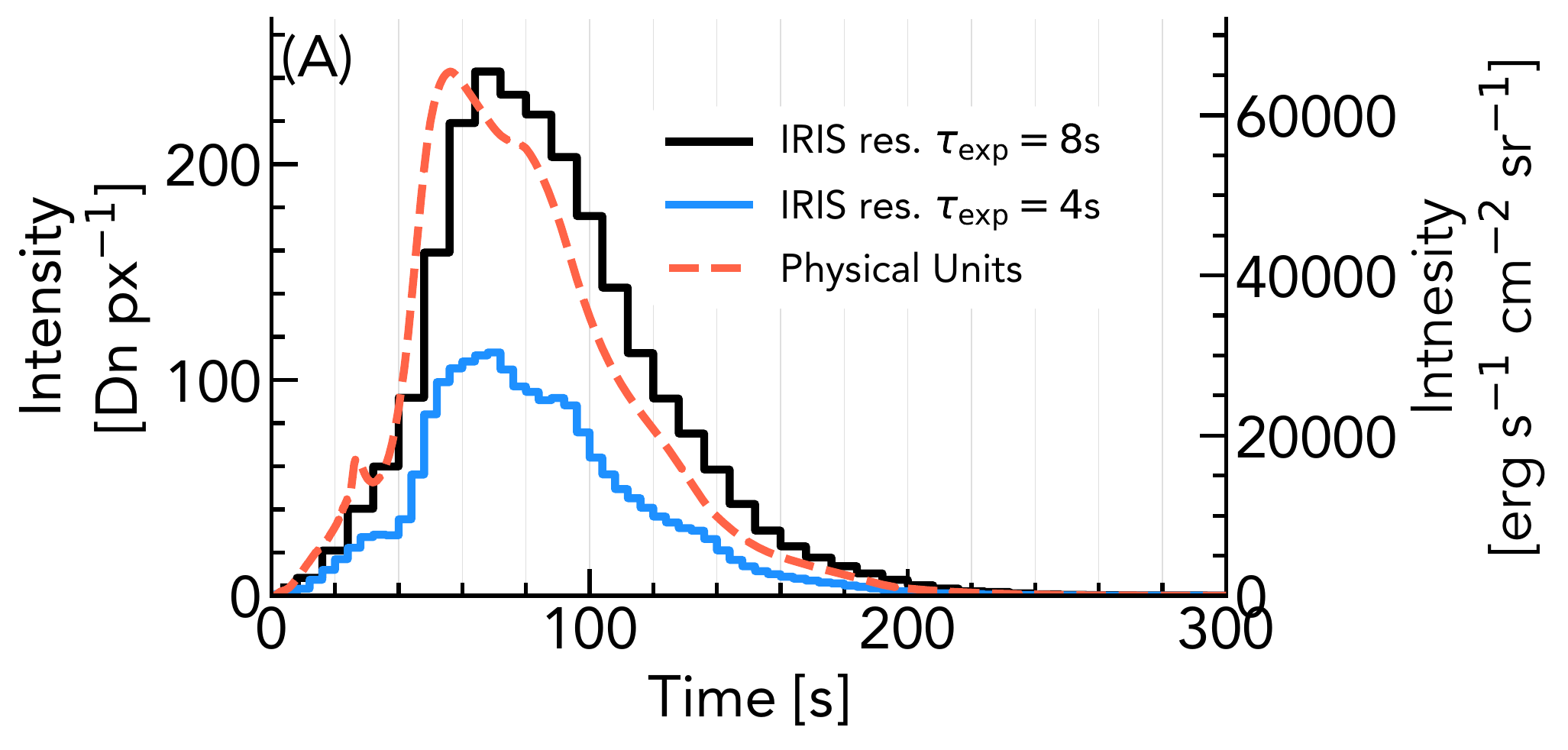}}	
	}
	\hbox{
	\subfloat{\includegraphics[width = .5\textwidth, clip = true, trim = 0.cm 0.cm 0.cm 0.cm]{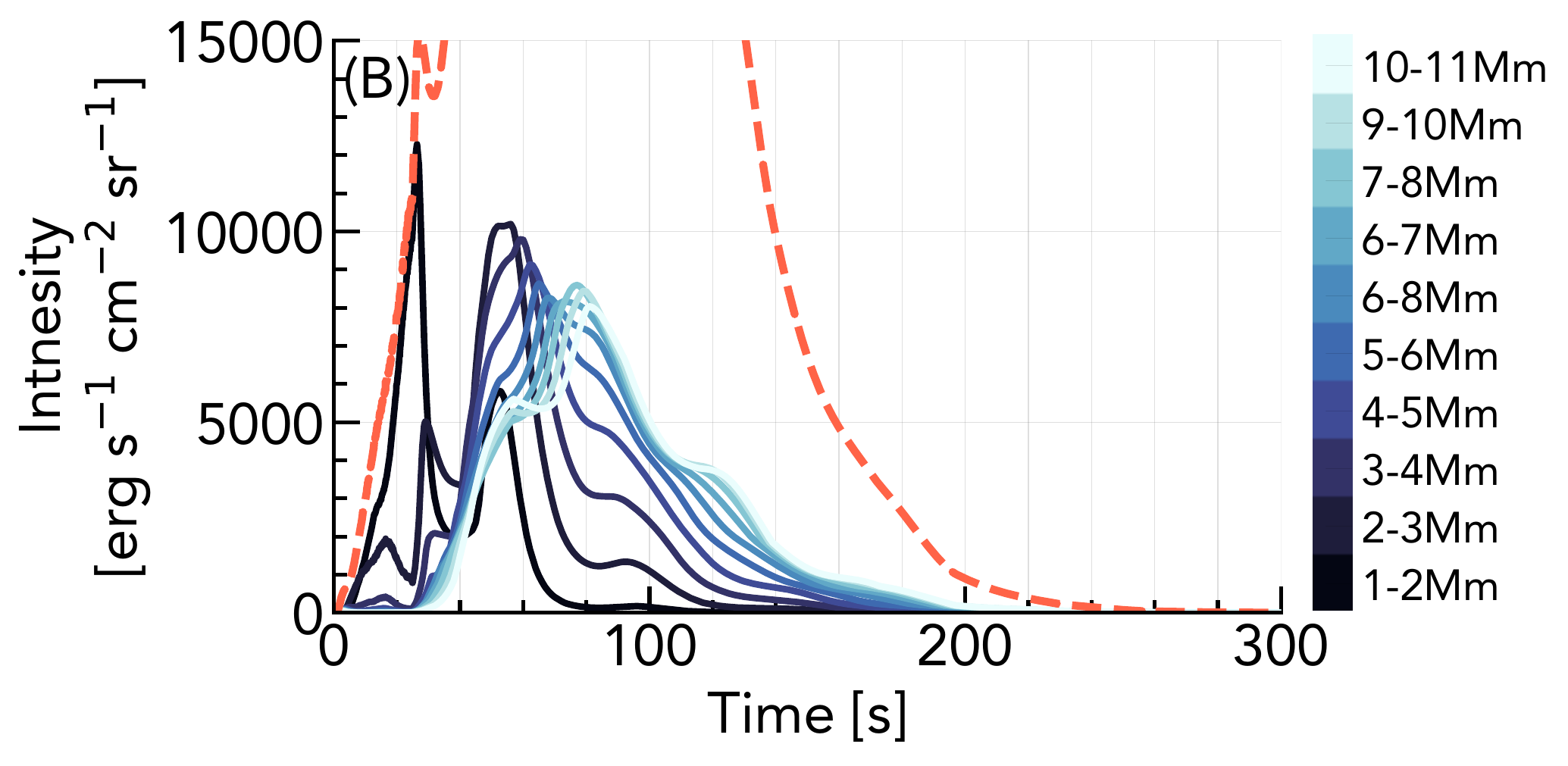}}	
	}
	\caption{\textsl{(A) Lightcurves of the Fe~\textsc{xxi} 1354.1\AA\ line from the 1D field-aligned model, both at native resolution and physical units (red dashed line, right hand axis), and as would be observed by IRIS at two typical exposure times (blue and black solid lines, left hand axis). These represent emission integrated through the full loop. Panel (B) shows the lightcurves broken down into height ranges (colored lines), with the total also indicated (red dashed line).}}
	\label{fig:1dfexxi_lcurve}
\end{figure}

Lightcurves are shown in Figure~\ref{fig:1dfexxi_lcurve}. Both at the native resolution and at IRIS resolution (spectral and temporal resolution, and convolved with the spectrograph effective area), with typical exposure times, $\tau_{\mathrm{exp}}= [4, 8]$~s, and poisson noise applied. These illustrate that observations made with longer exposures can obscure dynamics present in the simulation. The emission peaks several tens of seconds following cessation of the electron beam, as the corona takes time to become sufficiently hot and dense to produce the maximum Fe~\textsc{xxi} emission. These lightcurves represent emission summed over the full flare (a single flaring, if observed by IRIS). Also shown in Figure~\ref{fig:1dfexxi_lcurve} is a breakdown of the intensity within several height ranges, showing that initially the emission is stronger lower in the atmosphere near the TR, before gaining strength at higher altitudes. During the peak of emission Fe~\textsc{xxi} is forming over an extended range of heights.


\section{Modelling the Flare Arcade in 3D}

\subsection{Data-constrained Identification of Coronal Loops}
\cite{2018arXiv180700763A} performed data constrained 3D modelling of active region heating via nanoflares, using observations of active region AR11726 from SDO/AIA, \textsl{Hinode}'s Extreme-ultraviolet Imaging Spectrometer \citep[EIS;][]{2007SoPh..243...19C}, and the Extreme Ultraviolet Normal Incidence Spectrograph \citep[EUNIS;][]{2014ApJ...790..112B}. They constructed a 3D model of the magnetic field in AR11726 using the Vertical Current Approximation Non-linear Force Free Field (VCA-NLFFF) technique of \cite{2013ApJ...763..115A,2016ApJS..224...25A}. In that approach the photospheric magnetic field from SDO/HMI is extrapolated into the corona, with SDO/AIA observations of observed coronal loops used to ensure that the extrapolated magnetic field lines are co-aligned with actual coronal structures. We summarise some important features here but full details can be found in \cite{2018arXiv180700763A}. 

The 3D magnetic field within a volume that extended 0.5$R_\odot$ in the Cartesian $x-y$ plane and 1.5$R_\odot$ in the $z$ plane was obtained, tracing the magnetic field lines passing through each voxel of a $315\times315\times430$ heliocentric Cartesian grid. A total of 2848 field lines were traced. Area expansion of each loops into the corona was allowed, conserving magnetic flux. The cross-sectional area was defined as $A(s)B(s) = B(s=0)A(s=0)$, for a distance $s$ along the loop. A value of $220$~km was assumed for $A(s=0)$ based recent high-resolution observations \citep[e.g.][]{2014SoPh..289.4393K,2017ApJ...840....4A}. Within each voxel it is possible for loops to overlap due to the area expansion. Emission from overlapping loops within a voxel was averaged. Note that \texttt{RADYN} does not currently include area expansion but this is a planned upgrade to the code. 

\cite{2018arXiv180700763A} applied nanoflare heating to each loop and the time-averaged radiated emission within each voxel was computed, as was the time-averaged differential emission measure (DEM). DEMs are a commonly used tool to define the amount of material $n_{e} n_{H}$ (the electron density and hydrogen density) present in some certain temperature range $\delta T$, along the line of sight $h$: $\mathrm{DEM(T)} = n_{e} n_{h} \frac{\mathrm{d}h}{\mathrm{d}T}$ (with units of cm$^{-5}$~K$^{-1}$). Observationally, the DEMs can be derived from multi-wavelength observations and are a means to estimate the distribution of plasma within a temperature range. 

The heliocentric coordinates of each voxel were projected onto a 2D pixel grid (the observational planes of SDO, EIS or EUNIS), and the radiated emission or DEM in each voxel added to the appropriate pixel of the `image.'  If multiple voxels (and portions of multiple loops) corresponded to the same pixel then that emission was summed. In this manner the superposition of loops, the loop geometry, and the viewing angle in the observational plane were all self-consistently accounted for. 

Spectral information could then be convolved with instrumental responses to produce synthetic maps of EIS and EUNIS data, and the DEM maps could be convolved with AIA responses to produce synthetic AIA maps. A best-fit time-averaged DEM model of AR11726 was produced by \cite{2018arXiv180700763A}, from model-data comparisons of their nanoflare simulations to EIS \& EUNIS observations.

The 3D magnetic structure (the identified loops) and the observational pixel grid to which they were projected was used by us to construct a flare arcade model. The time-averaged DEM model produced by \cite{2018arXiv180700763A} was used as our $t = 0$~s, pre-flare, DEM. While this active region did not flare, we \rev{took advantage of} the \rev{existing 3D magnetic field construction} to begin our development of, and experiments with, flare arcade modelling.  \rev{Reproducing the work of \cite{2018arXiv180700763A} for a flaring active region is a non-trivial task, but } future efforts will involve \rev{the use of active regions that did flare}. 

The observational plane pixel grid had a pixel scale of $\delta x = \delta y = 0.6^{\prime\prime}$ (the AIA pixel scale). Our initial effort kept this same pixel grid even for forward modelling IRIS observables as sampling a finer grid would require remaking the voxels with smaller dimensions. While this is desirable, it is a time consuming process and for the demonstration of our new approach we believe keeping the original scaling is sufficient. Future efforts will explore the use of finer grids.

\subsection{Synthetic Flare Arcade Model}\label{sec:arcademodeldescript}
AR11726 was located fairly close to the solar limb. To make the line of sight projections more straightforward for this initial work the whole AR was translated by $45\degree$, around the $x-z$ plane, to near disk centre. This rotated region is referred to as AR11726rot, and was used to construct a flare arcade model.

\begin{figure*}
	\centering 
	{\includegraphics[width = .85\textwidth, clip = true, trim = 0.cm 0.cm 0.cm 0.cm]{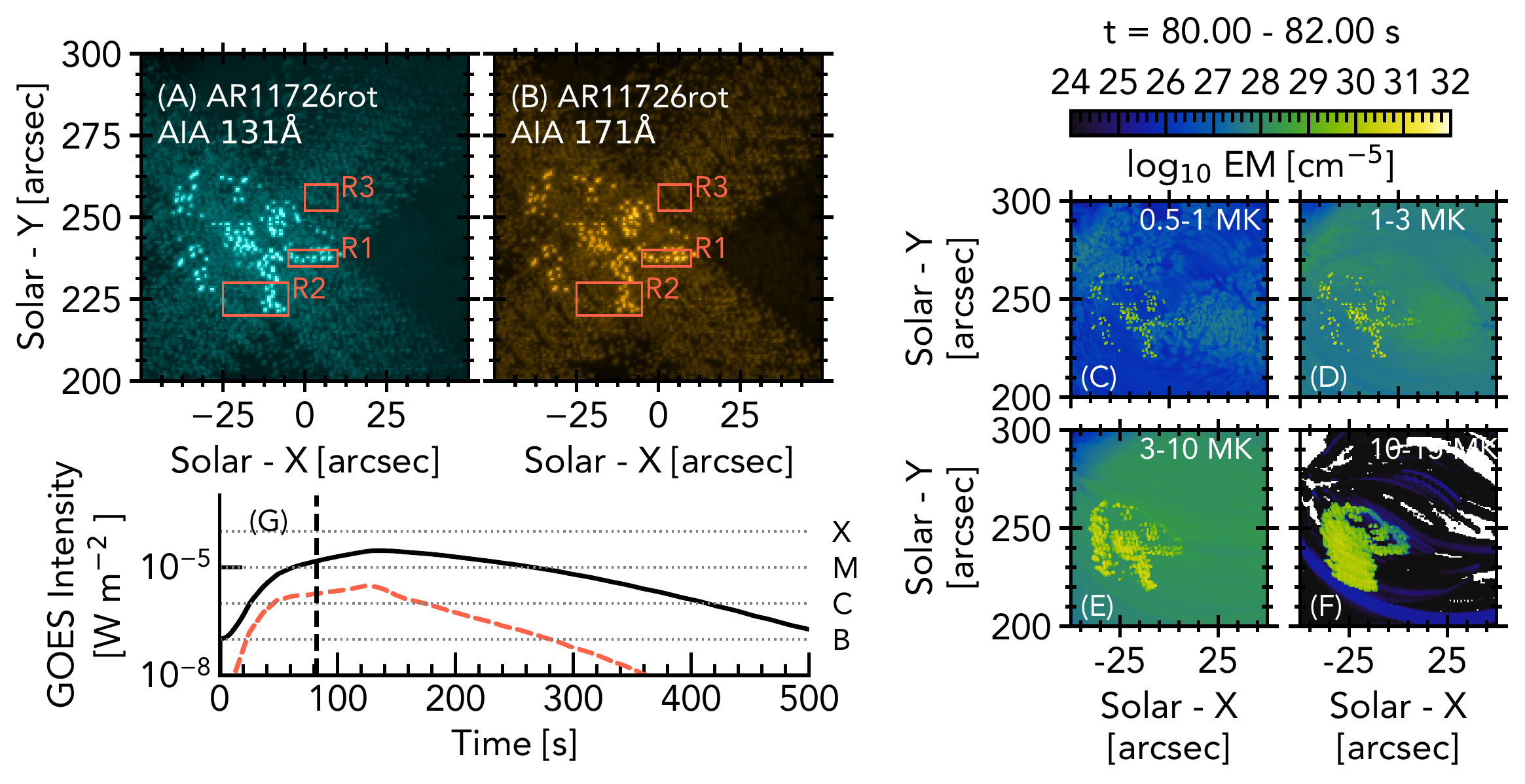}}	
	\caption{\textsl{The flare arcade as would be observed by AIA 131\AA\ and 171\AA\, at $t=80$~s into the flare arcade simulation (panels A \& B), and the EMs summed in various temperature ranges show the structures at present at different temperature regimes (panels C-F). The synthetic GOES $1-8$~\AA\ (black) and $0.5-4$~\AA\ (orange dashed) lightcurves are shown in panel (G), where the vertical dashed line indicates the current time, and the horizontal lines indicate flare class. The red boxes indicate sub-regions used study of spatially integrated AIA lightcurves and EMs. An animated version is available online.}}
	\label{fig:flare_aia_map1}
\end{figure*}
\begin{figure*}
	\centering 
	{\includegraphics[width = .85\textwidth, clip = true, trim = 0.cm 0.cm 0.cm 0.cm]{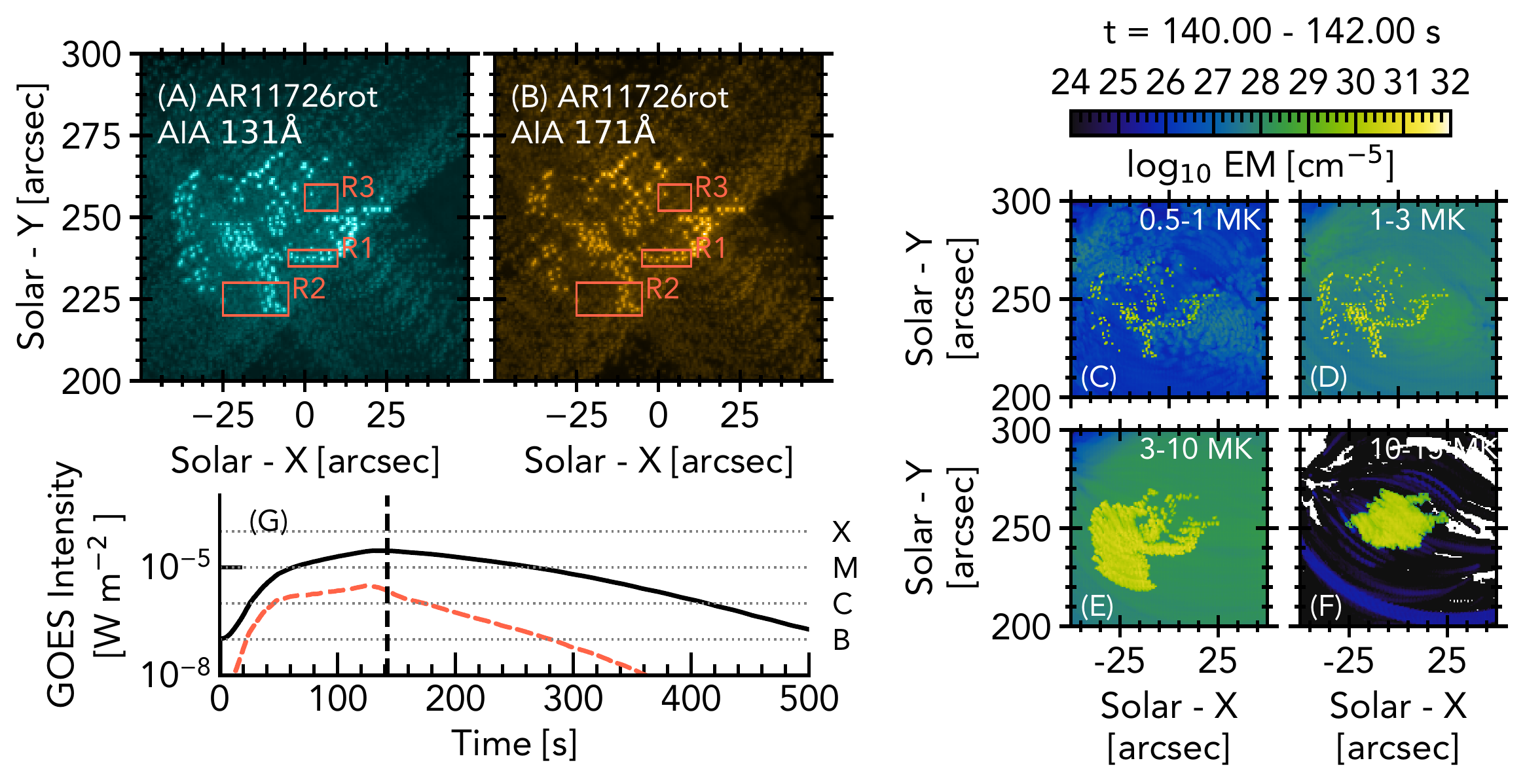}}	
	\caption{\textsl{Same as Figure~\ref{fig:flare_aia_map1}, but at $t=140$~s during the flare model. }}
	\label{fig:flare_aia_map2}
\end{figure*}
\begin{figure*}
	\centering 
	{\includegraphics[width = .85\textwidth, clip = true, trim = 0.cm 0.cm 0.cm 0.cm]{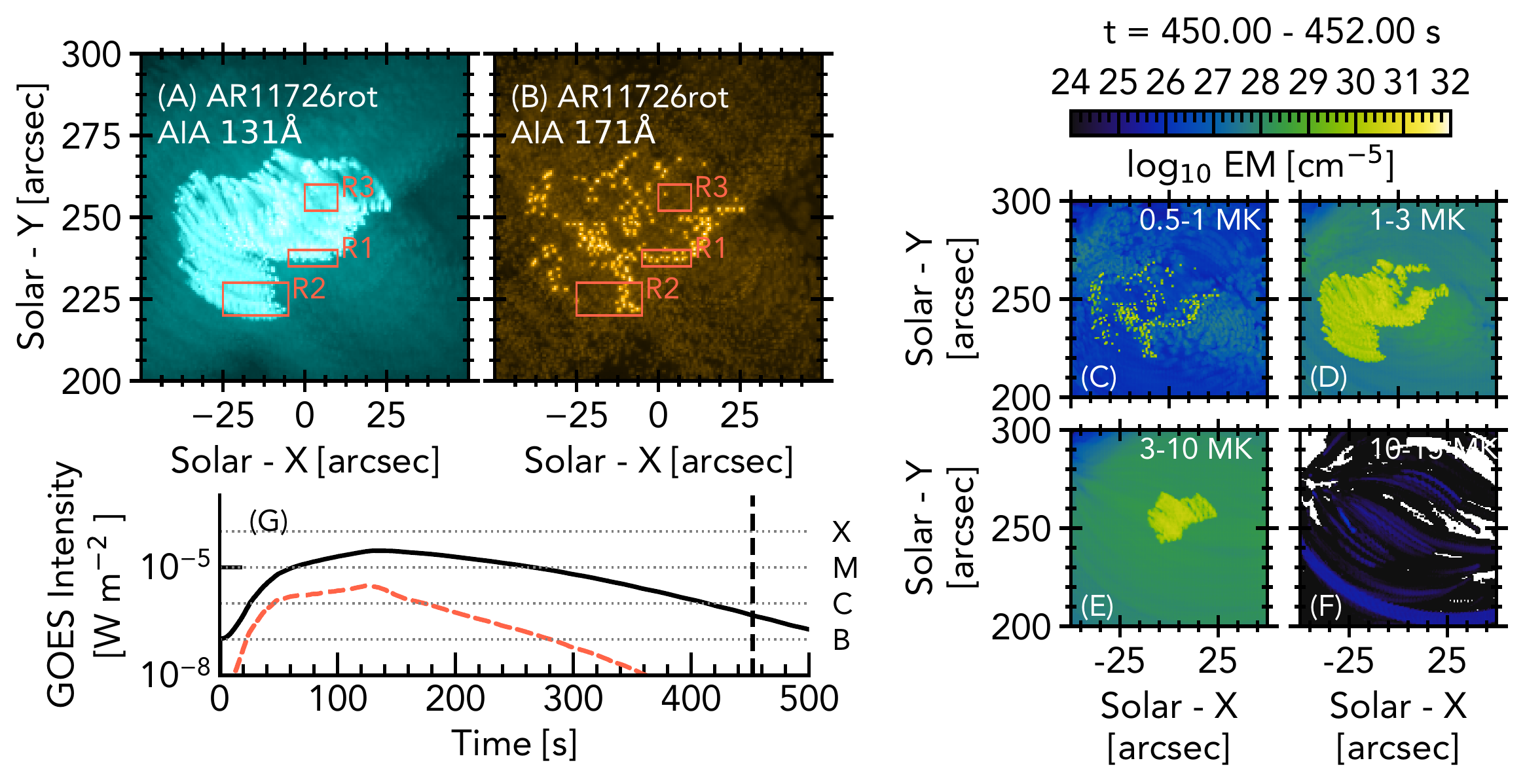}}	
	\caption{\textsl{Same as Figure~\ref{fig:flare_aia_map1}, but at $t=450$~s during the flare model. }}
	\label{fig:flare_aia_map3}
\end{figure*}

A subset of 180 loops were selected from the 2848 loop to form the flare arcade, chosen for their proximity to each other and for their loop lengths that were close to the 20~Mm long \texttt{RADYN} loops (recall that \texttt{RADYN} models one leg of a semi-circular loop, so that the total loop length would be 20~Mm, even though we only simulate 10~Mm from photosphere to corona). These loops were ordered by the distance of the loop apex from disk centre and activated in groups of $N_{\rm{loop}}=5$ every $\tau_{\rm{ac}} = 3$~s starting at $t=0$~s. Progressing at an arcade simulation cadence of $0.5$~s each voxel of the appropriate loop was filled with either the DEM or the velocity DEM \citep[VDEM,][which simply defines the amount of emission measure within that has a line-of-sight velocity in the range $v,v+\delta v$]{1995ApJ...447..915N} from the \texttt{RADYN} flare simulation. The same \texttt{RADYN} simulation was used for every loop, but since loops were activated at different times there were various stages of evolution during any one arcade snapshot. 

To produce synthetic images and broadband spectral responses (e.g. SDO/AIA or GOES soft X-rays) maps of the DEM in the 2D $x-y$ observational plane were produced. In each arcade snapshot the DEM, and the height grid on which it was defined, were interpolated from the \texttt{RADYN} simulation to $t_{\rm{sim}}$, the arcade simulation time. The arcade loops were described as distance, $s$, from one footpoint to the other with 200 cells per loop (the spatial resolution $\delta s$ varied). For each cell, $i$, within the loop the temporally interpolated DEM were spatially interpolated to $s_{i}$ \& $s_{i+1}$ and summed to find the total value in that cell. This was then divided by the distance $\delta s = s_{i+1}-s_{i}$ to obtain the DEM field \citep[see equation 5 in][though note in our case the DEM field is not time averaged]{2018arXiv180700763A}. When projected onto the 2D solar $x-y$ observational plane, each cell $i$ may span multiple [x,y] pixels. The pixels to which that cell should be projected were identified, and the DEM field multiplied by the appropriate line of sight to obtain the DEM. This DEM was summed with any DEM already projected onto that pixel either from the background or from another loop. The DEM maps $[x,y]$ were then convolved with the instrumental responses as described in Section~\ref{sec:synaia_goes}.

To produce synthetic spectra a similar method was used, with each voxel instead populated by the appropriate VDEM. After interpolating the VDEM between $s_{i}$ \& $s_{i+1}$  the spectrum over a passband $\Delta \lambda$ computed from the VDEM using Equations~\ref{eq:emiss} \& \ref{eq:emiss2}. The velocity information of the VDEM was used to Doppler shift the line where appropriate, and thermal broadening was applied based on the local temperature. Summing the intensity of the spectra between $s_{i}$ and $s_{i+1}$ provided the total intensity in cell $i$, $I_{\lambda,i}$. This was divided by $\delta s$, to provide the average emissivity in cell $i$, $j_{\lambda, i}$. The appropriate pixels into which $j_{\lambda, i}$  should be added were identified, and $j_{\lambda, i}$ multiplied by the `projected height' to yield intensity. As with the DEMs, this intensity was summed with any existing intensity in that pixel. Spectral maps $[\lambda, x,y]$ were then convolved with instrumental responses as described in Section~\ref{sec:syniris}

Summing the DEM or spectra within a pixel means that tahe projection of structures into the same pixel was taken into account and the effects of superposition of loops (with different velocity fields) is reflected in the output spectra. Both the synthetic images and spectroscopy explicitly assumes optically thin conditions and these methods are not suitable for modelling spectral lines or continua for which opacity is non-negligible. 

Any snapshot of the flare will show loops that were activated at some prior time (and thus at some time through their evolution), some that are newly activated, and some that are yet to be activated, with the progression of the arcade mimicking observations of flares (albeit, without ribbon separation in this initial effort). The parameters  $N_{\rm{loop}}=5$ and $\tau_{\rm{ac}} = 3$ were arbitrarily chosen \rev{(on the basis that they produced an M-class flare with soft X-ray lightcurve that exhibited a quick rise, with slow decay time)} for this \rev{proof-of-concept,} initial experiment, but when simulating a specific event these can be tailored. 

For each temperature bin the emission measure (EM) is EM(T) = DEM(T)$\times \delta T$, where $\delta T$ is the bin spacing. Figures~\ref{fig:flare_aia_map1},~\ref{fig:flare_aia_map2}~\&~\ref{fig:flare_aia_map3} show maps of the emission measure (EM) summed over various temperature ranges during both impulsive phase and gradual phase (an animated version is available online). 

\begin{figure*}
	\centering 
	{\includegraphics[width = .85\textwidth, clip = true, trim = 0.cm 0.cm 0.cm 0.cm]{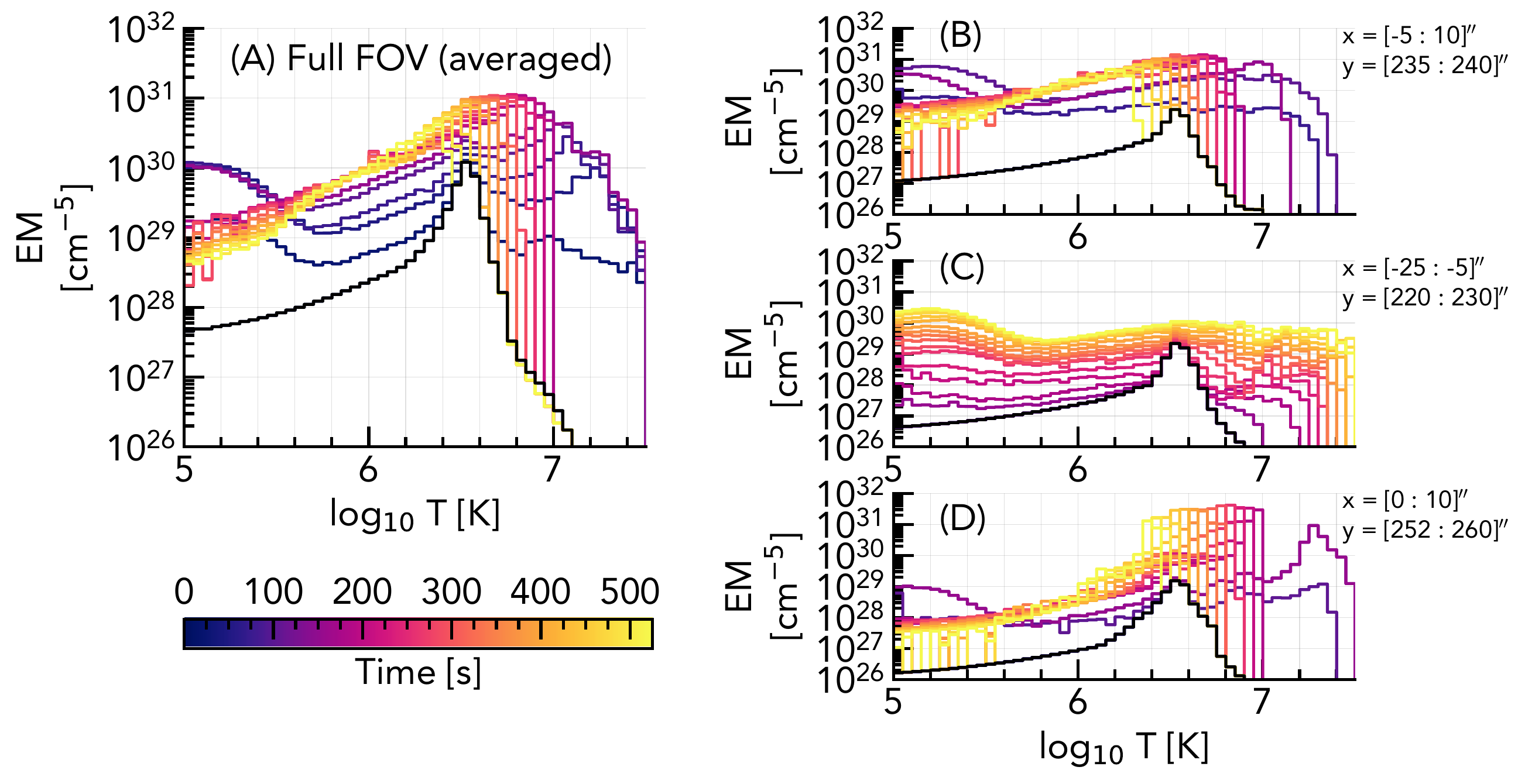}}	
	\caption{\textsl{EMs at various times in the arcade simulation. Panel A is the full field of view, and Panels B-D are the sub-regions identified in Figure~\ref{fig:flare_aia_map1}. The black line in each panel is $t = 0$~s.}}
	\label{fig:dem_arcade}
\end{figure*}

The EMs as functions of temperature are shown in Figure~\ref{fig:dem_arcade} for the full field of view and the sub-regions (identified in Figure~\ref{fig:flare_aia_map1}). In each case the EM was averaged over the appropriate area and the temporal evolution shown. 
Plasma in excess of $10$~MK is present at the earliest times in the flare, appearing as both loop and footpoint like sources. Plasma at temperatures $>25$~MK is present, albeit only for a short time and at low emission measure. The transition region steepens and narrows during the flare so that within a pixel of our synthetic observation there is both TR and coronal plasma. Foopoints therefore contain emission spanning tens of thousands to millions of Kelvin. Regions with mainly loop or looptop pixels (e.g. panel (D) on Figure~\ref{fig:dem_arcade}) show the EM peak at $T>10$~MK, falling steeply towards lower temperatures. At later times the EM increases in plasma at several MK while material cools. Regions that include footpoints have flatter EMs, with strong emission at cooler temperatures, extending from kK to MK.


\section{Synthetic SDO/AIA \& GOES Emission}\label{sec:synaia_goes}

To produce the AIA maps the DEMs were convolved with the temperature response of the coronal AIA filters. This was done at a cadence of $0.5$~s (the cadence at which we progressed the arcade model) with emission assumed to be unchanging over that time period. This provided intensity in DN~s$^{-1}$~pix$^{-1}$. Poisson noise was added, and the images convolved with the instrumental point spread function (PSF). Images were integrated over exposure times of $\tau_{\mathrm{aia}} = 2$~s, so that the final intensity was DN~pix$^{-1}$. Saturation was not taken into account, and it is likely that in reality these images would suffer from saturation and pixel bleeding effects, which are unfortunately common during flare observations.

Soft X-rays in the $[1-8]$~\AA\ range were synthesised to mimic GOES lightcurves. For each snapshot, the DEM was integrated over the field of view, with the spatial scale being $\delta x = \delta y = 0.6^{\prime \prime}$. For each temperature bin the emission measure was calculated EM = DEM$\times \delta T$, and the flux of thermal X-rays in the range $E = [1,50]$~keV (resolution $\delta E = 0.25$~keV) seen at Earth was calculated. This was done using the routine \texttt{f\_vth.pro} included in the SolarSoftWare package \citep[SSW,][]{1998SoPh..182..497F}, and included both lines and continuum. This flux was interpolated to wavelength grids of $[1,8]$~\AA\ and $[0.5,4]$~\AA, the GOES long and short passbands. Spectra were folded with the GOES-15 spectral response \citep[see][and \texttt{goes\_tf\_coeff.pro} in the SSW GOES tree]{2005SoPh..227..231W}. The GOES flux is shown on Figures~\ref{fig:flare_aia_map1}, with the flare peaking at GOES class M2.0 (this is largely a function of the flaring volume, so the number of loops we chose to activate in our model, since the same heating rate was applied to each loop). 

\begin{figure*}
	\centering 
	{\includegraphics[width = .75\textwidth, clip = true, trim = 0.cm 0.cm 0.cm 0.cm]{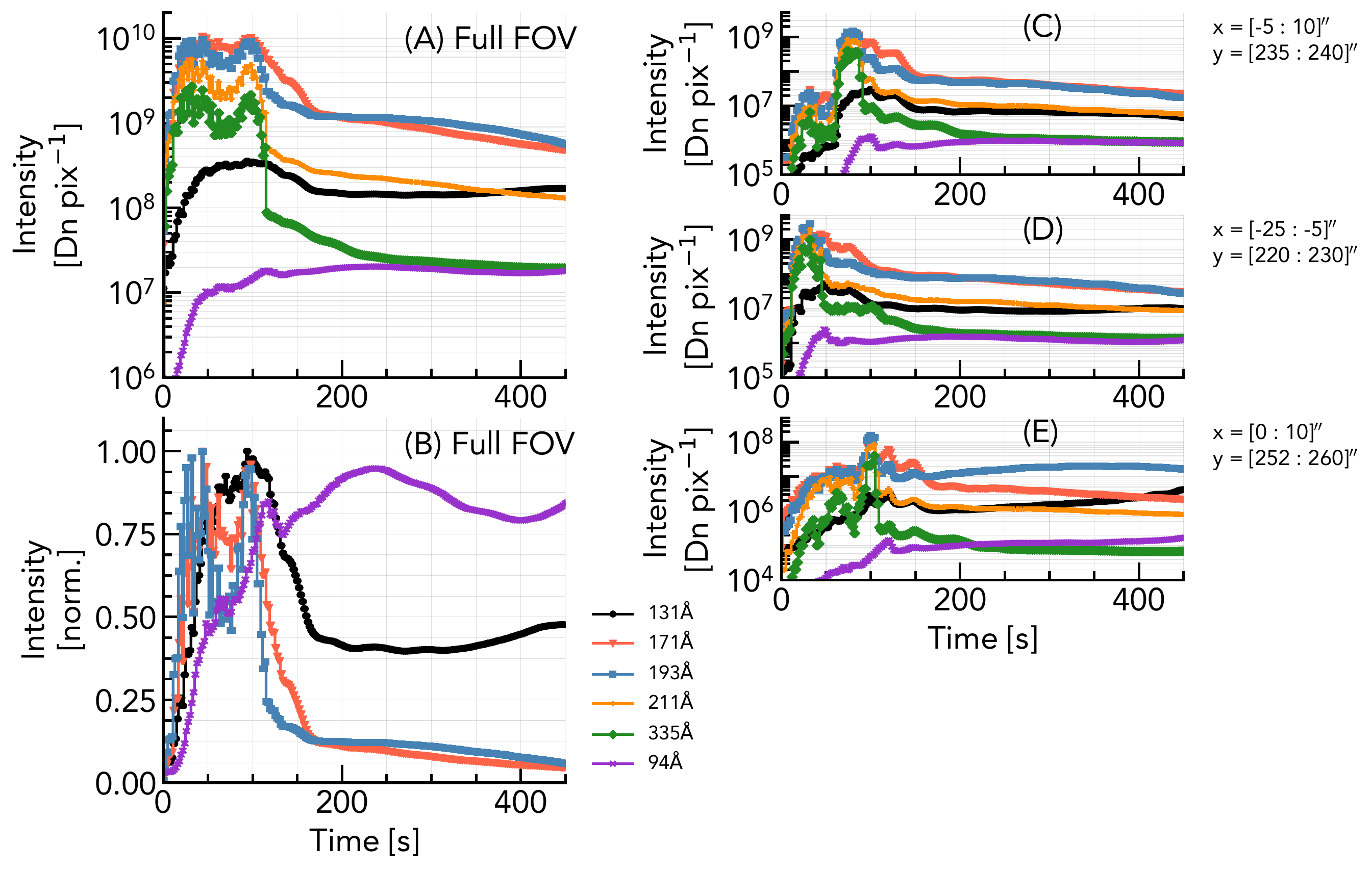}}	
	\caption{\textsl{AIA lightcurves from the arcade simulation. Panels A \& B are the full field of view, and Panels C-E are the sub-regions identified in Figure~\ref{fig:flare_aia_map1}.}}
	\label{fig:flare_aia_lcurves}
\end{figure*}

Maps of the flare as would be observed by AIA131 and AIA171 are shown in Figures~\ref{fig:flare_aia_map1},~\ref{fig:flare_aia_map2}~\&~\ref{fig:flare_aia_map3}. The footpoints (`ribbons') brighten significantly, followed by the loops as they fill with ablated material. We do not run the simulation past $t=550$~s but as the loops cool from $>10$~MK and material drains from the loops we would expect the loops to brighten in the channels that probe cooler plasma. Lightcurves of the full field of view, and of several sub-regions covering footpoint and loop sources are shown in Figure~\ref{fig:flare_aia_lcurves}. The AIA PSF results in the crosswise effects seen in the images, that artificially results in emission being present in the sub-regions before the flare actually appeared in those locations. In panels (C) \& (E) of Figure~\ref{fig:flare_aia_lcurves} there is emission several tens of seconds prior to the peak that is not present if the PSF is not applied. The 131\AA\ channel peaks somewhat after the cooler channels as the amount of plasma $>10$~MK increases (this channel also samples cooler emission $T\sim0.4$~MK). When plasma begins to cool towards the end of the flare the 94\AA\ channel peaks, as emission cools from $T>10$~MK to $T\sim6$~MK. 

While qualitatively similar to observations from AIA and GOES the timescales are too rapid in our model. The GOES flare is almost over within $\approx 10$ minutes, whereas observed flares have longer lifetimes \citep[e.g.][]{2013ApJ...778...68R}. 

It is common to estimate the (isothermal) temperature of the flaring soft X-ray emission from the ratio of the two GOES channels \citep[e.g.][]{2005SoPh..227..231W}. Though this isothermal assumption can lead to inaccurate temperatures \citep[e.g][]{2014SoPh..289.2547R}, this metric is still a useful one for studying the global flare. Studies have investigated flare heating and cooling timescales and characteristics based on these temperatures \cite[e.g.][]{2013ApJ...778...68R}. Since GOES is a sun-as-a-star observatory the observed flux is a combination of both the flare and the disk integrated emission. It is important, therefore, to remove this background emission before temperatures are derived. Determining this background is not always trivial, and the choice of background can have an impact on the resulting GOES temperatures \citep[e.g.][]{1990ApJ...356..733B,2012ApJS..202...11R}. 

\begin{figure*}
	\centering 
	{\includegraphics[width = .65\textwidth, clip = true, trim = 0.cm 0.cm 0.cm 0.cm]{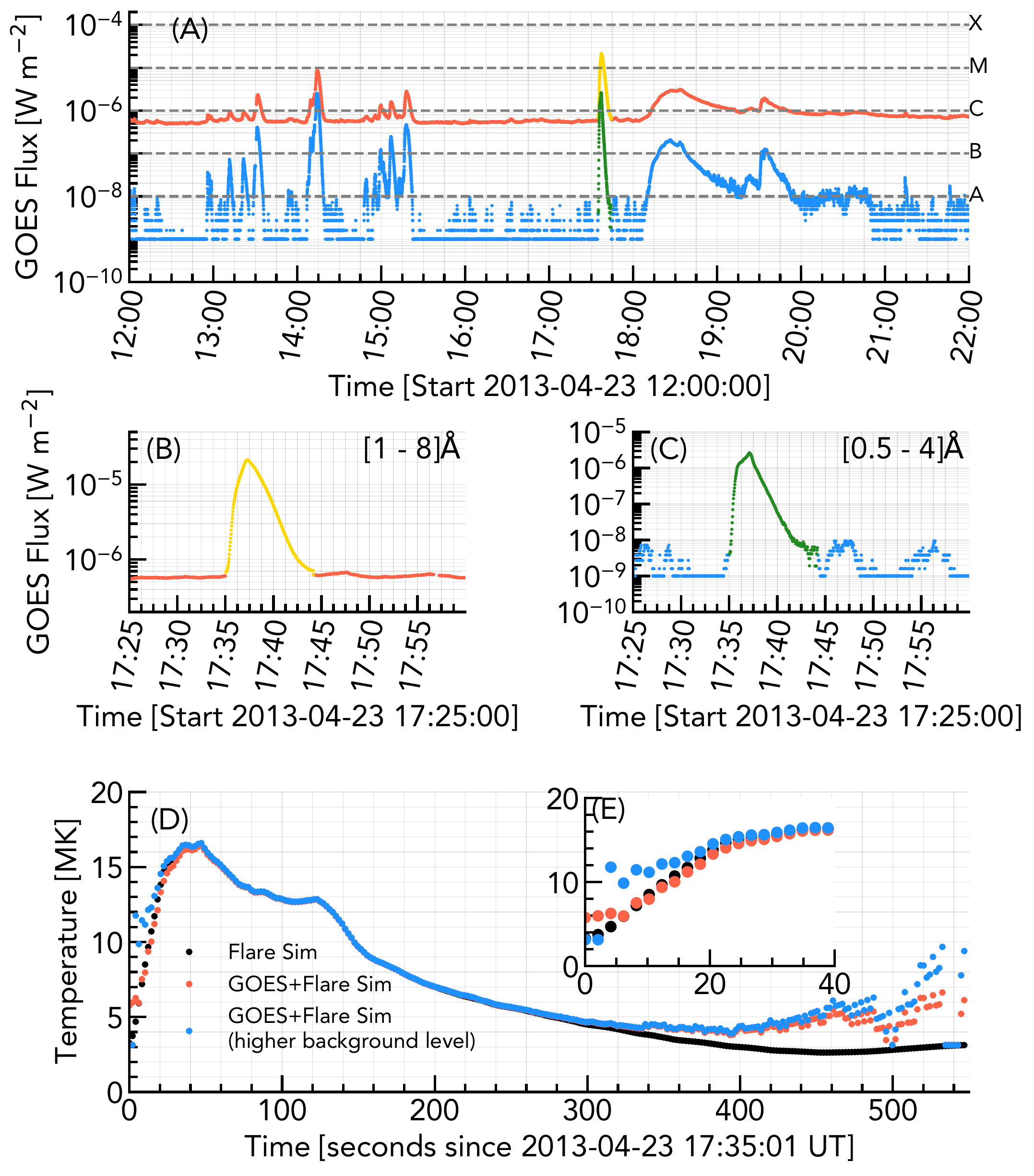}}	
	\caption{\textsl{Synthetic GOES SXR emission from the arcade model overlaid on observations from 23rd April 2013. Panel (A) shows a ten hour window of the observed GOES long (red) and short (blue) passbands with the synthetic flare emission added (yellow and green points, respectively). Panels (B) and (C) show a closer look at the time of the modelled flare. Panel (D) shows the derived isothermal temperatures with the inset panel (E) showing a more detailed view of flare onset. The black points are temperature derived from only the flare simulation, the orange is from the flare plus observed GOES emission with the average background subtracted, and the blue is from the same but with a higher background level subtracted.}}
	\label{fig:goestemp}
\end{figure*}

Our flare arcade simulation contains no contamination from background emission, so the temperature derived using our synthetic GOES fluxes ($\mathcal{F_\mathrm{flare}}$) represent the flare-only scenario. Of course GOES observations contain noise and an uncertain background, so to demonstrate the impact of this on our simulated results we combined our arcade model with actual GOES observations. The 550s of synthetic GOES fluxes, $\mathcal{F_\mathrm{flare}}$, were added to GOES observations from 2013-April-23 17:35~UT, $\mathcal{F_\mathrm{obs}}$, shortly after the time of the EUNIS observations, to give a total flux $\mathcal{F_\mathrm{tot}} = \mathcal{F_\mathrm{flare}}+\mathcal{F_\mathrm{obs}}$. Doing this allowed us to include a background level and also permitted us to include variations in the background due to solar sources and noise. Effectively, this is what GOES would have observed, had AR11726 actually flared. Note that we did not include photon counting statistics or the effects of digitization here \citep[see][]{2015SoPh..290.3625S}. The background level to subtract was measured by taking the mean of the GOES fluxes between 2013-April-23 [16-18]~UT, $\mathcal{F_\mathrm{back}}$.

Temperatures were computed for our arcade model for three cases: $\mathcal{F_\mathrm{flare}}$ (no background subtraction necessary),  $\mathcal{F_\mathrm{tot}} -\mathcal{F_\mathrm{back}}$, and $\mathcal{F_\mathrm{tot}} -1.15\mathcal{F_\mathrm{back}}$ (to demonstrate the impact of an uncertain background). 

From the appropriate fluxes the temperature was calculated using the SunPy V0.9.10 GOES software \citep{sunpy_community2020}. Figure~\ref{fig:goestemp} shows the temperatures derived from the three cases, along with the GOES lightcurves. Including the background and variability have a small impact at the start of the event but overall the behaviour is similar to the flare-only `clean' results. In both cases it takes $\sim20$~s for the temperature to reach $>10$~MK, and $\sim40-50$~s to reach the peak of $16.4$~MK. The decay is noisier for the case including a background and variability, as would be expected. Increasing the background level does have an impact in the initial stage of the flare. Here, the temperature exceeds $10$~MK after only 6s. This may not appear to be a significant difference since the timescales involved are short, but if one is interested in the very start of the flare then the choice of background be impactful.

The temporal behaviours of the GOES temperature, SXR flux, and EM in our model are qualitatively consistent with the picture of intense footpoint heating followed by chromospheric ablation which carries material into the flaring loops, increasing their density. The temperature is also consistent with observed GOES temperatures. However, the timescales are too short compared to observations \citep[e.g][]{2017ApJ...851....4R,2019ApJ...874...19S}. The FWHM was $\tau_{\mathrm{FWHM}} = 149.5$~s, and decay time was $\tau_{\mathrm{decay}} = 99.3$~s, where $\tau_{\mathrm{decay}} = \frac{-F_\mathrm{1-8\AA}(t)}{dF_{\mathrm{1-8\AA}}(t)/dT}\bigg|_{t=t_{\mathrm{end}}}$, following \citep{2017ApJ...851....4R}. While some observations show growth times, decay times, and FHWM on the order of those in our model, \cite{2017ApJ...851....4R} demonstrated that these would imply a ribbon separation close to $3-5$~Mm, which is smaller than our model. This discrepancy is likely partly because the individual loops cool too quickly, and partly due to the loop lengths used. Flare SXR timescales are correlated with ribbon separation, which \cite{2017ApJ...851....4R} showed is due to ongoing reconnection and loop expansion so that a range of loop lengths are involved in the flare. Loop length and cooling timescales will be explored with our model in a future work where we attempt to simulate an observed flare arcade.


\section{IRIS Fe~\textsc{xxi} Emission From the Flare Arcade Model}\label{sec:syniris}
\subsection{\rev{Examples of Fe~\textsc{xxi} Observations During Solar Flares}}\label{sec:observsec}
\begin{figure*}
	\centering 
	{\includegraphics[width = .75\textwidth, clip = true, trim = 0.cm 0.cm 0.cm 0.cm]{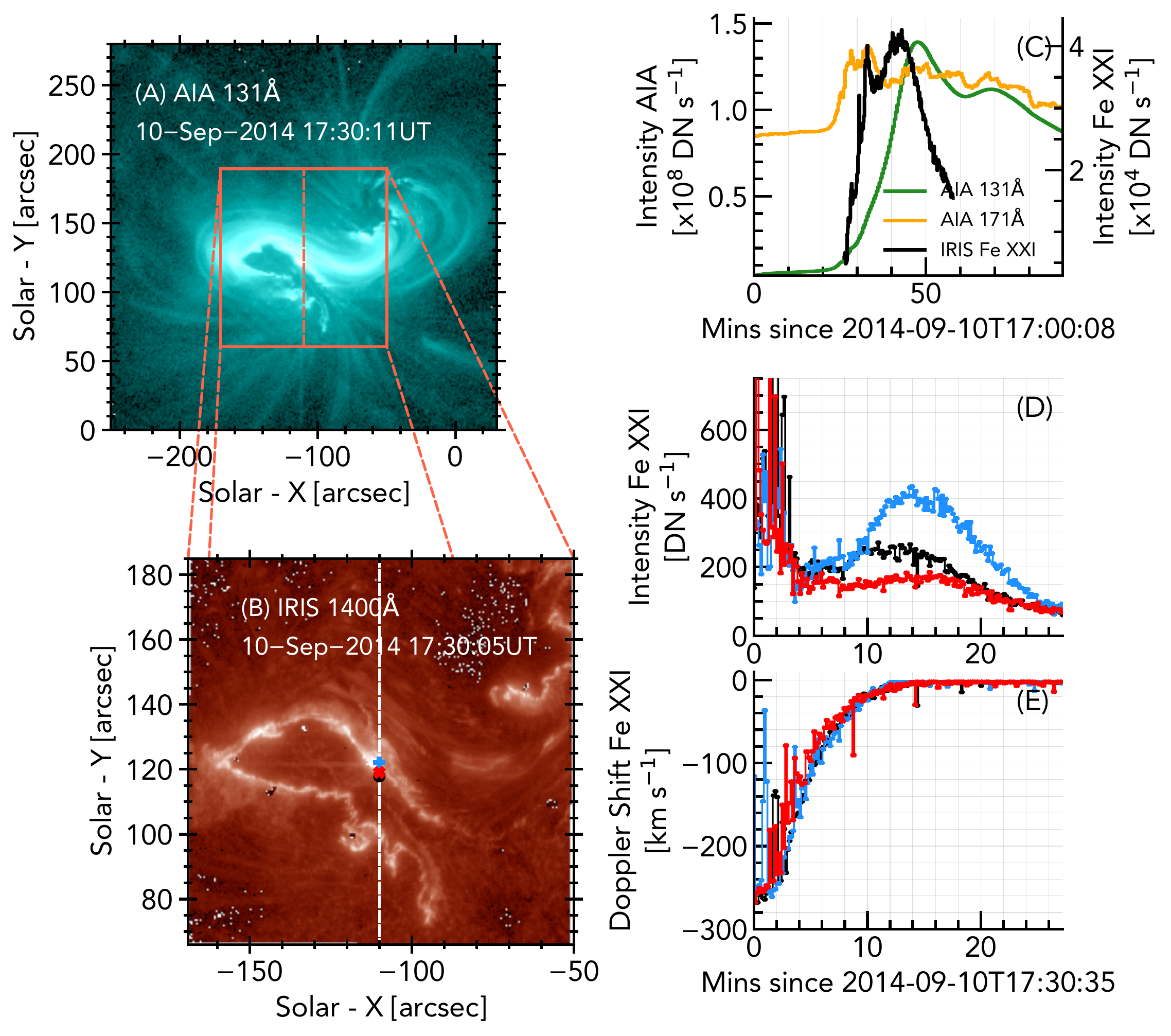}}	
	\caption{\textsl{Observations of the 2014-September-10 X class solar flare. Panel (A) shows AIA 131~\AA\ emission, with the field of view of the IRIS spacecraft indicated (the dashed line is the IRIS slit). Panel (B) shows the IRIS 1400~\AA\ SJI emission. Panel (C) shows the AIA 131~\AA\ (green), AIA 171~\AA\ (orange), and IRIS Fe~\textsc{xxi} (black) lightcurves, integrated over the full field of view. Panel (D) shows the  Fe~\textsc{xxi} lightcurves from the pixels indicated in panel (B), and panel (E) shows the Doppler shifts of those same pixels.}}
	\label{fig:obsoverview}
\end{figure*}
In recent years, thanks to the high spectral and spatial resolution afforded by IRIS, Fe~\textsc{xxi} 1354.1~\AA\ emission from flares has been studied in detail to probe flaring plasma properties and dynamics. This high temperature flare line offers excellent scope to interrogate model predictions of the coronal portion of flare loops. Prior to IRIS this line was observed on disk using \textsl{Skylab} \citep{1975ApJ...196L..83D}, and the UVSP instrument onboard the \textsl{Solar Maximum Mission} \citep{1986ApJ...309..435M}. Doppler shifts of up to $\sim200$~km~s$^{-1}$ were seen, line broadening was observed to decrease from flare maximum as the flare progressed, and profiles were usually quite asymmetric. However, owing to limits in spatial resolution these early results likely suffered from blending of Fe~\textsc{xxi} profiles from multiple sources, creating the asymmetries. 
\begin{figure}
	\centering 
	\hbox{
	        \subfloat{\includegraphics[width = 0.5\textwidth, clip = true, trim = 0cm 0cm 0cm 0cm]{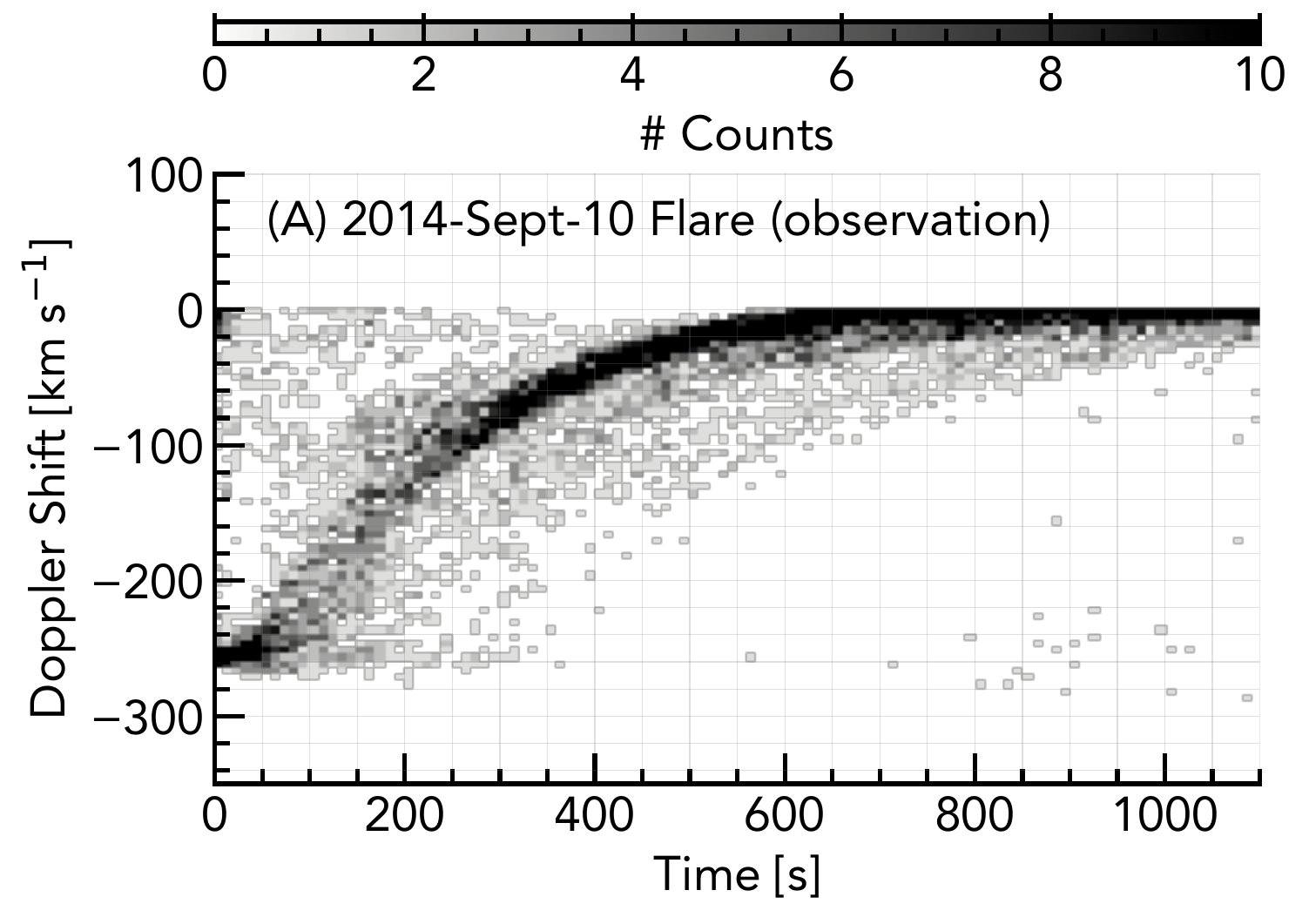}}
                   }
          \hbox{
	        \subfloat{\includegraphics[width = 0.5\textwidth, clip = true, trim = 0cm 0cm 0cm 0cm]{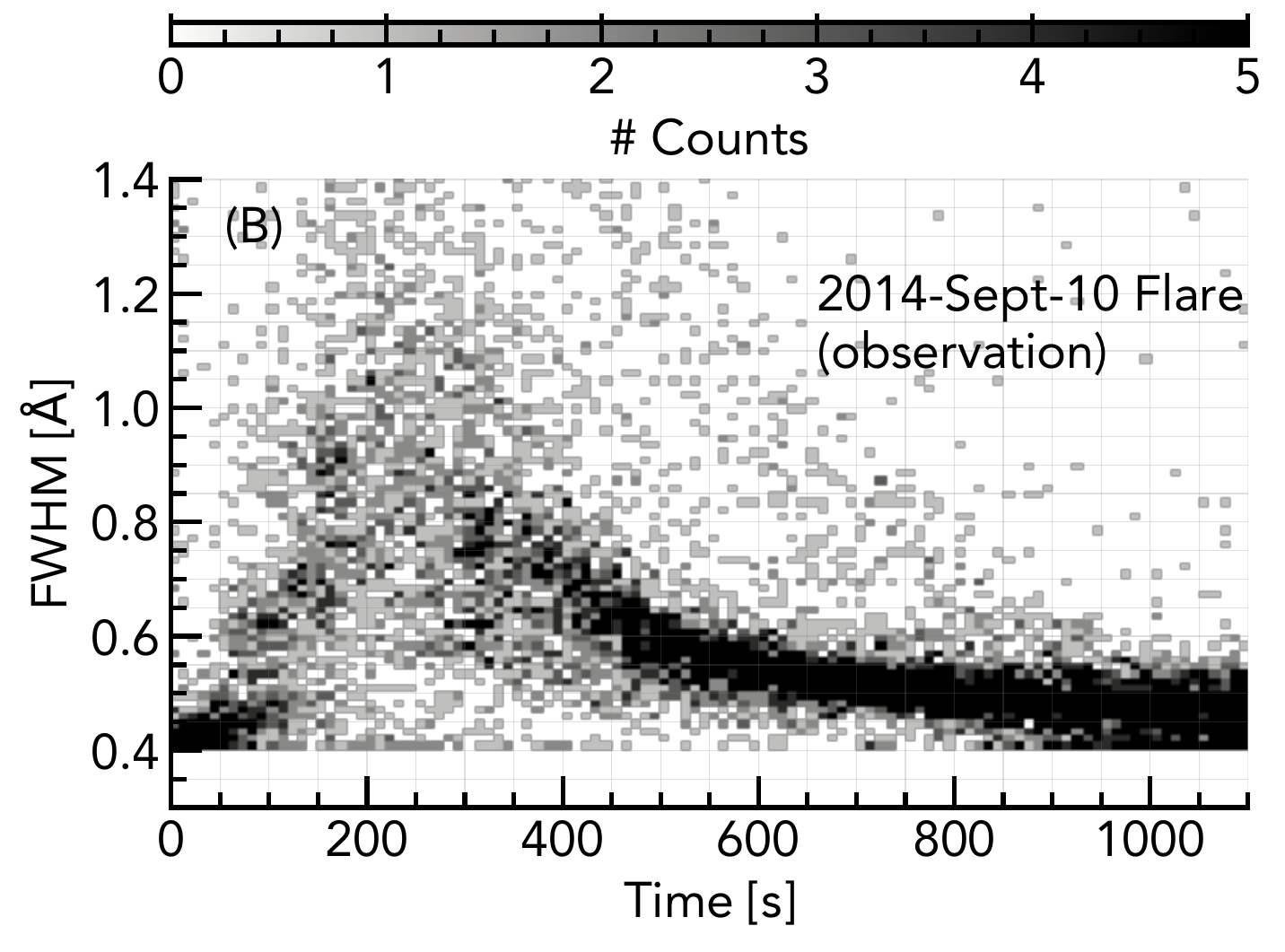}}	
                   }
          \caption{\textsl{ \rev{A superposed epoch analysis was applied to results from Gaussian fits to Fe~\textsc{xxi} line profiles observed during the 2014-September-10 X class solar flare}. Panel (A) shows the temporal evolution of Doppler shifts ($\delta t = 10$~s, $\delta v = 10$~km~s$^{-1}$). Panel (B) shows the line widths, defined as the FWHM of the Gaussian fits ($\delta t = 10$~s, $\delta W = 12.5$~m\AA). The temporal origin of each pixel in both cases was the time of first detection.}}
	\label{fig:superposedepoch_observed}
\end{figure}
With the advantage of improved spectral and spatial resolution IRIS observations indicated that the Fe~\textsc{xxi} profiles were fully blueshifted (with a single component), showed significant line broadening, with largely symmetric profiles \citep[e.g.][and references therein]{2014ApJ...797L..14T,2015ApJ...811..139T,2015ApJ...799..218Y,2015ApJ...803...84P,2015ApJ...807L..22G,2015ApJ...805..167S,2016ApJ...816...89P}. Generally the line profiles are observed to initially be weak, strongly blue shifted, broad, and symmetric \citep{2019ApJ...879L..17P}. As the flare progresses they strengthen in intensity, shift towards rest, and narrow. Intensities are around a few tens to a few hundred Data Number (DN) for IRIS exposure times of $\tau_{\mathrm{exp}}\sim4-8$~s. They first appear in ribbons/footpoints before spreading up loop legs to the loop apex (interpreted as chromospheric ablation). Some observations suggest that Fe~\textsc{xxi} sources are offset from the flare ribbon by $\sim0.3^{\prime\prime}$ \citep{2015ApJ...799..218Y}. However, an explanation for this could simply be that the signal is hidden by the bright continuum present in footpoints, so that by the first time it is detected the ribbon front has travelled some distance. 

Doppler shifts (almost exclusively blueshifts in the impulsive phase ribbons and loops) range from a few tens of km~s$^{-1}$ to $200-300$~km~s$^{-1}$, and take several hundred seconds to decay from the peak to rest. There is some evidence of small redshifts appearing in the gradual phase due to draining of flare plasma. Loop tops show profiles that are typically near stationary with negligible broadening. Line widths during the flare range from thermal width (nominally $0.43$~\AA, assuming ionisation equilibrium and a peak formation temperature of $11.2$~MK) at loop tops, to $\sim0.5-1$~\AA\ in ribbons and loops. The origin of the excess line width is not known with certainty and poses an interesting challenge for modelling to reproduce. 

Observations of the 10th September 2014 solar flare are presented in Figure~\ref{fig:obsoverview}, to place the lightcurves and Doppler shifts from our model in context. Shown in that figure are maps of AIA and IRIS SJI emission, lightcurves of the full field of view, and the intensity and Doppler shifts of representative pixels. The Fe~\textsc{xxi} \rev{observations} were de-blended and fit with a single Gaussian function. This same flare was studied in \cite{2015ApJ...807L..22G} and \cite{2019ApJ...879L..17P}, and we encourage the reader to consult those sources for a fuller discussion. 

\cite{2015ApJ...807L..22G} discovered a strikingly organised behaviour of the Fe~\textsc{xxi} line Doppler shifts. A superposed epoch analysis of the type performed by \cite{2015ApJ...807L..22G} is recreated here, to \rev{provide a comparison} against our modelled superposed epoch analysis \rev{presented in Section~\ref{sec:superposed}}. \rev{In addition to the Doppler shifts, we present a superposed epoch analysis of the observed line widths also.} The temporal origin is defined as the time at which a first time the line appeared clearly, with a peak of at least $10$~DN. This definition is somewhat subjective as it involved manually assessing movies of each of the 84 pixels used in the analysis. This was a difficult determination due to the very weak signal when the line first appears, compounded by the fact that it often drifts into the wavelength window of IRIS (meaning the peak blueshifts quoted here may be lower limits). The temporal binning is $\delta t = 10$~s and Doppler shift binning is $\delta v = 10$~km~s$^{-1}$.  Note the clustering of Doppler motions, shown in Figure~\ref{fig:superposedepoch_observed}(A), and the smooth decay to rest over several hundred seconds. This is less tightly clustered than \cite{2015ApJ...807L..22G}, likely due to a stricter determination of when the profile first appeared by \cite{2015ApJ...807L..22G}. We find that some pixels exhibit a brief rise phase. 

Figure~\ref{fig:superposedepoch_observed}(B) \rev{shows the analysis applied to the observed line widths}. The widths quoted here are the FWHM obtained from Gaussian fitting ($W = 2\sqrt{2 ln 2}\sigma$, for standard deviation of the Gaussian function $\sigma$), a combination of thermal width, instrumental width, and non-thermal width. In order to estimate the non-thermal widths one should subtract $0.43$~\AA\ from the the total widths. A line width binning of $\delta W = 12.5$~m\AA\ was used. At first some pixels show very little broadening (that we believe might be due to the fact that only partial profiles are detected initially, meaning the widths are underestimated), while others showed significant broadening with a large range of values present. In general there is more scatter in the line width distribution compared to Doppler shifts. 

\subsection{Forward Modelling IRIS Spectral Lines}
\begin{figure*}
	\centering 
	{\includegraphics[width = \textwidth, clip = true, trim = 0.cm 0.cm 0.cm 0.cm]{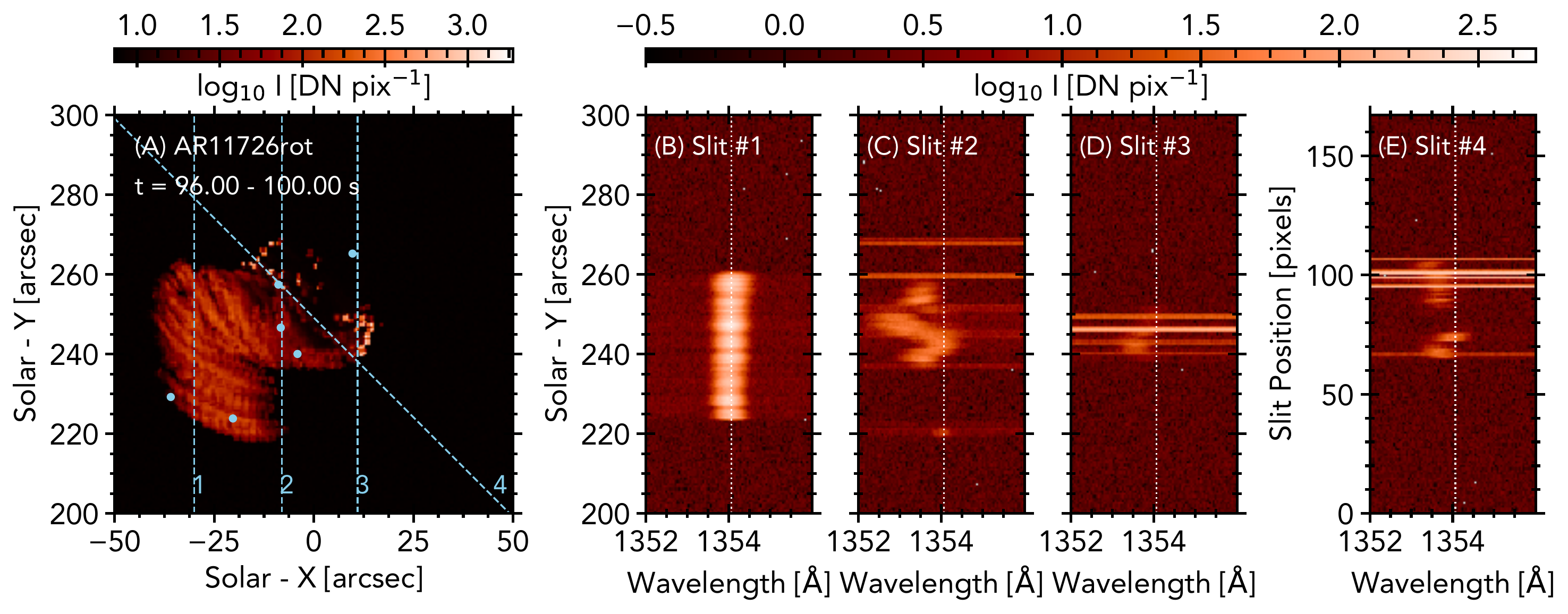}}	
	\caption{\textsl{The flare arcade AR11726rot at $t=98\pm2$~s. Panel (A) is a map of synthetic Fe~\textsc{xxi} emission, integrated over $1352-1356$~\AA, showing footpoint and loop flare sources at various stages of temporal evolution. The blue points indicate the sources for which the spectra are shown in Figure~\ref{fig:samplepixels}. Panels (B-E) show Fe~\textsc{xxi} from slits 1-4, respectively. Data has been degraded to IRIS resolution as described in the text, had poisson noise applied, and integrated with an exposure time of $4$~s. The broadband enhancement is caused by a strong flare continuum in the footpoint sources.}}
	\label{fig:flareiris2}
\end{figure*}
In our arcade model, the Fe~\textsc{xxi} 1354.1~\AA\ line was synthesised over the passband $[1352-1356]$~\AA\ using the approach described in Section~\ref{sec:arcademodeldescript}. Particularly in footpoint sources, the continuum is strongly enhanced during flares, which can drown out the Fe~\textsc{xxi} signal. To account for this effect we included the continuum in each cell. The continuum for the background AR was taken from the output of the \texttt{RH} radiative transfer code \citep{2001ApJ...557..389U}, solving for the FALC semi-empirical model atmosphere \citep{1993ApJ...406..319F}. For the flare continuum, the contribution function to the emergent intensity, $C_{\lambda,\mu}(z)$, \citep{1986A&A...163..135M,1998LNP...507..163C} was computed. The integral of  $C_{\lambda,\mu}(z)$ through height is the emergent intensity.  Included in  $C_{\lambda,\mu}(z)$ are various sources of emissivity, attenuated by optical depth. The sources of emissivity included here are various H processes (e.g. free-bound, free-free, H$^{-}$), scattering processes (e.g Rayleigh, Thomson), and background metals (in LTE). See \cite{2015SoPh..290.3487K,2017ApJ...836...12K} and \cite{Kerr_wlf_inprep} for further discussion of calculating continuum contribution functions. In each cell of the flaring loops  $C_{\lambda,\mu}(z)$ was interpolated to the appropriate time and position, and added to the emissivity before projecting into pixel $[x,y]$. This is appropriate as the continuum at these wavelengths is optically thin.

Spectra were converted from $I_\mathrm{erg}$ [erg~s$^{-1}$~cm$^{-2}$~sr$^{-1}$~\AA$^{-1}$] to $I_{\mathrm{phot}}$ [photons~s$^{-1}$~cm$^{-2}$~sr$^{-1}$~\AA$^{-1}$],  \rev{$I_{\mathrm{phot}}$ = $I_{\mathrm{erg}} \frac{\lambda}{hc}$}, and an exposure time of $\tau_{\mathrm{exp}} = 4$~s applied. Multiplying by the solid angle per pixel as viewed at 1 AU, smoothing with a spectral PSF \citep[assumed to be a Gaussian with FWHM of two FUV wavelength pixels, $\delta \lambda = 12.98$~m\AA~pix$^{-1}$,][]{2014SoPh..289.2733D}, multiplying by the IRIS effective area (calculated for 2014-September-10), and multiplying by the spectral dispersion, provided intensity in photons pix$^{-1}$. A background level of $B_{\mathrm{DN}} = 0.5$ DN~s$^{-1}$~pix$^{-1}$  \citep{2014SoPh..289.2733D} was converted to photon pix$^{-1}$, $B_{\mathrm{phot}} = 4B_{\mathrm{DN}} \tau_{\mathrm{exp}}$, where the factor is the number of photons DN$^{-1}$ \citep{2014SoPh..289.2733D}, and added to each exposure. Poisson noise was added and the intensity converted to DN~pix$^{-1}$. 

The 3D magnetic field structure obtained by \cite{2018arXiv180700763A}, and used by us here were originally designed to be projected onto an $x-y$ grid with pixel size equal to that of SDO/AIA: $0.6^{\prime\prime}$ pixel$^{-1}$. In order to obtain IRIS plate scales ($0.167^{\prime\prime}$ pixel$^{-1}$ we would be required to recreate the analysis of \cite{2018arXiv180700763A} using smaller voxel dimensions, a non-trivial exercise. Since this initial experiment is largely intended as a demonstration and proof-of-concept, have decided to keep the grid from \cite{2018arXiv180700763A}. Future efforts to model a specific flare will use the actual plate scale of the observations. 

\subsection{Synthetic Fe~\textsc{xxi} Line Profiles, Intensities and Temporal Evolution}
\begin{figure*}
	\centering 
	{\includegraphics[width = .75\textwidth, clip = true, trim = 0.cm 0.cm 0.cm 0.cm]{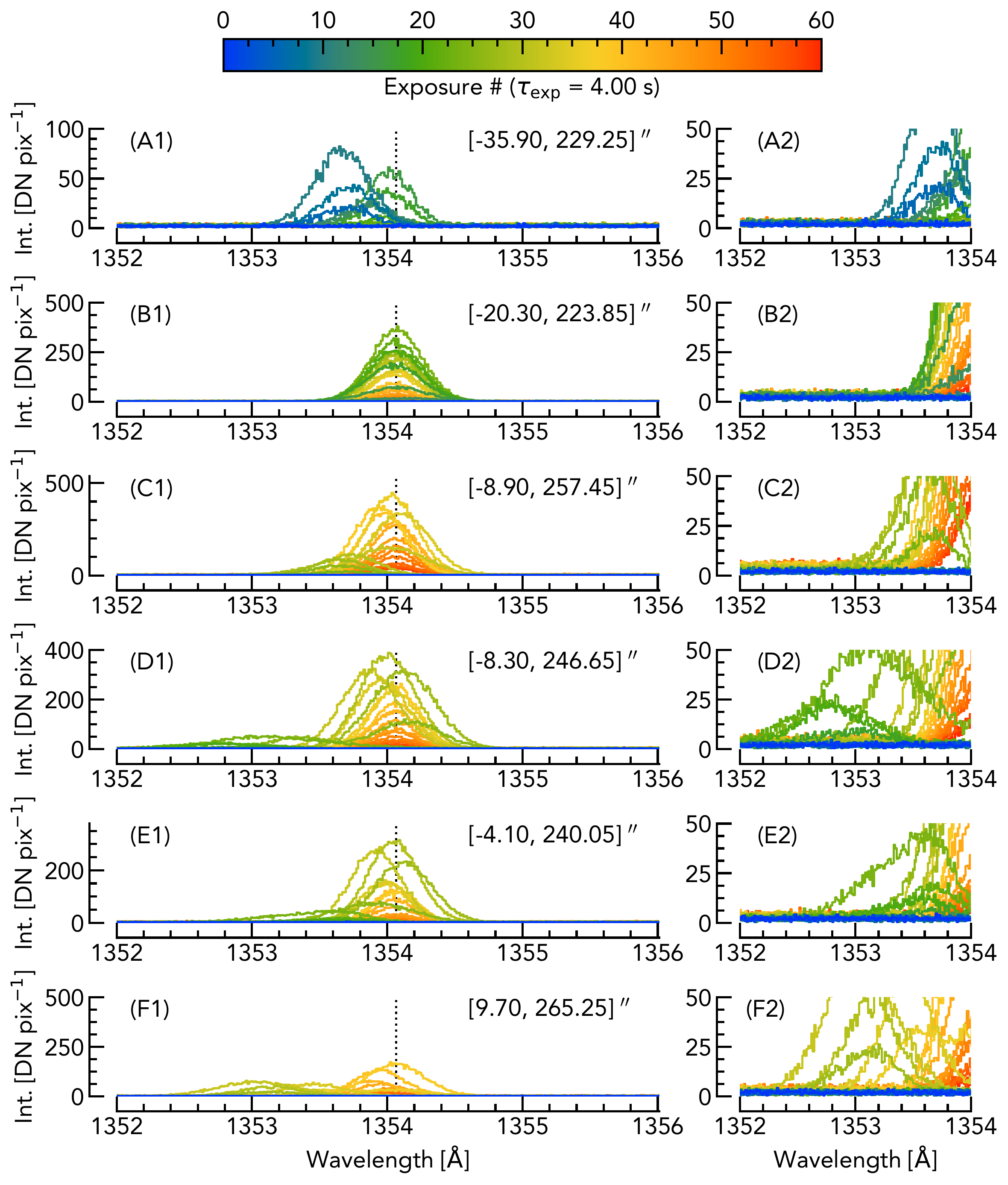}}	
	\caption{\textsl{Sample Fe~\textsc{xxi} line profiles from six pixels. Colour represents exposure number (time). Exposures 0 - 60 are shown, in steps of two. Panels A2-E2 show the same profiles, but zoomed in to see the weaker profiles in the initial impulsive phase of each pixel. An animated version is available online}}
	\label{fig:samplepixels}
\end{figure*}
\begin{figure*}
	\centering 
	{\includegraphics[width = .75\textwidth, clip = true, trim = 0.cm 0.cm 0.cm 0.cm]{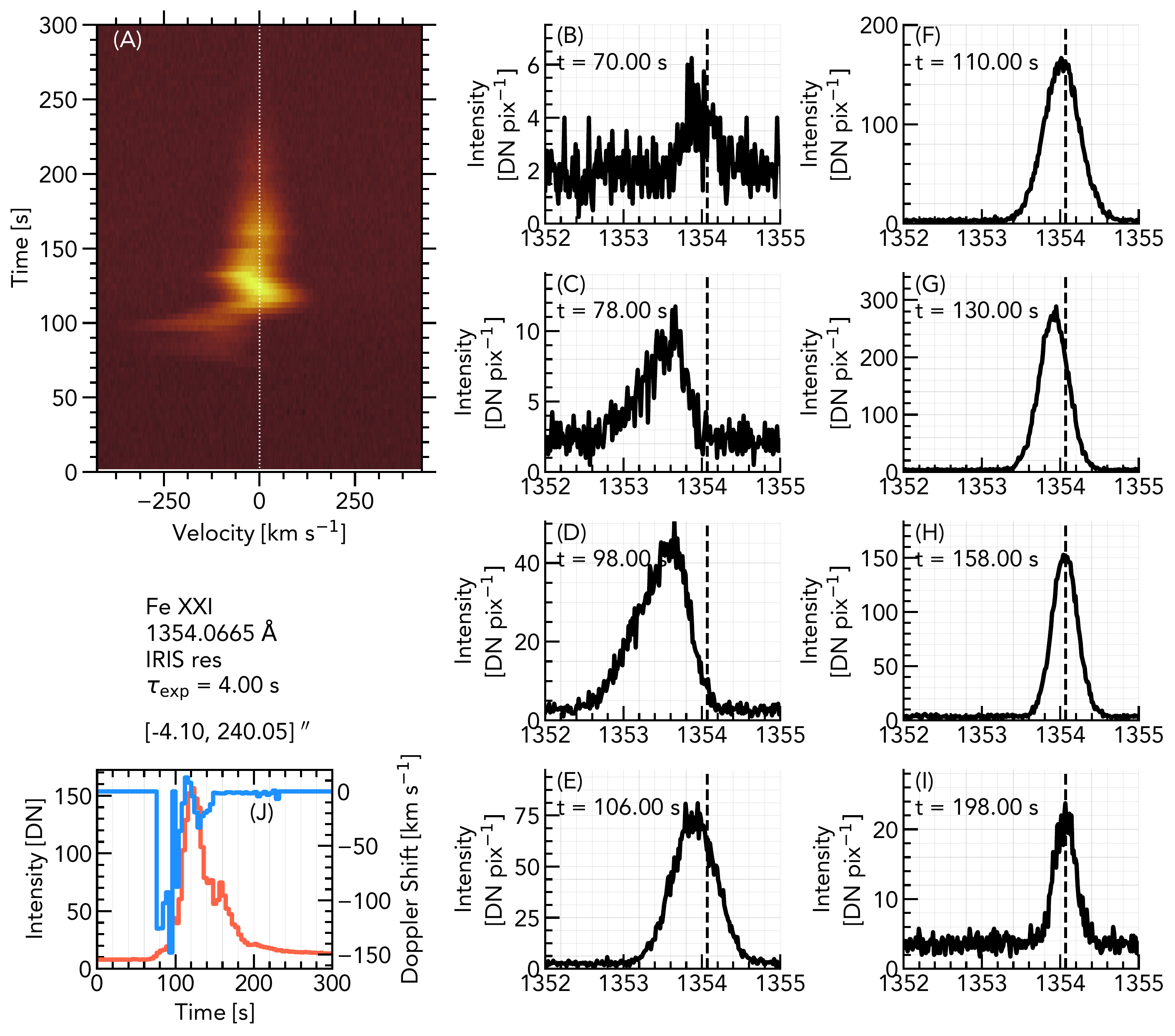}}	
	\caption{\textsl{A detailed overview of a single pixel ($x = [-4.10, 240.05]^{\prime \prime}$). Temporal behaviour is in panel A, and individual profiles at various times are in panel B-I. Lightcurves of intensity (orange) and Doppler shift (blue) are in panel J.}}
	\label{fig:samplepixels2}
\end{figure*}
\begin{figure*}
	\centering 
	{\includegraphics[width = .75\textwidth, clip = true, trim = 0.cm 0.cm 0.cm 0.cm]{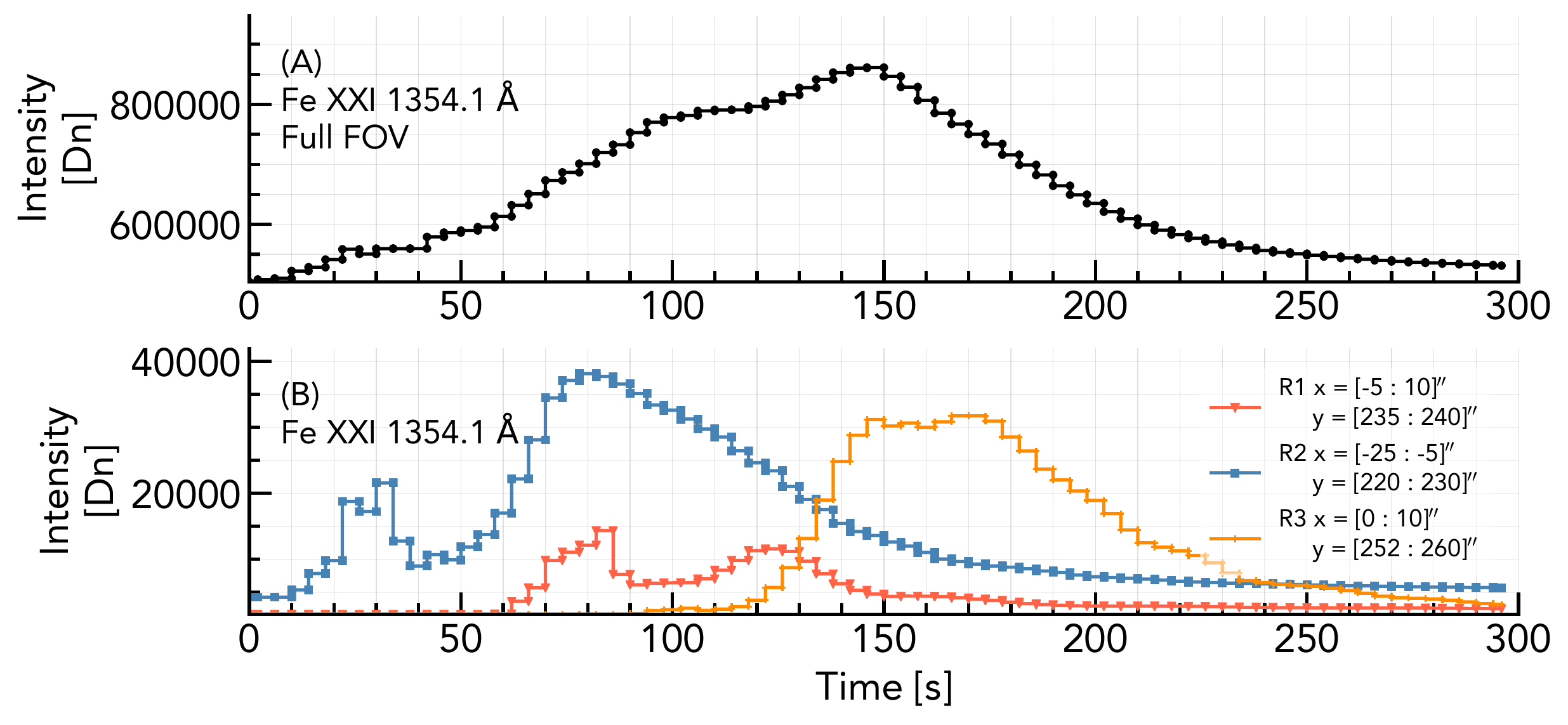}}	
	\caption{\textsl{Fe~\textsc{xxi} lightcurves from the full field of view (panel A), and from the three sub-regions (panel B) that are indicated on Figure~\ref{fig:flare_aia_map1}}}
	\label{fig:fexxi_arcade_lcurves}
\end{figure*}
Figure~\ref{fig:flareiris2} show a snapshot of synthetic IRIS Fe~\textsc{xxi} from AR11726rot. The maps are the emission integrated over the whole passband ($1352-1356$~\AA; recall that this is solely Fe~\textsc{xxi} plus continuum since we only generated emission from a single ion in this instance). Spectra from four representative slit positions show the varying intensities, widths, asymmetries and flows that result from different sources in the arcade. An animated version is available online. When the slit intersects a footpoint source the continuum brightens significantly. This can obscure the Fe~\textsc{xxi} signal, though it is possible to discern its presence in some pixels. When the continuum source decreases in intensity the Fe~\textsc{xxi} emission is much clearer. By this time the brightest continuum source in the footpoints has moved spatially along the slit. The Fe~\textsc{xxi} appears offset from the ribbon front for this reason. In some sources the Fe~\textsc{xxi} lines are easier to detect in the initial stages and a brief drift towards maximum Doppler shift is visible. These profiles are generally very weak (a few DN pixel$^{-1}$), and so the signal may become vanishingly small in real observations due to additional noise or the effect of spatial point spread functions. 

Figure~\ref{fig:samplepixels} shows individual line profiles from six pixels in AR11726rot (covering sources at footpoints, looptops, and on the loops). For each pixel, profiles from exposures 0 to 60 are shown in steps of two (effective cadence is then $8$~s), and zoomed in segment is included to make the weaker profiles easier to discern. An animated version is available online. Profiles are initially blueshifted and broad. Though sometimes symmetric, asymmetries are present. Over time the profiles drift back towards rest, increasing in strength, becoming narrower and more symmetric.  The characteristics and temporal evolution of the line profiles are qualitatively similar the behaviour seen in observations (with the exception of asymmetric profiles). The simulated intensities are of the correct magnitude compared to IRIS observations, suggesting that the temperatures and densities present in the model are consistent with the real flaring plasma.

Figure~\ref{fig:samplepixels2} shows in more detail the source at pixel $x = [-4.10, 240.05]^{\prime \prime}$, including the lightcurves of intensity and Doppler shift. The profile rapidly becomes blueshifted, but decays to rest over a short period of time of the order $\sim30$~s. Recall that the heating timescale was $t=25$~s. The lifetime of this source, from brightening to returning to near background level is on the order of $100$~s. This is shorter than the monolithic loop lightcurves (Figure~\ref{fig:1dfexxi_lcurve}) where the lifetime was closer to $200$~s.

Lightcurves of the full field of view, and of several sub-regions are shown in Figure~\ref{fig:fexxi_arcade_lcurves}.
Synthetic lightcurves of sub-regions containing footpoint emission (R1 and R2, red and blue curves) are qualitatively similar to the observed Fe~\textsc{xxi} lightcurve shown in Figure~\ref{fig:obsoverview}(C), which also contains footpoint emission. These exhibit a double peaked structure, the first peak being the footpoint sources low in the loop, and the second peak  due to ablation into the loop producing emission there. The region containing mainly loop or looptop sources (R3, orange line in Figure~\ref{fig:fexxi_arcade_lcurves}) only exhibits one peak. 
 
\subsection{Synthetic Fe~\textsc{xxi} Doppler Shifts}
To extract properties of the line profiles, Gaussian fits were made to the synthetic data. A single Gaussian function was fit to every pixel to determine the centroid, $\lambda_{\rm{c}}$, peak intensity, and standard deviation, $\sigma$. Photon counting noise was considered in each fit. A five term Gaussian function (background level with linear component, amplitude, centroid and standard deviation) was fit, to account for variations due to the continuum. The number of pixels ($251\times251\times75 = 4,725,075$ pixels for the $4$~s exposure data) precluded manually checking the quality of fit results, and so any fit with $\chi^{2} > 2$, or peak intensity $I_{\rm{peak}} < 5$~DN~pixel$^{-1}$ were omitted so as to avoid spurious data. Where $\chi^2 > 2$, a double Gaussian function was fit in case the single Gaussian fit was unsuccessful due to the presence of multiple components. Only a small proportion of pixels were deemed to be better fit by a double Gaussian function: $1.6$~\% in AR11726rot $\tau_{\mathrm{exp}} = 4$~s, and $2.2$~\% in AR11726rot $\tau_{\mathrm{exp}} = 8$~s. A somewhat larger proportion of longer exposure ($\tau_{\mathrm{exp}} = 8$) profiles exhibited double components, likely due to temporal smearing of profiles along the line of sight. Further, these profiles typically appeared when loops were more tightly clustered. 

Doppler shift(s) of the profiles were computed by $v_{\rm{Dopp}} = c~(\lambda_{\rm{c}} - \lambda_{\rm{rest}})/\lambda_{\rm{rest}}$ for speed of light $c$ and rest wavelength $\lambda_{\rm{rest}}=1354.0665$~\AA, the rest wavelength in CHIANTI v8.07, used for spectral synthesis in our arcade model. Note that while the rest wavelength has been suggested to be closer to $1354.1$~\AA\ \citep{2015ApJ...799..218Y}, based on IRIS observations, the CHIANTI value is still  $\lambda_{\rm{rest}}=1354.0665$~\AA.
 
For most of the duration of the flare in the \texttt{RADYN} simulation, plasma at temperatures that can form Fe~\textsc{xxi} exhibits mass flows acting against gravity. That is, mass motions are upflows along the flare loops that would be expected to produce blueshifted emission. Of course, the magnitude of this inferred flow would be modified by the inclination of loops and viewing angles. 

\begin{figure*}
	\centering 
	{\includegraphics[width = \textwidth, clip = true, trim = 0.cm 0.cm 0.cm 0.cm]{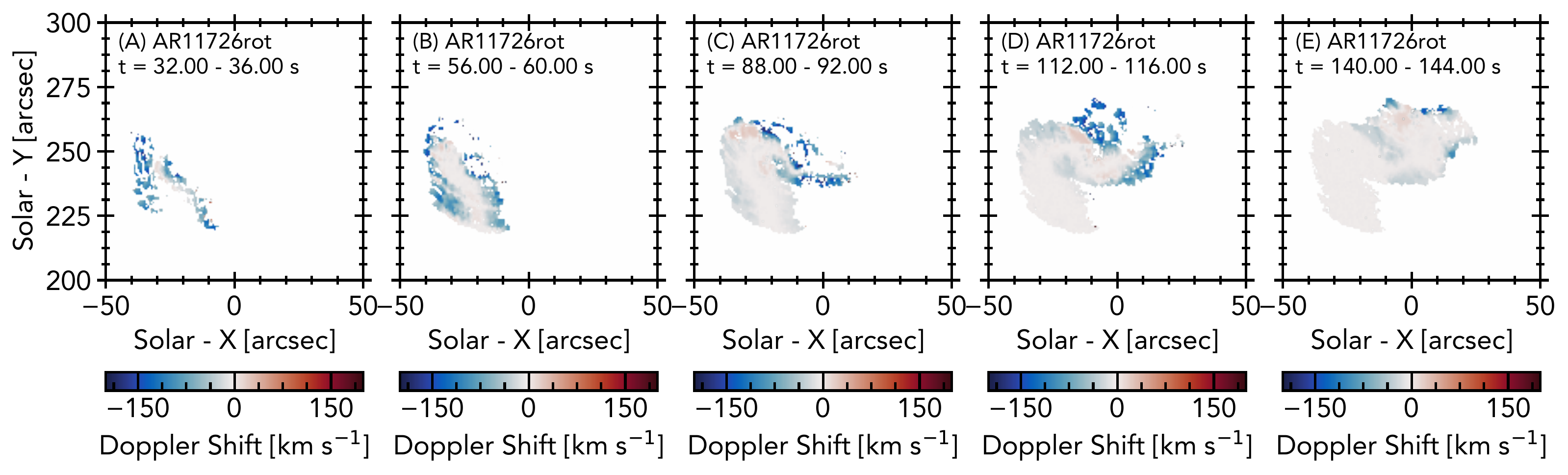}}	
	\caption{\textsl{Dopplergrams of Fe~\textsc{xxi} 1354.0665~\AA\ emission at various snapshots in the flare arcade simulation of AR11726rot. The background images are the integrated Fe~\textsc{xxi} line intensities (scaled by $I^{1/4}$, and $\alpha = 0.1$), to place the derived Doppler shifts in context. An animated version is available online.}}
	\label{fig:velmaps11726rot}
\end{figure*}

Both sides of the arcade model show blueshifted emission during the impulsive phase of each loop. Figure~\ref{fig:velmaps11726rot} shows several snapshots of the Doppler shift of the Fe~\textsc{xxi} 1354.1~\AA\ line. The integrated intensity of Fe~\textsc{xxi} is shown on each panel for context (intensity is scaled by $I^{1/4}$, and image $\alpha = 0.1$). Generally the strongest blueshifts are present around the edge of the flaring structure, so the footpoints of the loops, weakening with height along the loops. Flows appear first at the footpoints before the loops are filled in following chromospheric ablation. 

The magnitude of these blueshifts are consistent with observations of Fe~\textsc{xxi} and other high-temperature lines in flares, as is the morphology. Some small redshifts are present due to the reflecting upper boundary condition of the loops and draining of flare loops. They are associated with the gradual phase of each loop, and are small in magnitude. 

Though the field-aligned simulation contains larger upflows than those suggested by the Fe~\textsc{xxi} emission, the fastest upflows ($v>400$~km~s$^{-1}$ occur in hotter plasma where there is little or no Fe~\textsc{xxi}. The are also projection effects to take into consideration. The inclination of loops with respect to the line of sight effects the Doppler shift (and inferred upflow velocity). Even though each loop contained the same \texttt{RADYN} flare atmosphere there was a range of Doppler shifts present, apparent from the animated version of \ref{fig:velmaps11726rot} and the superposed analysis presented in Section~\ref{sec:superposed}. Having knowledge of the loop geometry can therefore be important for interpreting the Doppler shift results. 

\subsection{Synthetic Fe~\textsc{xxi} Line Widths}\label{sec:linewidths}
 
\begin{figure*}
	\centering 
	{\includegraphics[width = \textwidth, clip = true, trim = 0.cm 0.cm 0.cm 0.cm]{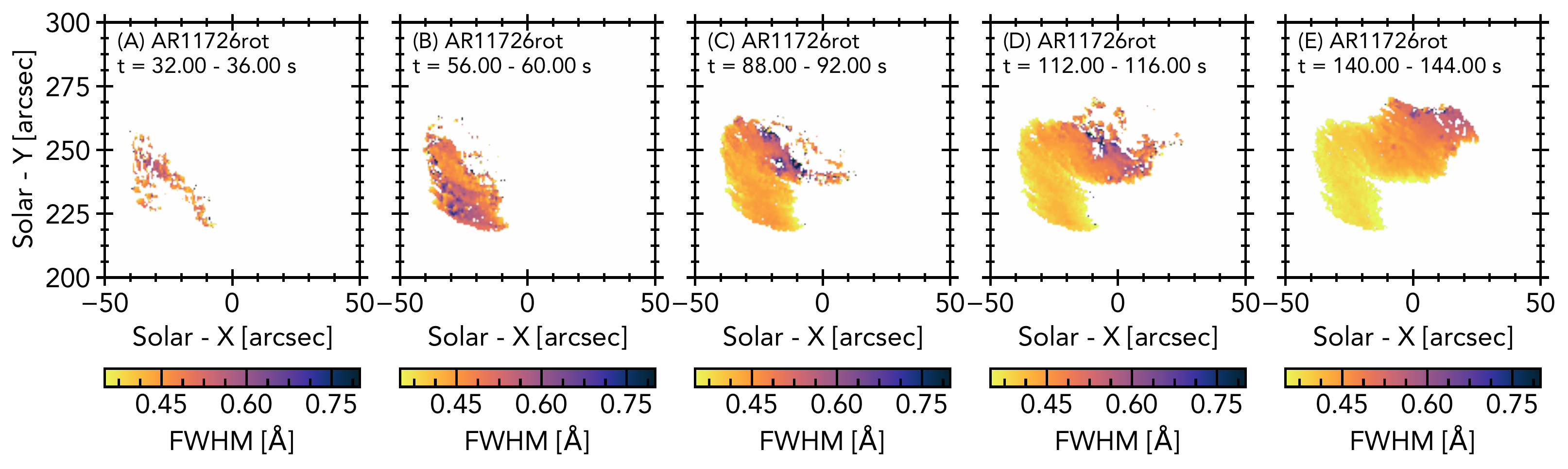}}	
	\caption{\textsl{Maps of Fe~\textsc{xxi} 1354.0665~\AA\ FWHM, obtained from Gaussian fitting, at various snapshots in the flare arcade simulation of AR11726rot. The background images are the integrated Fe~\textsc{xxi} line intensities (scaled by $I^{1/4}$, and $\alpha = 0.25$), to place the derived line widths in context.}}
	\label{fig:widthmaps11726rot}
\end{figure*}

\begin{figure}
	\centering 
	\hbox{
		\subfloat{\includegraphics[width = 0.5\textwidth, clip = true, trim = 0cm 0cm 0cm 0cm]{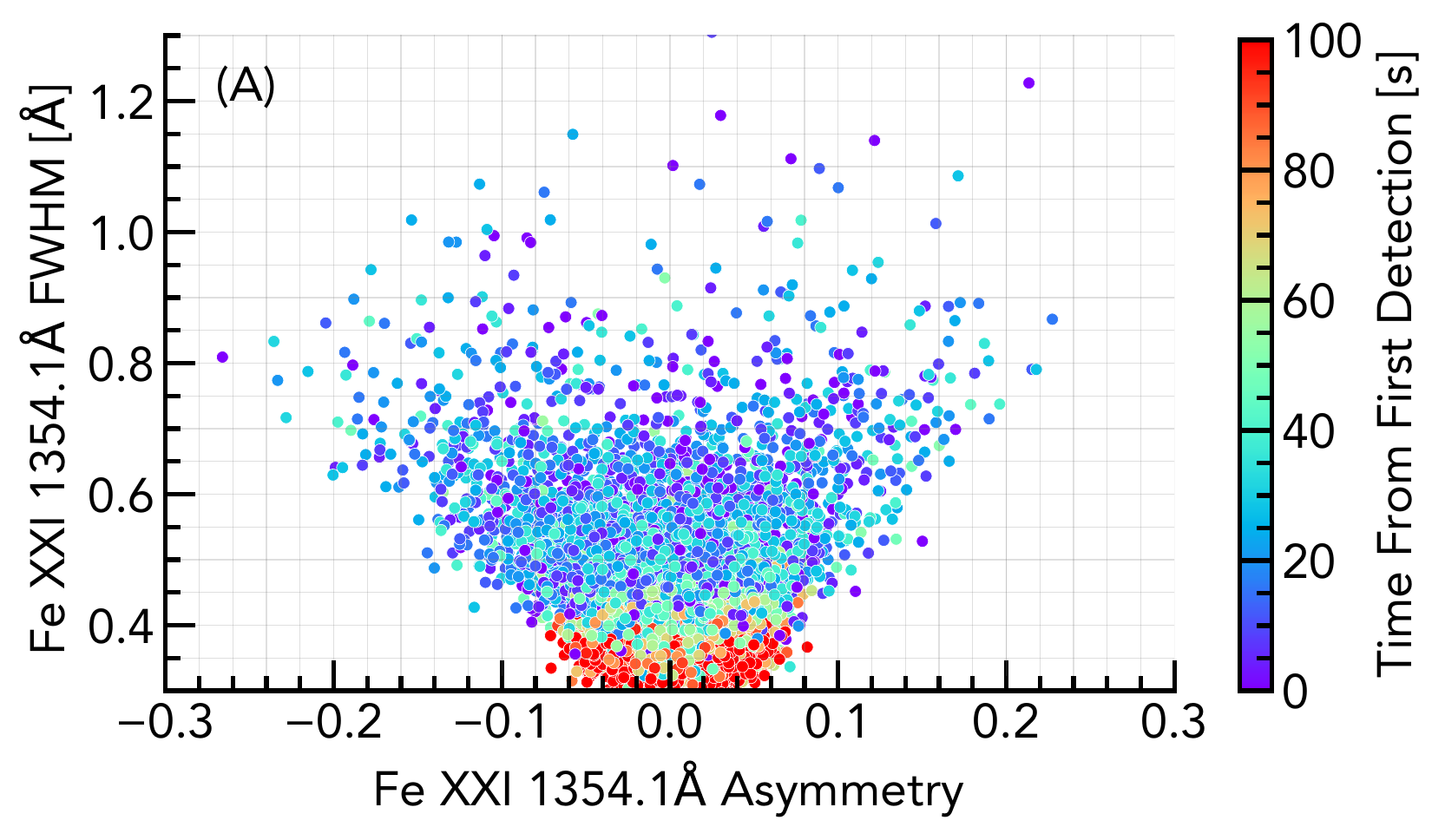}}	
		}
	\hbox{
	        \subfloat{\includegraphics[width = 0.5\textwidth, clip = true, trim = 0cm 0cm 0cm 0cm]{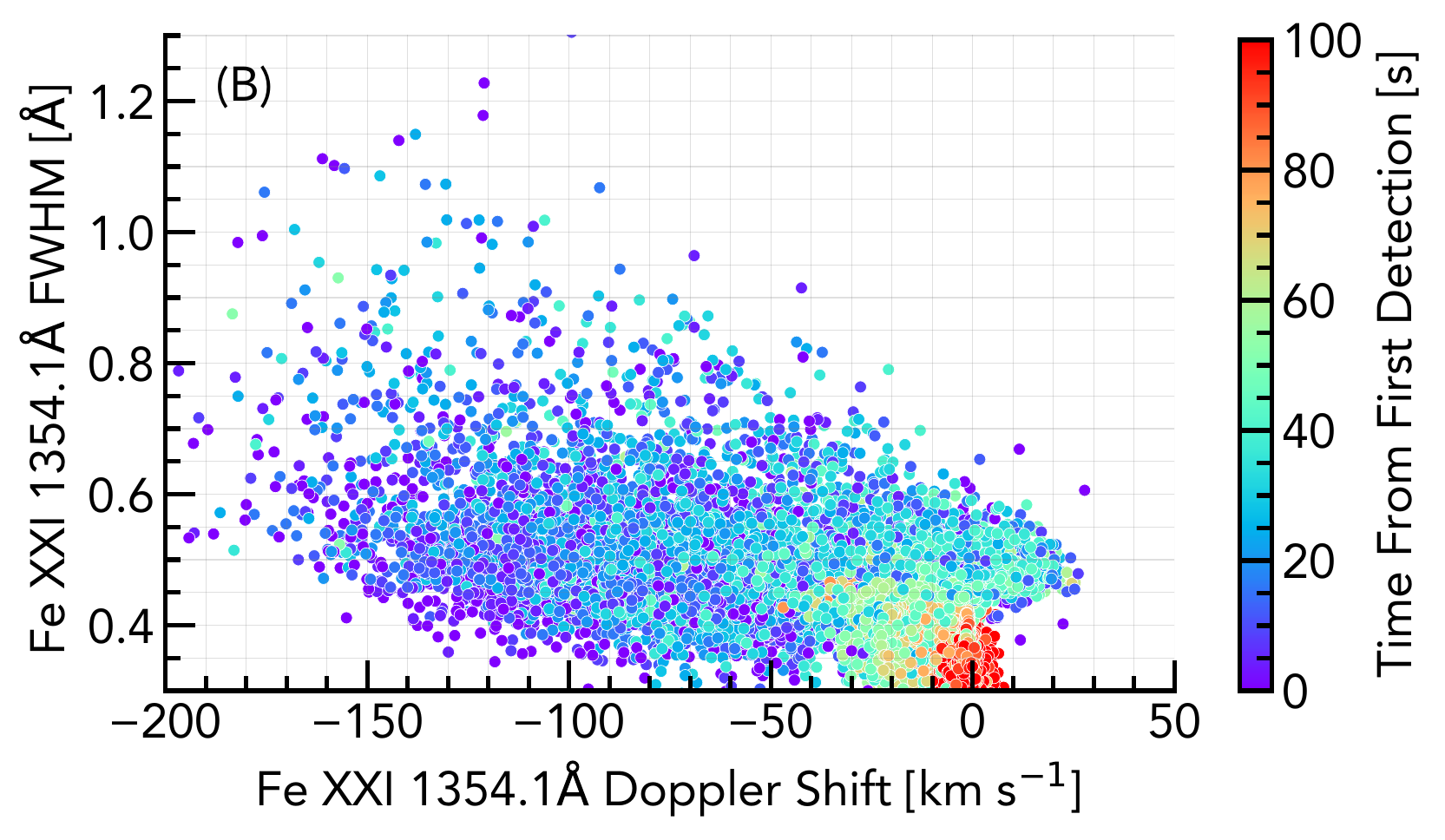}}
                   }
	\caption{\textsl{(A) The scatter of line width (FWHM) versus the asymmetry of the line, for AR11726rot and (B) the scatter of line width (FWHM) versus Doppler shift.Colour represents time since the line was first present with sufficient signal to perform a Gaussian fit.  }}
	\label{fig:widthcorrel}
\end{figure}
Presence of line widths in excess of the quadrature sum of the thermal and instrumental widths is usually referred to as non-thermal broadening. There are several candidates that can cause non-thermal broadening. Our model accounts for two broadening mechanisms: thermal and superposition of loops. For thermal broadening we assume that the electron and ion temperatures are equal, and we assume statistical equilibrium since we used equilibrium ionisation fractions in CHIANTI when calculating the contribution functions. If non-equilibrium ionisation effects are significant, and Fe~\textsc{xxi} forms in hotter plasma assumed by statistical equilibrium then we are likely underestimating thermal broadening. 

Though Fe~\textsc{xxi} has a contribution function peaking at $11.2$~MK, density increases following ablation of material into the flare loop means that there can be sufficient emission measure at temperatures in excess of this peak to produce detectable emission. This would increase the width of the line beyond the nominal thermal width. Similarly, Fe~\textsc{xxi} can form below this peak temperature so that the line width may drop below the nominal thermal width. Using the temperatures at which $G(n_{e}, T) > G_{\rm{peak}}/4$ as a guide then then a reasonable range of thermal widths can be on the order: $T\sim[8.08 - 16.70]$~MK, $W_{\rm{thm}} = [0.37 - 0.53]$~\AA, $\sigma = [0.16 - 0.23]$, and $v_{\rm{thm}} = [81.7 - 117.4]$~km~s$^{-1}$.

Superposition of multiple sources along the line of sight is accounted for by our arcade model. This will increase the line width as profiles experiencing different plasma motions will sum together. While this contributes towards enhanced line widths this effect is likely to produce asymmetric profiles, unless viewing angles and loop geometry were unusually ideal. Indeed, \cite{2019ApJ...879L..17P} demonstrated through loop modelling that accounted for superposition of loops, that this could broaden Fe~\textsc{xxi}, but not symmetrically. They were unable to produce both very broad and symmetric profiles. Another broadening candidate would be required to explain the symmetry. 

Gaussian FWHM are shown in Figure~\ref{fig:widthmaps11726rot}. A movis is available online. From these maps it is clear that while some pixels exceed values of $W \sim [0.8-1]$~\AA, the majority of profiles are broadened to values of $W\sim[0.5-0.6]$~\AA\ only. Qualitatively these maps do show what we expect. The broadest profiles are near footpoints, with width decreasing through the flare loop. 

Asymmetries, $A_{\mathrm{RB}}$, were measured using the same approach as \cite{2019ApJ...879L..17P} \citep[and following][]{2009ApJ...701L...1D,2011ApJ...738...18T}. $A_{\mathrm{RB}} = \frac{I_{\mathrm{R}} -  I_{\mathrm{B}}}{ I_{\mathrm{P}}}$, where $I_{\mathrm{P}}$ is peak intensity,  $I_{\mathrm{R/B}} = \Sigma_{+/-\lambda_{1}}^{+/-\lambda_{2}} I_{\lambda}/n$, $\lambda_{1}  = 50$~km~s$^{-1}$ and $\lambda_{2}  = 150$~km~s$^{-1}$. Figure~\ref{fig:widthcorrel}(a) shows the correlation between line width and asymmetry. While there is no strong correlation here (broad profiles can be both very asymmetric or symmetric) it is clear that when profiles are asymmetric they are broad, in agreement \cite{2019ApJ...879L..17P}. In that figure colour represents time since first detection. The broadest, more asymmetric profiles occur early in each loop. Maps of asymmetry show larger asymmetry in newly activated loops, when flows are strongest, consistent with observations analysed by \cite{2008ApJ...679L.155I} of cooler Fe~\textsc{xiv} lines. Figure~\ref{fig:widthcorrel}(b) shows the correlation between line width and Doppler shift. While there is a relation, the correlation is not very strong, similar to observations of \citep[e.g.][]{2011ApJ...740...70M}. 

We have also not considered turbulence, broadening by Alfv\'enic waves, or non-equilibrium ionisation \citep[where Fe~\textsc{xxi} could actually form in hotter plasma, e.g.][]{2017SoPh..292..100D}, which will feature in follow up work. \cite{2019ApJ...879L..17P} contains a more detailed summary of potential non-thermal broadening mechanisms.

\subsection{Synthetic Fe~\textsc{xxi} Superposed Epoch Analysis}\label{sec:superposed}
To characterise the response of all of the line profiles in the arcade simulation a superposed epoch analysis was performed. This can highlight any commonalities between the response of individual sources, and provide a statistical overview of the flare. Such an analysis can also be compared to observational examples presented in Figure~\ref{fig:superposedepoch_observed} and in \cite{2015ApJ...807L..22G}.

\begin{figure*}
	\centering 
	\hbox{
	        \subfloat{\includegraphics[width = 0.5\textwidth, clip = true, trim = 0cm 0cm 0cm 0cm]{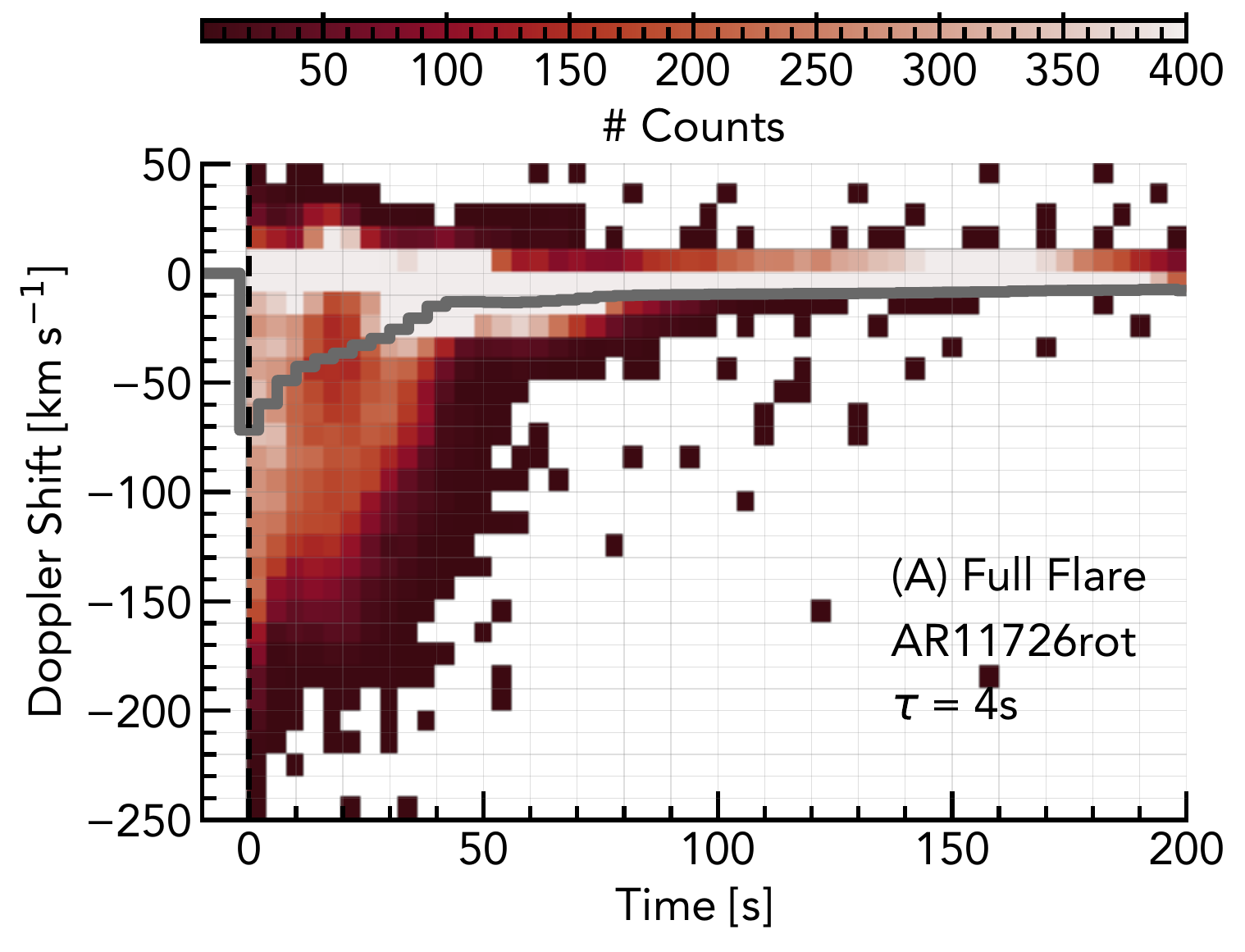}}
	        	\subfloat{\includegraphics[width = 0.5\textwidth, clip = true, trim = 0cm 0cm 0cm 0cm]{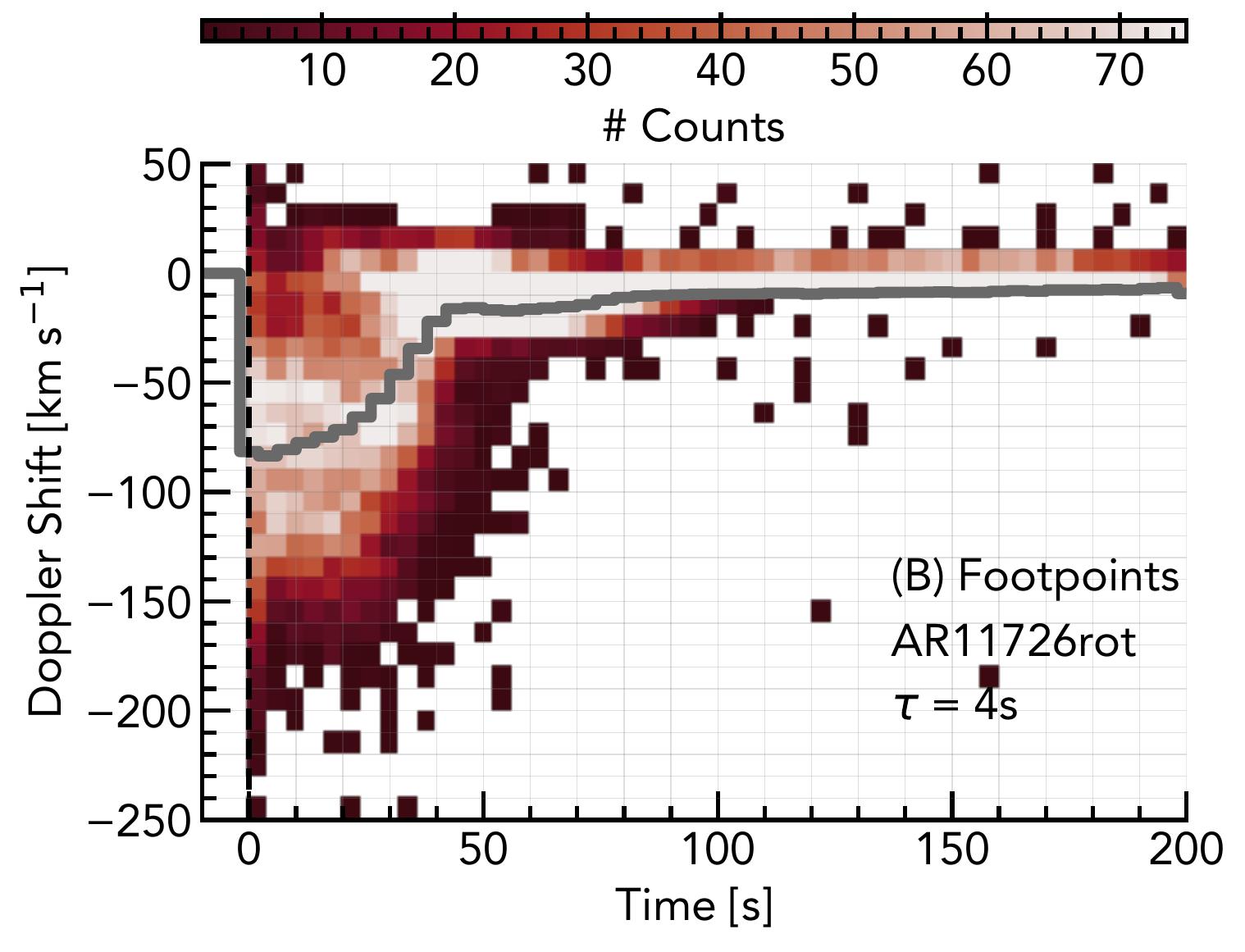}}	
                   }
          \hbox{
	        \subfloat{\includegraphics[width = 0.5\textwidth, clip = true, trim = 0cm 0cm 0cm 0cm]{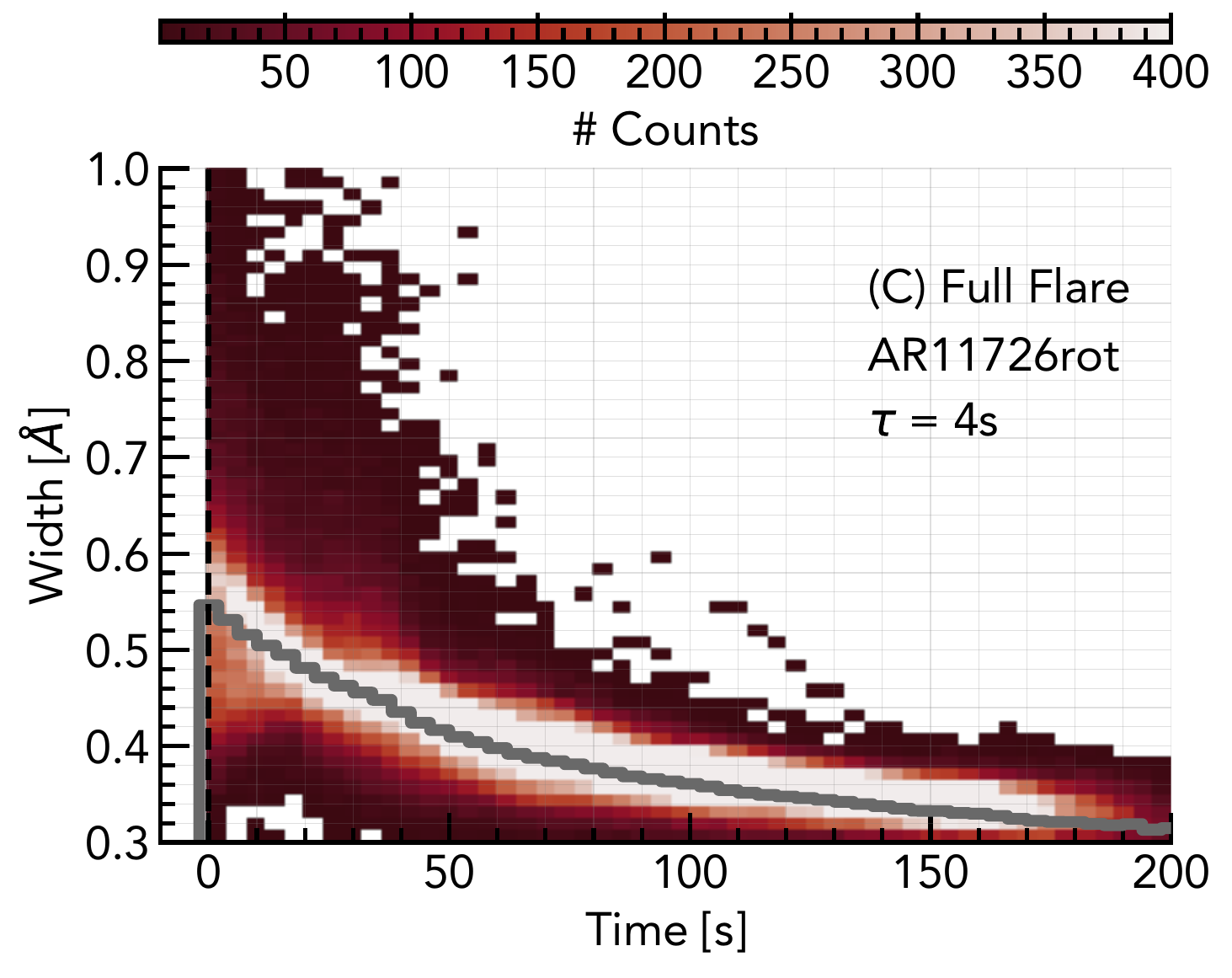}}	
	        	\subfloat{\includegraphics[width = 0.5\textwidth, clip = true, trim = 0cm 0cm 0cm 0cm]{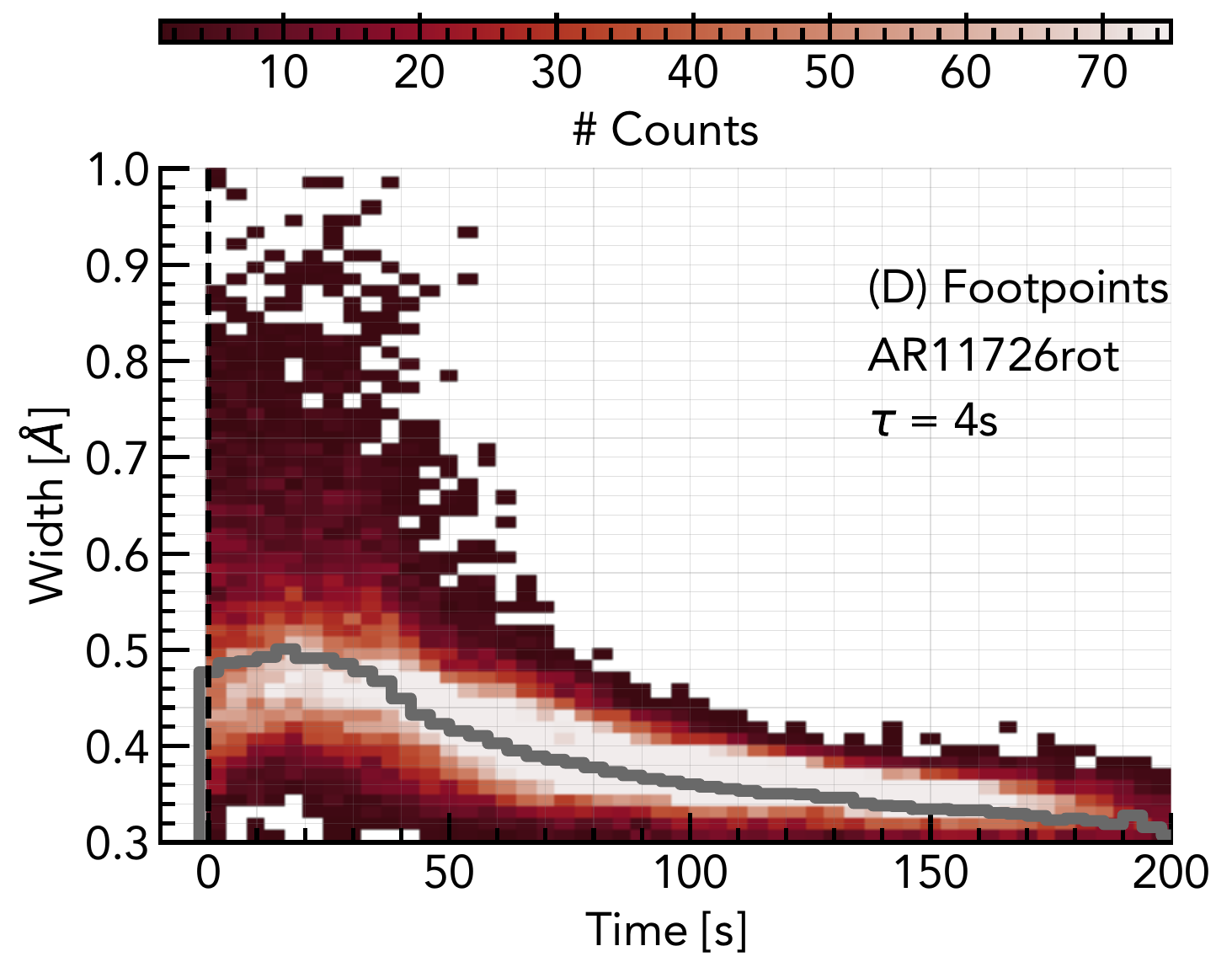}}	
                   }
          \hbox{
		\subfloat{\includegraphics[width = 0.5\textwidth, clip = true, trim = 0cm 0cm 0cm 0cm]{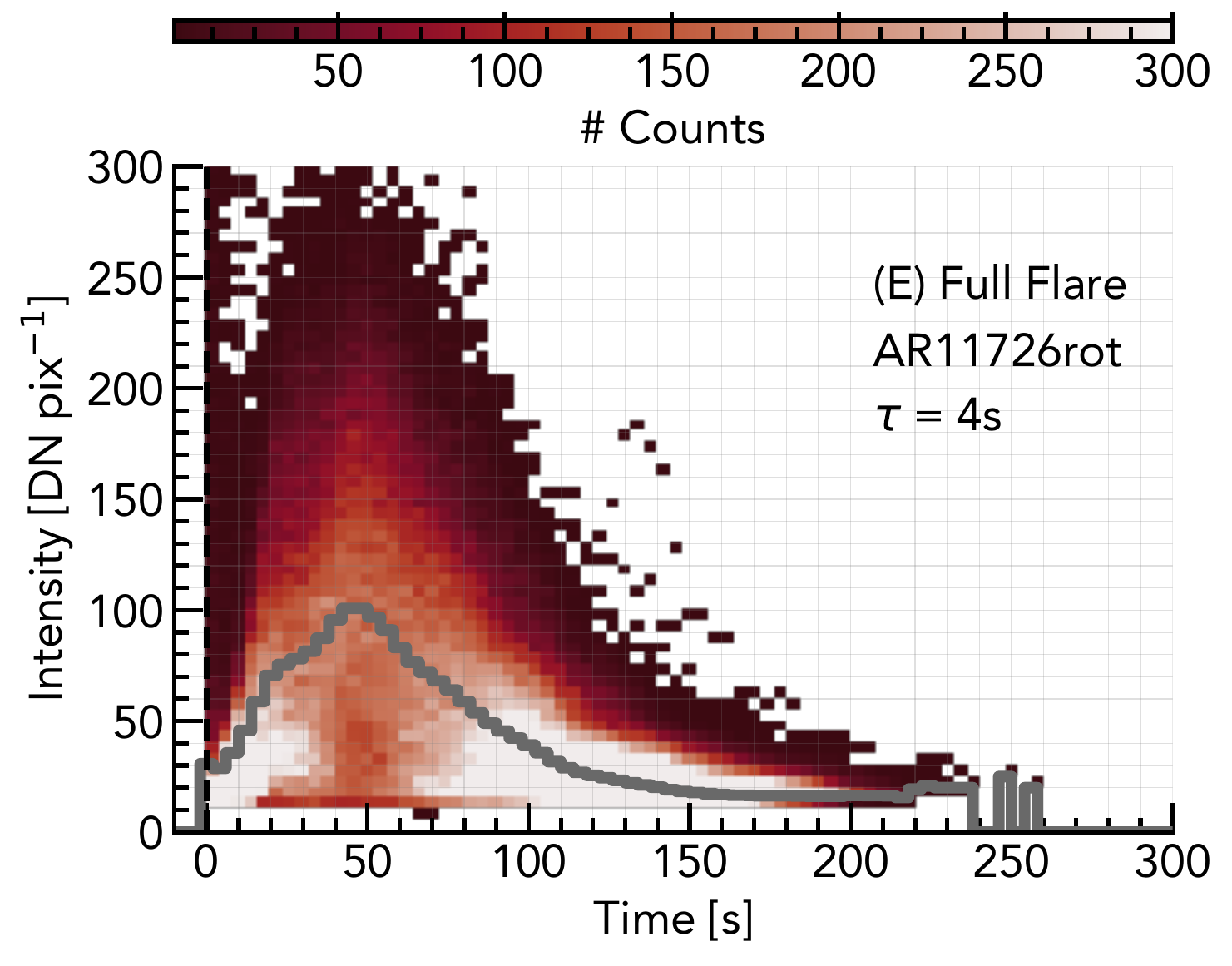}}	
		\subfloat{\includegraphics[width = 0.5\textwidth, clip = true, trim = 0cm 0cm 0cm 0cm]{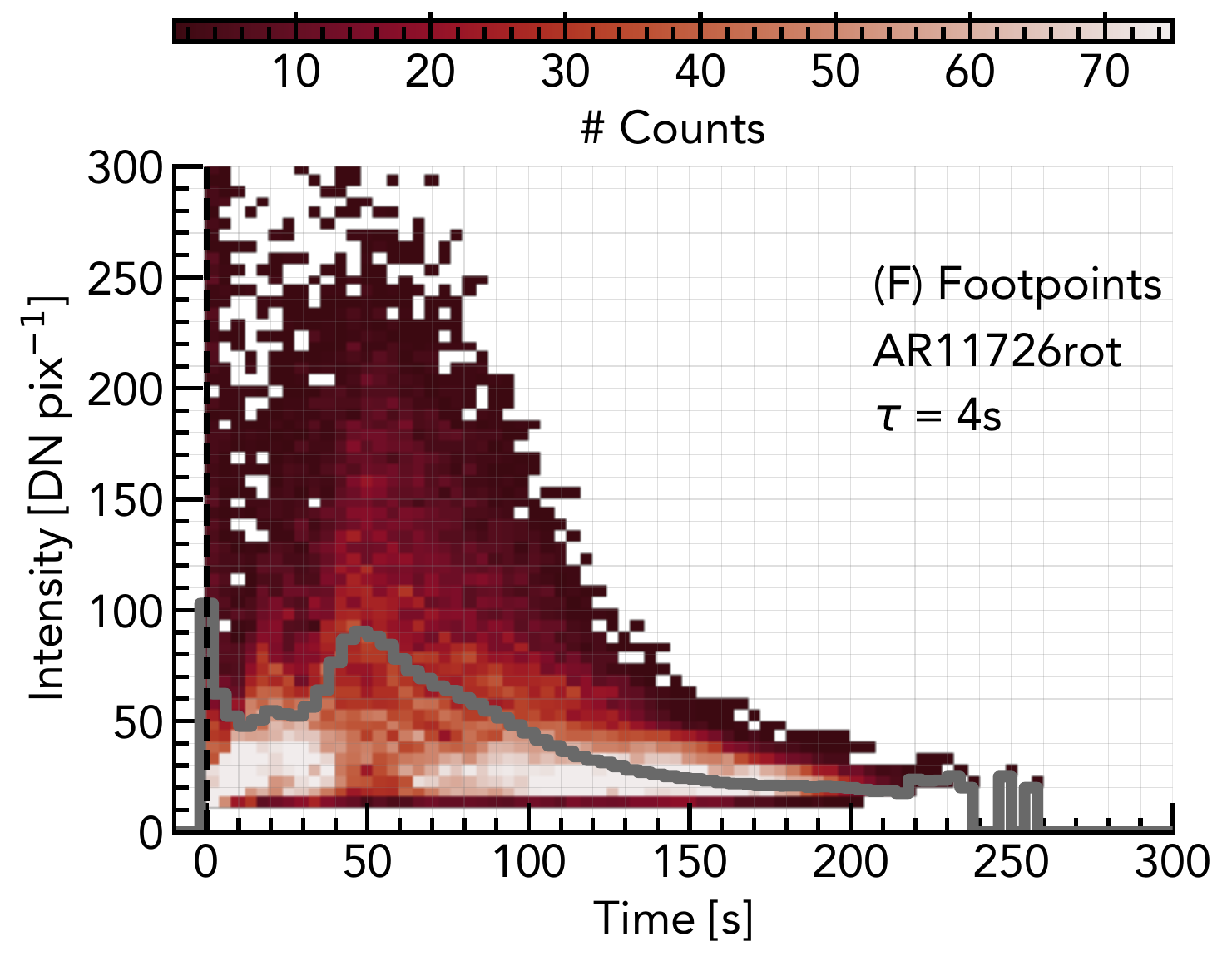}}	
		 }
	\caption{\textsl{Superposed epoch analysis of Fe~\textsc{xxi} line for the full flare (left hand column), and for footpoint sources only (right hand column). The first row is the Doppler shift, and the second row is the line widths, and the third row is integrated line intensity,. In each panel the weighted mean value is shown as a solid grey line.}}
	\label{fig:superposedepoch_flare}
\end{figure*}

The temporal origin of each pixel was the moment of first detection (first successful Gaussian fit to the data). This analysis was performed on the full flare (all flaring pixels) and separately on pixels identified as footpoints. Figure~\ref{fig:superposedepoch_flare} shows the superposed epoch analysis where the left hand column (A,C,E) are the full flare, and the right hand column (B,D,F) are footpoints only.  

For all line properties the time bins were $\delta t = 4$~s. The integrated line intensity used a binning of $\delta I = 10$~DN~pix$^{-1}$, the Doppler shift used a binning of $\delta v = 10$~km~s$^{-1}$, and the line widths used a binning of $\delta W = 12.5$~m\AA.

Our arcade model shows a larger spread of Doppler shifts than the observational analysis indicates. Further, observations have shown that blueshifts persist for several hundred seconds. In our model the Doppler shifts sharply decline over several tens of seconds and are only modest by 100s. Footpoint sources have a somewhat more sustained blueshift than the arcade overall. Mass flows from pixels in the loops are smaller in magnitude and shorter in duration than those from the footpoint pixels. The temporal behaviour of the line widths tracks that of the Doppler shifts. Footpoints exhibit an initially slower decay, followed by a decline to pre-flare values. When including loop and loop top sources the decay is very smooth. Together, the larger spread of Doppler shifts and rapid cessation of Doppler shifts suggests that we are missing some aspect of the heating process (either in construction of the arcade or the underlying \texttt{RADYN} modelling).

As expected the intensities take some time to peak relative to the Doppler shifts and line widths, about $t\sim50s$. The initial spikes there are due to the intense continuum in footpoint sources.  


\section{Summary and Conclusions}\label{sec:line_ratios}

We have created a data-constrained flare arcade model by grafting the results of a state-of-the-art flare loop simulation onto observed active region loop structures. This bridges the gap between advanced 1D loop models that can capture the NLTE, non-local, radiation hydrodynamics on spatial scales appropriate for flares, and 3D models that captures effects such as loop geometry, superposition of loops, and viewing angle on the solar disk. While this initial arcade model is rather simplified, it sets a framework for us to investigate both individual flare sources, as well as the global flare and stellar flares. 

Synthetic observables from SDO/AIA, GOES/XRS, and the IRIS spacecraft were forward modelled taking into account instrumental effects where appropriate. This illustrates the utility of this arcade model as a means to facilitate a more accurate model-data comparison for coronal emission in flares. 

Some specific summary points are: 

\begin{itemize}

\item The morphological characteristics of flares are well represented by our arcade model. While increases in the emission measures at temperatures in excess of $>10$~MK first appeared near loop tops, these were too small to produce observable SDO/AIA or IRIS radiation. Radiation sampling plasma $>10$~MK is therefore first observed near the footpoints of loops. Following ablation of material into the the loop and loop tops, emission measures became significantly stronger and radiation was observed in those locations. The synthetic AIA movies show the ablation process. As loops cool they become visible in passbands that sample cooler plasma. The brightest emission is initially the footpoints/ribbons, switching to the post-flare loops in the gradual phase.

\item Synthetic GOES/XRS lightcurves were qualitatively similar to observations, with an a steep, impulsive rise phase with a slower, more gradual decay phase. The temperature and EM from GOES are what we would expect from flares. However the decay timescales imply a very close ribbon distance \citep[based on ][]{2017ApJ...851....4R}. Our model does not include ribbon separation or increasing loop length, both of which could lengthen the decay phase. Combined with synthetic AIA lightcurves, it is clear that the cooling timescale of our flare loops from \texttt{RADYN} is too rapid. 
 
\item Synthetic Fe~\textsc{xxi} emission also shows a qualitative match to observations, both the images and spectral behaviour. The magnitude and location of Doppler shifts, and line intensities, were consistent with observations. However, the lifetime of Doppler shifts was too short. Compared to observed superposed epoch analysis, there is a much larger spread of Doppler shifts in the model. Observations are instead tightly clustered. Though line broadening occurred, the line widths were too narrow suggesting that additional physics is required in the model to broaden the lines. 

\end{itemize}

\rev{Based on our experience running electron beam driven simulations (and that of others, c.f. works referenced in Section~\ref{sec:intro}) we do not believe that varying the non-thermal electron beam parameters (flux, low-energy cutoff, spectral index) will by themselves produce upflow durations more consistent with observations. Instead the answer likely lies in the modelling approach, or in improvements to the physics of the model}. Planned improvements over this initial work include modelling the loop structures and flare evolution from an actual flare, inclusion of ribbon separation, including varied loop lengths, varying the electron beam properties injected onto the loops, and performing multi-threaded modelling \citep[e.g][]{2018ApJ...856..149R}. Going beyond the standard model electron beam scenario, we will include return currents and proton beams \citep[using the recently developed \texttt{FP} code merged with \texttt{RADYN},][]{FP_inprep}).

Multi-threaded modelling by \cite{2018ApJ...856..149R} has been able to achieve Doppler shifts with durations more consistent with observations. However, this requires sustained energy injection (of 60--200s) into a single atmospheric volume, over many individual threads within a single IRIS pixel. There is no clear explanation as to why energy deposition into one location would last this long, given reconnection timescales. Indeed, the timescales associated with redshifts of chromospheric spectral lines (chromospheric condensations) forward modelled in single threaded \texttt{RADYN} simulations show much closer consistency with observations than the timescales of upflows \citep[e.g.][]{2020ApJ...895....6G}. \rev{Further, \cite{2020ApJ...895....6G} used IRIS ultraviolet observations to show that the energy injection timescale was on the order of $\approx20$~s}.  We therefore believe that as well as pursuing a multi-threaded approach, we should investigate the physics of the flare heating and cooling also. 

Recent updates to non-thermal particle transport in \texttt{RADYN} include a self-consistent treatment of the beam induced return current \citep{FP_inprep}. Initial results by \citep{FP_inprep} indicate that the heating rates can be significantly modified by the return currents. Additionally, \cite{2018ApJ...865...67E} have showed that thermal conduction can be suppressed by turbulence and non-local effects, and provided a mechanism to include this in code such as \texttt{RADYN}. This could have implications for both flare impulsive and gradual phase dynamics and associated timescales. We have began to adapt \texttt{RADYN} to include suppression of conduction. Post-impulsive phase heating has been suggested as an explanation for the longer than expected flare cooling times \citep[e.g.][]{2016ApJ...820...14Q,2018ApJ...856...27Z}. This can be investigated in our simulations in combination with suppression of conduction. 

With regards to line broadening, non-equilibrium ionisation \rev{would result in a different Fe ionisation stratification that that predicted by our assumption of ionisation equilibrium}. \rev{This could} result in ions forming in plasma significantly hotter than the equilibrium formation temperatures, with a correspondingly larger thermal width. These effects can be investigated, for example by using the minority species version of \texttt{RADYN} \citep{2019ApJ...871...23K,2019ApJ...885..119K} to obtain non-equilibrium Fe ion fractions, \rev{or by applying our framework with flare atmospheres from other codes, such as \texttt{HYDRAD} \citep{2003A&A...401..699B,2019ApJ...871...18R} that can model NEI ion fractions}. Including ad-hoc micro-turbulence, and investigating the potential of broadening via Alfv\'enic waves \citep[estimating broadening using the models of ][]{2016ApJ...827..101K,2018ApJ...853..101R} are other avenues to pursue, as suggested by \cite{2019ApJ...879L..17P}.

While we have used data-constrained loop structures here, and plan to do this for a flaring active region, there is nothing to preclude our model being used with artificial loop structures either from a toy-model or originating from magnetohydrodynamic (MHD) flare and CME models. We would be keen to collaborate with MHD modellers, to simulate observables from those codes with the more accurate thermodynamics ((and non-thermal particle beams) available from using \texttt{RADYN} and our arcade modelling approach. \\

\textsc{Acknowledgments:} \small{GSK was funded by an appointment to the NASA Postdoctoral Program at Goddard Space Flight Center, administered by USRA through a contract with NASA. JCA acknowledges funding support from the Heliophysics Supporting Research and Heliophysics Innovation Fund programs. VP acknowledges support by NASA grant 80NSSC20K0716. {\it IRIS} is a NASA small explorer mission developed and operated by LMSAL with mission operations executed at NASA Ames Research center and major contributions to downlink communications funded by the Norwegian Space Center (NSC, Norway) through an ESA PRODEX contract. We thank Dr. J. Reep for assistance with the GOES-15 spectral responses. This research benefited from discussions held at a meeting of GSK's and VP's International Space Science Institute team: `Interrogating Field-Aligned Solar Flare Models: Comparing, Contrasting and Improving'}

\bibliographystyle{aasjournal}
\bibliography{Kerr_etal_Flare3D_FeXXI}

\end{document}